\documentclass{article} % For LaTeX2e
\usepackage{iclr2027_conference,times}

\usepackage{graphicx}
\usepackage{booktabs}
\usepackage{amsmath}
\usepackage{amssymb}

\usepackage{array}
\usepackage{multirow}
\usepackage{tabularx}
\usepackage{makecell}
\usepackage{longtable}
\usepackage{titletoc}
\usepackage{adjustbox}

\usepackage{algorithm}
\usepackage{algorithmic}

\usepackage{newfloat}
\usepackage{listings}

\usepackage{xurl}
\usepackage[hidelinks]{hyperref}
\usepackage{url}

\usepackage{microtype}
\usepackage{wrapfig}

\usepackage[most]{tcolorbox}

\definecolor{contribbg}{RGB}{230,240,252}
\definecolor{contribframe}{RGB}{60,110,180}

\newtcolorbox{contributionbox}{
  colback=contribbg,
  colframe=contribframe,
  boxrule=0pt,
  arc=3mm,
  left=3mm,
  right=3mm,
  top=2mm,
  bottom=2mm,
  boxsep=0pt,
  before skip=6pt,
  after skip=6pt,
}

\title{Vision-Language Model Confidence Is Not a Property of the Answer}

\author{
\begin{tabular}{@{}l@{}}
\\[-2pt]
\multicolumn{1}{c}{} \\[-2pt]
{\bfseries Reza Khanmohammadi}\textsuperscript{\normalfont 1,*}
\quad
{\bfseries Erfan Miahi}\textsuperscript{\normalfont 2}
\quad
{\bfseries Ivan Brugere}\textsuperscript{\normalfont 3}
\quad
{\bfseries Charese H. Smiley}\textsuperscript{\normalfont 3}
\\
{\bfseries Simerjot Kaur}\textsuperscript{\normalfont 3}
\quad
{\bfseries Kundan Thind}\textsuperscript{\normalfont 4}
\quad
{\bfseries Mohammad M. Ghassemi}\textsuperscript{\normalfont 1}
\\[4pt]
{\normalfont
\textsuperscript{1}Michigan State University
\quad
\textsuperscript{2}Independent AI Researcher
}
\\
{\normalfont
\textsuperscript{3}JPMorgan AI Research
\quad
\textsuperscript{4}Henry Ford Health
}
\\[3pt]
{\normalfont\small
\texttt{\{khanreza,ghassem3\}@msu.edu}
\quad
\texttt{mhi.erfan1@gmail.com}
}
\\
{\normalfont\small
\texttt{\{ivan.brugere,simerjot.kaur,charese.h.smiley\}@jpmchase.com}
}
\\
{\normalfont\small
\texttt{kthind1@hfhs.org}
}
\quad
{\normalfont\small
\textsuperscript{*}Corresponding author
}
\end{tabular}
}

\iclrfinalcopy % Uncomment for camera-ready version, but NOT for submission.
\begin{document}

\maketitle

% ========================= MAIN SECTIONS =========================
\vspace{-15pt}
\begin{abstract}
Vision-language models are increasingly deployed behind a confidence gate: the system reads how confident the model is in its answer and defers when that confidence is low. This makes the confidence signal itself worth attacking. We show that a white-box adversary who perturbs only the input image, within an $L_\infty$ budget of $8/255$ and under the constraint that the model's answer stay byte-identical, can invert the confidence ranking, lowering it on correct answers and raising it on wrong ones until the signal points the wrong way. The answer never changes, and most of the inversion persists even when the whole next-token distribution is held near the clean one, so the answer does not determine the confidence attached to it. The quantity we hold fixed and the quantity that inverts are different: confidence is a separate signal read from the same network, and it can be corrupted on its own. A gate reading it is turned against itself, made to reject good answers and to accept the wrong ones it was built to catch. Across four vision-language models and three visual question-answering benchmarks, the attack drives the model-internal readouts we test below chance in 83 of 84 readout-by-cell profiles under an adversary that knows which answers are correct; for the two readouts carrying a disjoint calibration reference, it falls below chance under an adversary that does not. Training a probe on the model's frozen hidden states does not fix this: the robustness such a probe gains against the attack is paid for with the information that made it useful. Nor does reading confidence from a separate, independently trained model, which holds up only until the attacker reaches it and then falls into the same regime as the internal signals. How far an answer-preserving adversary can reach a signal governs where it survives; we offer this as a characterization of where the signal holds, and whether a robust and informative readout can be built remains open. For deployment, a gate under this attack can admit most of the wrong answers it would otherwise catch and, corrected for how often the model is wrong, can leave the system worse off than using no gate at all. 

\par\noindent\makebox[\linewidth][c]{%
\includegraphics[height=1em]{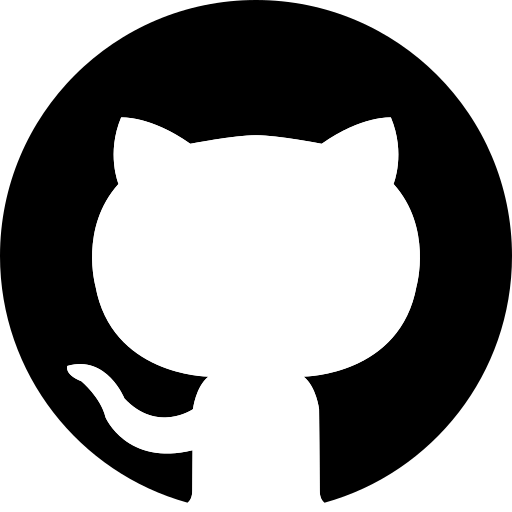}%
\hspace{0.35em}%
\textbf{Code: }%
\href{https://github.com/Ledengary/vlm-confidence-attacks}{%
\texttt{github.com/Ledengary/vlm-confidence-attacks}}%
}
\end{abstract}
\vspace{-10pt}
% \begin{center}
% \includegraphics[height=1em]{figures/github-logo.png}
% \textbf{Code: }
% {\ttfamily
% \href{https://github.com/Ledengary/vlm-confidence-attacks}
% {\small https://github.com/Ledengary/vlm-confidence-attacks}}
% \end{center}
% \vspace{-5pt}
\begin{figure}[h]
  \centering
  \includegraphics[width=\linewidth]{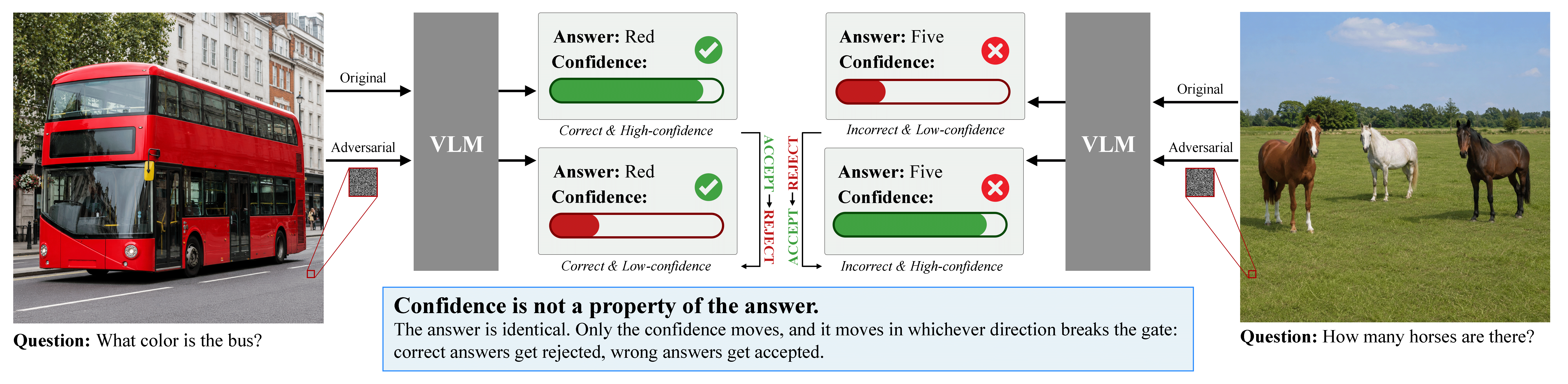}
  \vspace{-20pt}
  \caption{An answer-preserving confidence attack. A perturbation to the input image, bounded within an $L_\infty$ budget and constrained to leave the model's answer byte-identical, moves the confidence the model reports in that answer. On a correct answer the attack drives confidence down until a gate that would have accepted the answer rejects it; on a wrong answer it raises confidence so the gate accepts it. The answer does not change, and that is enough to reverse the gate's decision.}
  \label{fig:teaser}
  \vspace{-10pt}
\end{figure}

\section{Introduction}
\label{sec:intro}

{\looseness=-1
A vision-language model returns an answer, but a deployed system rarely takes that answer on faith. It reads off a confidence score and uses it to decide whether to accept the answer, defer to a human, or abstain. This gate is only worth having if the score it reads is hard to fool. We show that it is not. An adversary who perturbs only the input image, within an $L_\infty$ budget of $8/255$, and who is forbidden from changing the model's answer by a single token, can still invert the confidence ranking: driving it down on answers that happen to be correct, so the gate rejects them, and up on answers that are wrong, so the gate waves them through. The answer is held byte-identical throughout. Only the model's stated trust in that answer moves, and it moves far enough to reverse the gate's decision (Figure~\ref{fig:teaser}).
\par}

With white-box gradients one can move any differentiable quantity, so the point is what stays fixed while the score moves. The answer is a hard constraint, fixed exactly across a full generated sequence, yet the confidence attached to it can be pushed down on correct answers and up on wrong ones until the two are anticorrelated. A quantity that can be steered this freely while the behavior it describes stays fixed is a separate signal, read from the same network, that can be corrupted on its own. Prior work established that confidence can be manipulated with the prediction held fixed, for image classifiers \citep{galil2021ace,obadinma2024calibration} and, through recoverable and steerable confidence directions, for text language models \citep{miao2026closing,ji2025verbal}. That a fixed-prediction confidence attack exists is therefore settled, and we do not add one. The question those results leave open is which confidence signals survive such an attack and why. In open-ended vision-language generation, with the entire answer preserved instead of a single class label, we show that confidence is separable from the answer it reports, identify the two mechanisms through which an answer-preserving attack corrupts it, and measure how the natural external remedy degrades as the attacker's access grows.

The readouts read from inside the model fail almost without exception. Across four models and three benchmarks, the attack drives the internal readouts below chance in 83 of 84 readout-by-cell profiles under an adversary with correctness labels, and the two readouts that carry a disjoint calibration reference collapse under an adversary without them. Training a sturdier signal fails in a way that explains why: what makes a probe more robust to the attack is the loss of the information that made it useful, so robustness and informativeness never arrive together. Moving the check into a separate, independently built model only relocates the problem. Such a verifier survives while the attacker cannot reach it, then falls as the attacker gains query access and then gradient access, until with full access it collapses as completely as the internal signals. Reachability, how far an answer-preserving adversary can move a signal, is the axis along which we locate where the separable signal survives.

{\looseness=-1
Two further results fix what the inversion is and what it costs. Holding the answer byte-identical pins only the top token at each position, so one might suspect the attack lives in the slack the rest of the next-token distribution leaves free; bounding the KL divergence of the full distribution at every answer position to $0.2$ nats leaves $62\%$ of the inversion intact. The cost is measured on a gate a deployment would run: a hidden-state gate with its threshold fixed to a $10\%$ false-accept rate on clean data admits up to $84.8\%$ of the wrong answers it had rejected, and once each benchmark is reweighted to the model's natural error rate, the gated system is less accurate than accepting every answer; on the balanced evaluation pool, accepted accuracy on InternVL/POPE falls from $0.744$ to $0.182$.
\par}

\begin{contributionbox}
\textbf{We make three contributions.} (1) An answer-preserving attack that inverts model-internal confidence below chance across four vision-language models, three benchmarks, and seven readouts, with a label-free variant that never consults ground-truth correctness for the two readouts carrying a disjoint calibration reference. (2) An exact account of the collapse: attacked discrimination decomposes into a term from the merging of items that share an answer string and a term from the movement of a single item's readout on its own, and no probe we train on frozen hidden states, adversarially trained ones included, is both informative and robust. (3) The property followed outside the model: an independent verifier's reliability falls as the attacker gains access to it, from transfer to query to gradient, and a confidence gate under this attack accepts most of the wrong answers it would otherwise reject and, corrected for how often the model is wrong, leaves the system worse off than using no gate at all.
\end{contributionbox}
% \clearpage
\section{Related Work}
\label{sec:related}

\noindent\textbf{Confidence as a deployed safety signal.} Selective prediction trades coverage for reliability by accepting confident predictions and deferring the rest \citep{elyaniv2010selective}, and failure detection frames calibration, misclassification detection, and out-of-distribution detection as one problem in which a score must rank correct predictions above failures \citep{jaeger2023failure}. This ranking is what a gate depends on, and it is what we measure: discrimination between correct and incorrect answers, with the answers themselves held fixed. Prior work asks whether the ranking is informative on clean or shifted inputs. We ask whether it stays informative when the input is chosen adversarially and the answer it describes is unchanged.

\noindent\textbf{Prediction-preserving confidence attacks.} The closest predecessor is ACE \citep{galil2021ace}, which holds an image classifier's prediction fixed and uses correctness labels to lower confidence on correct items and raise it on incorrect ones, defeating softmax confidence, ensembles, MC dropout, and SelectiveNet. Calibration attacks extend this to bidirectional over- and under-confidence across several uncertainty channels \citep{obadinma2024calibration,kopetzki2021evaluating}, and the adaptive-attack literature has shown repeatedly that detectors and defenses must be tested against an adversary aimed at the specific mechanism \citep{carlini2017detection,athalye2018obfuscated,tramer2020adaptive}. We take both lessons as given and evaluate every readout with an attack aimed at it. These works show that particular estimators can be broken. We ask which confidence signals survive such an attack and why, in a setting where the preserved object is a complete variable-length generation held byte-identical instead of a single class label, and we find that none of the readouts we test remains both informative and robust.

\noindent\textbf{Where model confidence is read, and how it has been stressed.} The readouts we attack span the ways confidence is estimated in practice: output likelihood and self-evaluation \citep{kadavath2022know}, learned probes over hidden states \citep{azaria2023internal,beigi2024internalinspector}, and stability under internal perturbation \citep{khanmohammadi2025ccps}. Multi-pass estimators that obtain multiple responses, either by resampling from the same prompt or by querying semantically equivalent perturbations, and estimate uncertainty from their semantic distribution \citep{farquhar2024semantic,zhang2025vluncertainty} are a further family, which we do not attack; self-consistency appears only as a channel in Appendix~\ref{app:scope-threat}. A parallel line shows that confidence and truth are recoverable as approximately linear directions that can be steered at inference \citep{miao2026closing,cho2026manifold,ji2025verbal,park2024linear,marks2024geometry}, and recent work has begun to stress language-model uncertainty directly, through a training-time backdoor on multiple-choice distributions \citep{zeng2024fragile} and through prompt transformations that degrade estimator ranking \citep{bakman2025wild}. None of these holds a complete open-ended generation fixed, perturbs the image against a frozen model at test time, or asks what the estimator families share once the answer is pinned, which is where their common vulnerability appears.

\noindent\textbf{Neighboring guarantees and attacks.} Certified-uncertainty methods bound worst-case class probability or robustly separate out-of-distribution inputs within a neighborhood \citep{meinke2020ccu,bitterwolf2020good,meinke2022prood}, and adversarially robust conformal methods guarantee coverage of a prediction set under perturbation \citep{gendler2022rscp,yan2024rscpplus}. These certify different objects; none holds the prediction fixed while certifying that correct and incorrect answers stay separable. The amplitude certificate we test in Appendix~\ref{app:mech-certificate} is a sufficient condition for a readout to survive, and no readout we test meets it. Adversarial work on generative VLMs targets the generated content, transferring targeted responses or eliciting unsafe behavior \citep{zhao2023vlmrobustness,bailey2024imagehijacks,qi2024visualjailbreak}; there the answer is what changes. Our attack holds the answer exact and moves only the confidence attached to it.
\section{Threat Model and Measurement}
\label{sec:setup}

\noindent\textbf{The answer-preserving image attack.} We assume a white-box adversary that perturbs only the input image within an $L_\infty$ budget of $8/255$, and we work throughout in valid pixel space: every optimization step is projected onto $[x-\epsilon,x+\epsilon]\cap[0,1]$, and the returned endpoint is serialized as an 8-bit image and verified after the fact. The defining constraint is that the model's answer must not change. During optimization every emitted answer token must lead its strongest competitor by a fixed logit margin, and the returned image is then checked by regenerating the full answer with greedy decoding and requiring it to be byte-identical to the clean answer; any perturbation that alters a single token is rejected. Items whose answer changes are excluded, which removes at most $1.3\%$ of any cell; an item for which no answer-preserving perturbation moves the score in the harmful direction keeps its clean image and clean score, which weakens the measured collapse (Appendix~\ref{app:phenomenon-full}). Holding the answer fixed is what makes the measurement meaningful: whatever the confidence score does, it does while the behavior it describes stays constant, so any change is a change in the monitor alone. The full setup, attack algorithm, and verification checks are in Appendix~\ref{app:setup}.

\noindent\textbf{What we measure.} For a given readout and item, we run the attack in both directions, once to push the score as low as it will go and once as high, subject to the answer-preservation constraint. The attack is projected gradient descent (PGD) on the image, a single answer-preserving descent per item and direction, that optimizes the readout's score, may pass through iterates that break answer preservation, and returns the best iterate that preserves it. The two directions trace out the range of scores the readout can be made to take while the answer is held fixed, its answer-preserving reachable range, witnessed by the extremal images the attack finds. The quantity of interest is discrimination: whether the score still ranks correct answers above incorrect ones, measured by AUROC over a balanced pool with the answers and their correctness labels held fixed. We report it in two forms. The worst-case form asks how far discrimination can be driven when the adversary knows which answers are correct and pushes each in the harmful direction; this is an integrity audit and an upper bound on damage. The operational form asks the same of an adversary with no correctness labels, who chooses the push direction from the model's own clean confidence, and it is the form that speaks to a real deployment. The worst-case form upper-bounds the damage the operational one can do.

\noindent\textbf{Models, benchmarks, and readouts.} We evaluate four vision-language models spanning four architectures and $2$B to $12$B parameters: InternVL3.5-2B \citep{internvl}, Molmo2-4B \citep{molmo}, LLaVA-OneVision-7B \citep{llava}, and Gemma-3-12B \citep{gemma}. We use three visual question-answering benchmarks: GQA \citep{GQA}, which is in-distribution for the trained probes, and VQAv2 \citep{VQAv2} and POPE \citep{POPE}, which are out-of-distribution. Each model and benchmark contributes a canonical balanced pool of $1000$ items, split evenly between correct and incorrect answers. We attack seven confidence readouts chosen to span how confidence is estimated in practice: answer-span token probability \citep{litcab}; two hidden-state probes, $P(\text{IK})$ \citep{kadavath2022know} and SAPLMA \citep{azaria2023internal}, read at the prompt-final and answer-final tokens; a perturbation-stability estimator, CCPS \citep{khanmohammadi2025ccps}, with a differentiable variant trained on its features as its attack surrogate; and two aggregation readouts used as defenses, FULL, which combines the individual readouts, and COUPLED, an answer-coupled probe. Verbalized confidence \citep{verbalized-ce} and self-consistency \citep{self-consistency} are reported in the supplement and excluded from the headline results. Correctness labels come from a vision-capable judge, which agrees with blind human adjudication on $99.0\%$ of a stratified $300$-item sample. All results use one canonical item draw and a fixed seed. Forward passes are bit-reproducible for InternVL (eager attention); for the three SDPA-attention models, floating-point non-determinism means a re-run reaches a different valid endpoint for the same item, with stable scores and verdicts (Appendix~\ref{app:phenomenon-multiseed}).

\noindent\textbf{Scope.} Because the answer is held byte-identical, every result that follows is about the trust signal in isolation: an accept-or-defer decision that reverses under attack does so because the monitor was corrupted, not because the answer changed. Sections~\ref{sec:core} and~\ref{sec:impossibility} concern readouts of the attacked model's own outputs and internal state. A confidence check placed in a separate, independently trained model that reads the question, image, and answer is a different design, and Section~\ref{sec:ladder} takes it up directly.
\section{Answer-Invariant Confidence Inversion}
\label{sec:core}

\begin{table}[t]
\centering
\caption{The answer-preserving attack inverts the model-internal readouts, and an exact decomposition shows why one profile resists. Left (seven-readout median): discrimination between correct and incorrect answers (AUROC), median over the seven readouts per cell, clean and under attack, with the interquartile range across readouts; the label-free column is the median over the two eligible readouts. \emph{WC-oracle} chooses each item's push direction from ground-truth correctness; \emph{label-free} uses only the model's own clean confidence, over the two readouts carrying a disjoint calibration reference. The attacked median is below chance ($0.5$) in every cell. Right (binding readout): the decomposition of Eq.~\eqref{eq:decomp} for each cell's binding readout (COUPLED, except P(IK) on InternVL/GQA and VQAv2), with $w$ the same-answer pair weight and $\rho$ the same-answer range overlap. $^{\dagger}$Molmo2-4B/POPE is the sole profile above chance ($A_{\mathrm{adv}}=0.520$), with the highest $w$ and lowest $\rho$ in the matrix.}
\label{tab:core}
\resizebox{\textwidth}{!}{% Section 4 combined core table -- MEDIAN VARIANT of the aggregated left block (compare to core_decomp.tex).
% Left block = MEDIAN over the 7 readouts: Clean(median) and WC-oracle(median) from T4; IQR = Q3-Q1 of the
%   7 readouts' WC-oracle (unchanged spread); label-free(median) over the 2 eligible readouts P(IK),SAPLMA
%   (T6) -- with n=2 the median equals their mean, so these values match the mean variant.
% Right block (binding-estimator decomposition w, A_same, A_cross, A_adv, rho) UNCHANGED; A_adv == current
%   max-attacked column; rho per-binding-readout (survivor rho=0.359). dagger = Molmo2-4B/POPE. INNER TABULAR ONLY.
% Requires: \usepackage{booktabs,multirow}
\small
\setlength{\tabcolsep}{4pt}
\begin{tabular}{llcccc@{\hspace{8pt}}ccccc}
\toprule
 & & Clean & \multicolumn{3}{c}{Attacked AUROC} & \multicolumn{5}{c}{Decomposition} \\
\cmidrule(lr){4-6}\cmidrule(lr){7-11}
Model & Dataset & (median) & WC-oracle & IQR & label-free & $w$ & $A_{\mathrm{same}}$ & $A_{\mathrm{cross}}$ & $A_{\mathrm{adv}}$ & $\rho$ \\
\midrule
\multirow{3}{*}{InternVL3.5-2B} & GQA              & 0.797 & 0.016 & 0.005 & 0.291 & 0.040 & 0.000 & 0.022 & 0.021 & 1.000 \\
                           & VQAv2            & 0.724 & 0.015 & 0.011 & 0.391 & 0.045 & 0.000 & 0.029 & 0.028 & 1.000 \\
                           & POPE             & 0.779 & 0.000 & 0.000 & 0.308 & 0.498 & 0.002 & 0.002 & 0.002 & 0.998 \\
\midrule
\multirow{3}{*}{Molmo2-4B} & GQA              & 0.780 & 0.009 & 0.003 & 0.353 & 0.066 & 0.450 & 0.279 & 0.291 & 0.441 \\
                           & VQAv2            & 0.787 & 0.039 & 0.029 & 0.367 & 0.051 & 0.514 & 0.236 & 0.250 & 0.384 \\
                           & POPE$^{\dagger}$\, & 0.795 & 0.000 & 0.000 & 0.287 & 0.525 & 0.567 & 0.467 & 0.520 & 0.359 \\
\midrule
\multirow{3}{*}{LLaVA-OV-7B} & GQA              & 0.828 & 0.007 & 0.008 & 0.327 & 0.034 & 0.037 & 0.049 & 0.049 & 0.958 \\
                           & VQAv2            & 0.758 & 0.020 & 0.015 & 0.354 & 0.043 & 0.056 & 0.052 & 0.052 & 0.931 \\
                           & POPE             & 0.745 & 0.000 & 0.001 & 0.304 & 0.241 & 0.032 & 0.038 & 0.037 & 0.962 \\
\midrule
\multirow{3}{*}{Gemma-3-12B} & GQA              & 0.745 & 0.060 & 0.050 & 0.432 & 0.070 & 0.097 & 0.349 & 0.331 & 0.883 \\
                           & VQAv2            & 0.774 & 0.086 & 0.020 & 0.435 & 0.053 & 0.152 & 0.346 & 0.336 & 0.793 \\
                           & POPE             & 0.733 & 0.008 & 0.022 & 0.362 & 0.465 & 0.044 & 0.295 & 0.178 & 0.942 \\
\bottomrule
\end{tabular}
}
\end{table}

\noindent\textbf{The model-internal readouts invert below chance under the attack.} We run the answer-preserving attack against each of the seven readouts, in both directions, across all four models and three benchmarks, and measure the discrimination that survives. Table~\ref{tab:core} reports the median over the seven readouts in each cell: clean AUROC ranges from $0.72$ to $0.83$, and under the worst-case attack the median falls to $0.09$ or lower in every one of the twelve cells, with a narrow spread across readouts. An AUROC near zero means the readout now ranks incorrect answers above correct ones almost perfectly: the model reports its highest confidence where it is wrong. Seven of the seven readouts fall below chance in eleven of the twelve cells, and six of seven in the last, whether the readout is a likelihood, a hidden-state probe, or a perturbation-stability measure. The single exception, the answer-coupled readout on Molmo/POPE, is explained by the decomposition below. The collapse is stable across three attack seeds, with a largest across-seed standard deviation of $0.046$ and every below-chance cell below chance on every seed; the per-readout results are in Appendix~\ref{app:phenomenon} and the multi-seed evidence in Appendix~\ref{app:phenomenon-multiseed}.

\noindent\textbf{A label-free attacker, with no correctness oracle, inverts the readouts too.} A real attacker has no correctness labels. The label-free column replaces the oracle with the model's own clean confidence: the attacker treats a high clean score as a stand-in for a correct answer, and pushes the confident-looking items down and the unconfident-looking ones up. This drives discrimination below chance in all twelve cells, to between $0.26$ and $0.46$, for the two readouts that carry a disjoint clean-calibration reference (Appendix~\ref{app:phenomenon-labelfree}). The oracle inverts the ranking almost completely while the label-free attacker only pushes it past chance, but the conclusion for a deployment is the same under both: a gate reading this signal can be made to rank wrong answers above right ones by an attacker with no knowledge of which is which. The restriction to two readouts is one of evaluation hygiene, since a proxy fit on the evaluation pool would contaminate the measurement; it is not a limit of the attack.

\begin{figure}[t]
  \centering
  \includegraphics[width=\linewidth]{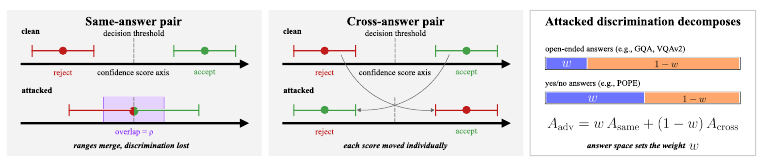}
  \vspace{-20pt}
  \caption{The exact decomposition. A confidence readout must rank correct against incorrect answers, and these pairs split into two kinds: same-answer pairs, whose two items produce the identical answer string, and cross-answer pairs, whose answers differ. Attacked discrimination is the weighted sum of the two, $A_{\mathrm{adv}} = w\,A_{\mathrm{same}} + (1-w)\,A_{\mathrm{cross}}$, with the same-answer weight $w$ setting which mechanism dominates. The one surviving profile, Molmo/POPE, has high $w$ and low $\rho$: its same-answer term is large and the attack cannot merge it.}
  \vspace{-10pt}
  \label{fig:decomp}
\end{figure}

{\looseness=-1
\noindent\textbf{Attacked discrimination splits exactly into a same-answer and a cross-answer term.} Partition the correct-incorrect pairs a readout must rank into those whose two answers are the same string and those whose answers differ, and let $w$ be the fraction that share an answer (Figure~\ref{fig:decomp}). AUROC is a mean over these pairs, so it separates without residual into a same-answer part and a cross-answer part, \par}
\begin{equation}
\label{eq:decomp}
A_{\mathrm{adv}} \;=\; w\,A_{\mathrm{same}} \;+\; (1-w)\,A_{\mathrm{cross}},
\end{equation}
where $A_{\mathrm{same}}$ and $A_{\mathrm{cross}}$ are the attacked discrimination on same-answer and cross-answer pairs. The identity is exact: across all $84$ profiles the right side reconstructs the measured $A_{\mathrm{adv}}$ to within $6\times10^{-17}$ (Appendix~\ref{app:decomp}). It names the two ways the attack corrupts the ranking. On same-answer pairs the answer string carries no information, so the readout's discrimination lives entirely in the score, and the attack destroys it by pushing two items that share an answer until their reachable score ranges overlap and no threshold separates them; we measure this overlap per cell as $\rho$, the fraction of same-answer correct-incorrect pairs whose reachable ranges overlap. On cross-answer pairs the answers differ, and the attack corrupts the ranking by moving each item's score on its own. The weight $w$ decides which term dominates a given cell. Appendix~\ref{app:mechanisms} develops both mechanisms in full.

\noindent\textbf{The same-answer weight sets which term dominates, and explains the one surviving profile.} Where answers are rarely shared the collapse is almost entirely cross-answer: on LLaVA/GQA, $w$ is $0.034$ and the cross-answer term is $97\%$ of $A_{\mathrm{adv}}$. Where answers are often shared the same-answer term carries real weight, and whether it collapses turns on $\rho$. Two POPE cells show this at nearly fixed weight. On InternVL/POPE, $w=0.498$ but the overlap is near-total, $\rho=0.998$, so the same-answer intervals merge, $A_{\mathrm{same}}$ falls to $0.002$, and the cell collapses with the rest. On Molmo/POPE, $w=0.525$ but the overlap is the lowest in the matrix, $\rho=0.359$: the same-answer intervals do not merge, $A_{\mathrm{same}}$ stays at $0.567$, and because same-answer pairs make up half the cell, the readout holds at $A_{\mathrm{adv}}=0.520$, marginally above chance in point estimate, with a bootstrap interval that includes $0.5$ on every seed (Appendix~\ref{app:phenomenon-multiseed}). With the weight nearly fixed across these two cells, the overlap alone accounts for the difference. Low overlap alone is not enough: Molmo's other two cells also have low $\rho$ but small weight, and collapse to $0.29$ and $0.25$ as the cross-answer term dominates. Survival needs both a same-answer term the attack cannot merge and enough weight for it to carry the cell, and only Molmo/POPE has both. Attacked discrimination admits a measured upper bound $c^{\star}+(1-\rho)\,w$, where $c^{\star}$ is the cell's in-sample answer-string-prior AUROC; it bounds $A_{\mathrm{adv}}$ in all $84$ profiles (Appendix~\ref{app:mech-ceiling}), and no readout meets a uniform amplitude certificate (Appendix~\ref{app:mech-certificate}).

\noindent\textbf{The answer does not determine the confidence, even when the full distribution is constrained.} Holding the answer byte-identical pins the top token at every position but leaves the rest of the next-token distribution free, and the attack does move it. We therefore add a bound on the KL divergence of the full next-token distribution at every answer position, enforced by rejecting any step that violates it, and re-run the attack. At a bound of $0.2$ nats a mean $62\%$ of the full inversion survives for the SAPLMA readout, which stays below chance in all twelve cells, and P(IK) retains a comparable share, $69\%$ (Appendix~\ref{app:distribution}). For three of the four models this bound is $22$ to $68$ times the divergence induced by a single-greylevel pixel dither, so the surviving majority is not an artifact of maximal distributional freedom. Gemma is the decisive case. Its output distribution is so sensitive to input change that a $0.2$-nat bound holds its sampled continuations at the clean noise floor, behaviorally indistinguishable from the clean model, and there $83$ to $93\%$ of the inversion still survives. The sensitivity is a property of Gemma's image-to-logit map, not of input scale (Appendix~\ref{app:distribution}). Tightening to $0.05$ nats is where the picture splits. For InternVL, Molmo, and LLaVA that bound is still five to seventeen times their benign-dither divergence, and they return toward chance with that slack unused; for Gemma it is below its own dither floor, and Gemma stays inverted. The answer does not determine confidence at any bound; confidence depends in part on the distribution's fine structure, and the erosion from $62\%$ to $36\%$ measures how much of it lives there.

\noindent\textbf{A confidence gate under attack becomes worse than no gate.} We simulate a gate directly: the FULL hidden-state gate, with its threshold fixed to a $10\%$ false-accept rate on clean data and held fixed under attack, as in deployment. Transferring the token-probability attack to this gate, with the answer unchanged, turns previously rejected wrong answers into accepted ones, up to $84.8\%$ in the worst cell. Because the relevant comparison depends on how often the model is wrong, we reweight each benchmark to its natural error rate. Under that accounting, the attacked gate accepts answers at lower accuracy than accepting every answer in $8$ of $12$ cells under transfer and all $12$ under an attack aimed at the gate, and this holds across the error rates we consider. On InternVL/POPE, accepted accuracy on the balanced pool falls from $0.744$ on clean images to $0.182$ under attack, and at natural prevalence the gated system sits below the un-gated one (Appendix~\ref{app:deployment}): the gate, meant to raise reliability, leaves the system worse than no gate. The area under the risk-coverage curve rises from a clean mean of $0.30$ to $0.80$ under attack. Appendix~\ref{app:deployment} gives per-cell harm, threshold sensitivity, and the full risk-coverage analysis.
% \clearpage
\section{No Internal Readout Escapes on Short Answers}
\label{sec:impossibility}

\begin{figure}[t]
\centering
\begin{minipage}[c]{0.55\linewidth}
    \centering
    \includegraphics[width=\linewidth]{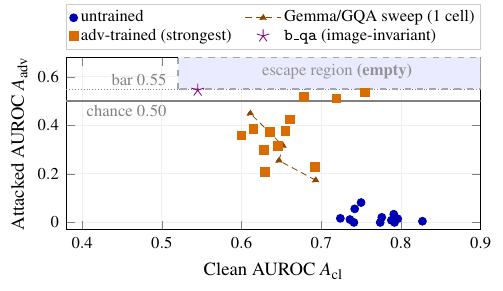}
\end{minipage}%
\hspace{0.015\linewidth}%
\begin{minipage}[c]{0.42\linewidth}
    \caption{No trained readout is both informative and robust.
    Each point is a readout for one of the twelve cells, by clean ($x$)
    and attacked ($y$) discrimination. Blue: untrained, informative but
    collapsing. Orange: adversarially trained at each cell's strongest
    setting. The shaded escape region, clean above the answer-string prior
    and attacked above $0.55$, is empty. \texttt{b\_qa}, the pixel-free
    control, cannot be moved by the attack but sits below the prior:
    unreachable yet too weak to gate.}
    \label{fig:frontier}
\end{minipage}
\vspace{-10pt}
\end{figure}

% \begin{wrapfigure}[21]{r}{0.46\textwidth}
% \centering
% \vspace{-10pt}
% \includegraphics[width=0.46\textwidth]{figures/frontier.pdf}
% \vspace{-25pt}
% \caption{No trained readout is both informative and robust. Each point is a readout for one of the twelve cells, by clean ($x$) and attacked ($y$) discrimination. Blue: untrained, informative but collapsing. Orange: adversarially trained at each cell's strongest setting. The shaded escape region, clean above the answer-string prior and attacked above $0.55$, is empty. \texttt{b\_qa}, the pixel-free control, cannot be moved by the attack but sits below the prior: unreachable yet too weak to gate.}
% \label{fig:frontier}
% \end{wrapfigure}
\noindent\textbf{Adversarial training gains robustness only by giving up informativeness.} On each of the twelve cells we train a family of confidence probes on the same hidden states with adversarial training that rewards invariance to the answer-preserving attack, sweeping the robustness weight over $\{0.3, 1, 3, 10\}$ with three seeds at the strongest setting, and then attack each trained probe with the same answer-preserving attack and $L_\infty$ budget of $8/255$ used throughout. Across all twelve cells and every weight, and for a separate gradient-penalty variant, no probe is both informative on clean inputs and robust under attack. The two properties trade off along a frontier (Figure~\ref{fig:frontier}). On Gemma/POPE, the lightest-penalty probe has clean AUROC $0.865$ and attacked AUROC $0.061$; raising the penalty until attacked AUROC reaches its maximum of $0.385$ drags clean AUROC down to $0.629$. Robustness is bought with the discrimination the probe was meant to provide, and the escape region of the figure, which traces the Gemma/GQA sweep, is empty in all twelve cells. The frontier is a property of post-hoc probes over frozen representations: a head cannot move the representation it reads, so invariance can only be bought by ignoring the directions the attack moves. End-to-end adversarial fine-tuning of the encoder or projector is outside our scope. The collapse is also stable across attack configurations and under an answer-margin constraint (Appendix~\ref{app:frontier-invariance}).

{\looseness=-1
\noindent\textbf{The closest probe misses a pre-specified escape bar.} We fixed the criterion for an escape before running the sweep: at the same operating point, clean AUROC above the answer-string prior and attacked AUROC above $0.55$. One cell, LLaVA/POPE, comes closest. Its best probe reaches clean AUROC $0.755$ and attacked AUROC $0.536$ as three-seed means (the seed-23 values are $0.768$ and $0.544$), the highest attacked value anywhere in the sweep, but $0.536$ sits below the bar, and the three-seed standard deviation of $0.008$ places it there reliably. None of the defenses we trained, at either budget, escaped; whether a better-resourced defense could remains open. The full per-cell sweep is in Appendix~\ref{app:frontier}.
\par}

\noindent\textbf{Long, unique answers switch off same-answer merging and isolate the cross-answer term.} On short answers the two terms of Eq.~\eqref{eq:decomp} run together, because answers are often shared and both are active at once. When every answer is distinct there are no same-answer pairs, $w$ goes to zero, and only the cross-answer term remains, so long, unique answers let us ask whether that term is attackable on its own. Because these benchmarks do not elicit long answers under a neutral prompt, we induce them with a reasoning prompt, giving median answers of $77$ to $97$ tokens on three models, and use a fourth, Molmo, whose answers are natively long ($25.5$-token median under a neutral prompt). Holding a roughly ninety-token answer byte-identical is a severe constraint: only about $3\%$ of attack steps preserve it, and between $8$ and $16$ candidate perturbations per item are rejected for breaking it, so any collapse here is not an artifact of preservation being easy on short answers.

\noindent\textbf{With merging gone, the attack still inverts internal readouts on two of four models.} On Gemma, P(IK) and SAPLMA fall from clean $0.67$ and $0.60$ to $0.38$ and $0.40$, and on Molmo, whose long answers are native, from $0.68$ and $0.62$ to $0.43$ and $0.43$. The token-probability readout collapses on all four models, as low as $0.08$ on InternVL. The attack can therefore move a single fixed answer's confidence far enough to corrupt the ranking with no shared-answer structure to exploit, which is the second mechanism shown in isolation. On InternVL and LLaVA the trained probes resist, with P(IK) and SAPLMA staying between $0.58$ and $0.62$ on both models. None of reasoning training, native answer length, backbone, size, the strength of the answer-final correctness signal, or the input-to-logit sensitivity of Appendix~\ref{app:distribution} predicts the split: the natively long model collapses, the two resisters carry the strongest answer-final signal yet keep it, and the least and most dither-sensitive models both collapse. What separates the groups is how far a model's readout moves under an answer-preserving perturbation, which this analysis measures but does not explain (Appendix~\ref{app:mech-cross}).
\section{Outside the Model: The Reachability Ladder}
\label{sec:ladder}

\begin{table}[t]
\centering
\caption{An independent verifier only relocates the attack: its reliability falls as the attacker gains optimization power over it. Verifier AUROC averaged over the twelve base$\to$verifier pairs, at increasing levels of attacker access. \emph{Clean} is the verifier's unattacked AUROC and \emph{no access} is its transfer AUROC against the answer-preserving attack aimed at the base model; \emph{decision-only} (HopSkipJump) and \emph{score-only} (SPSA) query the verifier under a fixed query budget, and \emph{white-box} attacks its gradients directly. Discrimination falls from clean through score-only to white-box; the two intermediate rungs sit close together. The three rows report all twelve pairs, the eleven excluding the one under-powered white-box cell, and the nine whose verifier admits an exact differentiable path. Per-pair values, confidence intervals, and caveats are in Appendix~\ref{app:ladder}.}
\label{tab:ladder}
% Main-text (Section 6) compact reachability-ladder AGGREGATE. STRICT SUBSET of ladder_full.tex (report 222,
%   data/snowglobe/verifier_escape/span12_ladder_results.json): the three aggregate mean rows only, values
%   copied verbatim so the body and appendix cannot disagree. Per-pair rows, CIs, set column, and notes live
%   in the appendix table. No \label here (the main-text wrapper supplies tab:ladder); no footnote here.
% Requires: \usepackage{booktabs}
\small
\begin{tabular*}{\textwidth}{@{\extracolsep{\fill}}lrrrrr@{}}
\toprule
 & clean & no access & decision-only & score-only & white-box \\
\midrule
All 12 pairs & 0.655 & 0.571 & 0.601 & 0.480 & 0.287 \\
Excluding under-powered cell & 0.646 & 0.558 & 0.590 & 0.468 & 0.246 \\
Bit-exact verifiers (9) & 0.688 & 0.589 & 0.620 & 0.496 & 0.223 \\
\bottomrule
\end{tabular*}

\vspace{-10pt}
\end{table}

\noindent\textbf{An independent verifier resists the transferred attack only while the attacker has no access to it.} If every readout inside the answering model is reachable, one can instead hand the image, question, and answer to a second, independently trained model and ask it to judge correctness. The attacker is white-box on the answering model but, at first, has no access to this verifier. In that setting the verifier survives: across the twelve base-verifier pairs its discrimination under the transferred attack stays around $0.57$ on average, while the answering model's own internal readout on the same attacked images sits near $0.05$. The images that invert the internal signal do not transfer to an independent one, and cross-model transfer of the perturbations is likewise weak (Appendix~\ref{app:scope-threat}). This looks like a defense; the rest of the section shows it relocates the attack.

\noindent\textbf{Verifier discrimination falls as the attacker gains optimization power over the verifier.} We grant progressively more verifier access and measure what survives (Table~\ref{tab:ladder}). With score-only access, the attacker queries the continuous score and runs gradient-free search; mean discrimination falls to $0.48$, already below chance. With white-box access to the verifier's gradient, it collapses to $0.06$--$0.35$ across the nine pairs with an exact differentiable verifier path (Appendix~\ref{app:ladder-parity}), a mean of $0.223$, matching an internal readout, and white-box is the strongest attack in ten of twelve pairs. Throughout the white-box rung the base model's answer is held byte-identical under the same constraint as the internal attack: a perturbation is accepted only when a fresh greedy decode of the base model reproduces the committed answer, with the clean image as fallback, so the verifier's collapse is measured on images the base model still answers identically. The two intermediate rungs, transfer and decision-only, sit close together at $0.57$--$0.60$ and are not separably ordered: a label-aware transfer oracle is already a strong baseline that hard-label boundary search does not reliably beat. The movement that matters is across rungs that add real optimization power, from clean through score-only to white-box.

\noindent\textbf{The verifier's survivors are saturated items, already decided on the clean image.} The items a query attacker cannot flip are the ones the verifier was already most certain about, whose scores sit too far from the decision boundary for a bounded perturbation to cross. Across the twelve pairs about half of the items are decision-flippable under hard-label access alone, so the surviving discrimination rests on the unflipped remainder, the items that began farthest from the boundary. On the strongest pair the mechanism is explicit: the unflippable items sit at mean distance $0.335$ from the decision boundary against $0.047$ for the flipped ones, and the attack needs a median of $132$ queries to reach its first flip, against $52$ to $63$ on weaker verifiers. A strong verifier is harder to attack because more of its answers begin beyond what a bounded perturbation can move, and survival tracks clean quality: the verifiers that retain the most discrimination under attack are those whose answers start farthest from the boundary, while a verifier near chance on clean data has nothing left to lose. Removing the answer-preservation constraint from the query loop and rerunning both attacks on the identical items changes almost nothing, moving final AUROC by at most $0.033$ and the unflippable residue by at most one item, so the residue is saturation and not a consequence of holding the answer fixed. That concession is local to the query loop; the internal attack of Sections~\ref{sec:core} and~\ref{sec:impossibility} has no oracle to saturate, and there the constraint is the phenomenon.

\noindent\textbf{An external check relocates the phenomenon.} Reachability, how far an answer-preserving adversary can move the signal, connects the verifier's survival to the internal results. Inside the model, a readout collapses because the attack reaches it directly, and training cannot move it out of reach without emptying it of the information that made it useful. Outside the model, an independent verifier collapses once the attacker can reach it, by query or by gradient, and survives only where reach fails: at no access, or where an item sits so far from the verifier's decision boundary that the budget cannot cross it. This connection has a limit. Inside the model, reachability is co-measured with the collapse, because moving the score and succeeding are the same act, so we do not offer it as a forecast of which internal readouts fall. For the verifier's saturated survivors it does more: their clean distance from the decision boundary is fixed before any perturbation, so there reachability is measurable ahead of the attack and predicts what survives.

% \begin{figure}[t]
%   \centering
%   \includegraphics[width=0.62\linewidth]{figures/ladder.pdf}
%   \caption{\textbf{The reachability ladder.} Verifier discrimination (AUROC) against a correct-incorrect ranking, as the attacker gains access to the verifier, for four base-verifier pairs. Survival falls monotonically from no access, through decision-only and score-only queries, to full white-box access, where it collapses to the level of an internal readout. The one pair that survives decision-only access is the one with the strongest clean verifier.}
%   \label{fig:ladder}
% \end{figure}
\section{Conclusion}
\label{sec:conclusion}

\noindent\textbf{Summary.} Confidence is increasingly the signal that decides whether a model's answer is trusted, deferred, or shown to a user, and this paper shows that when the signal is read from inside the model, an adversary who leaves the answer untouched can turn it to the opposite of its purpose. The failure is quiet: the answer a user sees is unchanged, and only the decision about whether to trust it has been corrupted. Anyone building on model-internal confidence should treat it as an attackable surface. Reading the check from a separate model does not fix this: an independent verifier holds up only while the attacker cannot reach it, and degrades to the same collapse once it can. Under an adversary who can reach the check, on the short-answer benchmarks that make up the core matrix, we find no readout, inside the model or outside it, that stays both informative and reliable; on long answers two of four models hold above chance, which is the open question below. A confidence signal that can be steered while the answer it describes is held exact is a channel of its own, corruptible on its own.

\noindent\textbf{Limitations.} Four limits bound the claim. The perturbation is small but not invisible: at $8/255$ it is comparable on LPIPS to a two-pixel crop-and-resize, and on a minority of items is perceptible on close inspection (Appendix~\ref{app:setup-norms}). The attack is per-item and white-box; it does not transfer across models, and no universal perturbation carries it (Appendix~\ref{app:scope-threat}). The threat model is therefore an adversary with gradient access to the answering model and control of the input image, as in open-weight deployment; a server-side re-encode outside the attacker's control is partly covered by the purifier results in Appendix~\ref{app:scope-defenses}, where JPEG at quality 50 removes between a quarter and two thirds of the manipulable range. Label-free inversion is shown only for the two readouts with a disjoint calibration reference, and we do not impute it to the other five. Two defense families, input purification and randomized smoothing, are only partially evaluated: smoothing appears to hold under a transferred attack and fails under an adaptive one on the cells we ran, and we do not extend that result to untested models (Appendix~\ref{app:scope-defenses}). The defense frontier is measured on post-hoc probes over frozen hidden states, not on end-to-end fine-tuning, and the attacked readouts are single-pass; multi-pass estimators are not evaluated. Full scope and negative results are collected in Appendix~\ref{app:scope}.

\noindent\textbf{Open question.} Whether a robust and informative readout can be built remains open, and the clearest lead comes from the one place the pattern is not uniform. On long answers, whether an internal readout resists the attack is decided by how movable it is item by item, and that movability varies across models in a way none of the properties we examined predicts. Understanding what makes one model's confidence reachable and another's not is where a full account of this phenomenon, and any principled defense against it, will have to begin.

% \clearpage
% ==================== END OF 9-PAGE MAIN TEXT ====================
\subsection*{AI use statement}

In this work, we used generative AI tools to aid with editing, language refinement, and polishing of the main manuscript, and to assist in drafting selected portions of the appendix. All AI-assisted manuscript content was manually reviewed, revised where necessary, and checked by the authors for technical accuracy and consistency with the underlying experiments, analyses, and reported results. During research execution, generative AI assistants were also used to support programming and implementation tasks; any AI-assisted code incorporated into the experimental workflow was manually reviewed, executed, and tested by the authors before use, and the resulting analyses and reported numerical results were checked against the experimental outputs. Generative AI tools were not used for retrieval or discovery of related work, and other disclosure categories listed in the ICLR 2027 AI policy were either not used or were not applicable to this work. We have reviewed all AI-assisted work and take responsibility for the final content of this work, including text, claims, code, analyses, and other artifacts produced with the aid of generative AI.
\subsection*{Ethics statement}

This work studies the robustness of confidence mechanisms used to decide
whether vision-language model outputs should be trusted, deferred, or rejected.
The adversarial techniques developed for this purpose are dual-use: although
they are intended to expose weaknesses in confidence-based safety mechanisms,
similar methods could be used to manipulate such mechanisms in deployed
systems. We therefore frame the attacks as robustness and integrity evaluations
and explicitly characterize the assumptions, scope, and limitations of the
threat models considered. We also evaluate defensive approaches and report
negative results alongside successful attacks in order to distinguish the
demonstrated vulnerabilities from broader claims not supported by our
experiments. Our goal is to support the development and evaluation of confidence
mechanisms that remain reliable under adversarial manipulation.
\subsection*{Reproducibility statement}

The supplementary material contains the full code release together with a representative data sample and the artifact stores used to regenerate the reported results. The release reproduces all tables and the data-derived figure in the paper from the provided artifacts and includes a manifest linking each result to the corresponding script and inputs. The codebase contains the complete experimental pipeline, including dataset preparation and the committed seed-23 item identifiers, model inference and hidden-state extraction, the answer-preserving attack, probe training, the verifier experiments, correctness evaluation, and result aggregation. The exact model checkpoints, software environment, hyperparameters, sweep ranges, and selection criteria are documented in Appendix~\ref{app:setup} and Table~\ref{tab:app-hyper}; the decomposition and supporting derivations are given in Appendix~\ref{app:decomp}; and the principal verification checks on answer preservation, pixel validity, perturbation magnitude, and correctness-label agreement are reported in Appendices~\ref{app:setup-preservation}--\ref{app:setup-judge}. The accompanying data sample contains 300 items stratified across all four models and three benchmarks, with the corresponding clean and adversarial images, questions, model answers, correctness labels, and clean and attacked scores for all seven confidence readouts. Together, the released code, artifacts, configuration, and documented evaluation protocol are intended to support reproduction of both the main results and the analyses reported in the appendices.

\subsubsection*{Acknowledgments}
This work was supported by the JPMorgan Chase AI Research Faculty Research Award. The authors are solely responsible for the contents of this paper; the opinions expressed do not necessarily reflect those of the funding organizations. The authors also acknowledge the use of Large Language Models to assist in polishing the language and grammar of this manuscript.
\subsection*{Disclaimer}

This paper was prepared for informational purposes by the Artificial Intelligence Research group of JPMorgan Chase \& Co and its affiliates (``JP Morgan''), and is not a product of the Research Department of JP Morgan. JP Morgan makes no representation and warranty whatsoever and disclaims all liability, for the completeness, accuracy or reliability of the information contained herein. This document is not intended as investment research or investment advice, or a recommendation, offer or solicitation for the purchase or sale of any security, financial instrument, financial product or service, or to be used in any way for evaluating the merits of participating in any transaction, and shall not constitute a solicitation under any jurisdiction or to any person, if such solicitation under such jurisdiction or to such person would be unlawful.

\bibliography{iclr2027_conference}
\bibliographystyle{iclr2027_conference}

% ========================== APPENDICES ==========================
\clearpage
\appendix
\startcontents[appendices]
\begin{center}
{\Large\bfseries Appendix Contents}
\end{center}
\vspace{-10pt}
\noindent\rule{\textwidth}{0.4pt}
\vspace{-20pt}
\printcontents[appendices]{}{1}{}
\noindent\rule{\textwidth}{0.4pt}
\section{Setup and Verification}
\label{app:setup}

This appendix records the experimental setup and the checks that make the main results interpretable: the implemented attack, verification that answers are preserved byte for byte, that the perturbed images are valid, that the perturbation is small, that the automated correctness judge agrees with human adjudication, and that no result was obtained by tuning against it. Every table is regenerated from the released stores.

\subsection{Models, environment, and the implemented attack}
\label{app:setup-attack}

All results are tied to exact checkpoints and a fixed environment. The four vision-language models are InternVL3.5-2B, Molmo2-4B, LLaVA-OneVision-7B, and Gemma-3-12B, obtained from the repositories listed in the released artifact manifest; model-family names do not identify weights uniquely, so the repository identifiers define the systems evaluated here. Attacks, probe training, and hidden-state extraction ran on NVIDIA H200 GPUs under a pinned software stack, and every randomized procedure uses global seed 23. Seeding fixes every sampled quantity but not the floating-point reduction order inside SDPA attention, so for Molmo, LLaVA, and Gemma the attack's 8-bit endpoint is not bit-reproducible across runs; the released code documents which endpoints reproduce exactly and which reproduce as a different valid endpoint with matching answer preservation and budget. All hidden-state probes read the final transformer layer, fixed in advance and not selected on validation: SAPLMA at the last answer token, P(IK) at the last prompt token, and CCPS and CCPS-D on final-layer answer tokens; only the hidden width differs across models (2048, 2560, 3584, 3840). Probes are trained on a 20,000-item GQA training split at its natural class ratio (between 15,832/4,168 and 16,961/3,039 correct/incorrect across models) with class-balanced binary cross-entropy, positive weight $n_-/n_+$, and early-stopped on a disjoint 5,000-item validation split. Aggregation and table generation reproduce from the released artifacts on CPU.

The attack is projected gradient descent through the frozen model, one descent per item and direction, returning only answer-preserving iterates. For an item whose answer is to be held fixed, it takes signed-gradient steps on the image against the target estimator, with the harmful direction set by item correctness: the score is pushed down when the answer is correct and up when it is wrong. At every step the attack operates on the raw RGB image in $[0,1]$ and projects onto $[x-\epsilon, x+\epsilon] \cap [0,1]$, so the iterate stays within both the budget and the valid pixel range; endpoints are stored as $8$-bit images (Appendix~\ref{app:setup-clip}). At each iterate every emitted answer token must lead its strongest competitor by at least a fixed margin $\tau_{\mathrm{safe}}=1.0$, which is stricter than the argmax requirement; among feasible iterates the attack keeps the one with the largest harmful score movement. Optimization may pass through infeasible points, but only a feasible iterate is returned, and the final answer is regenerated by fresh greedy decoding and checked for byte identity before the item enters any analysis. An item that never finds a feasible harmful iterate keeps its clean score and zero perturbation, which weakens the measured collapse. All parameters, including the ones swept, are given in Table~\ref{tab:app-hyper}.

The procedure above is the answer-fixed attack: one descent per item and direction that holds the emitted answer byte-identical and commits the single feasible iterate with the largest harmful movement. The fully coordinated attack is the worst-case reachability oracle of Appendix~\ref{app:mech-reachable}: for each item it selects, over that item's answer-preserving candidate endpoints, the one that most damages the label-directed ranking, so same-answer items are driven into a common reachable overlap. The two share the objective and the byte-identity constraint and differ only in selection, so the oracle's within-answer discrimination is at most that of the answer-fixed attack. The exact decomposition (Table~\ref{tab:app-decomp-full}) uses the coordinated attack; the within-answer diagnostic (Table~\ref{tab:app-within}) uses the answer-fixed attack, which is why on the survivor, COUPLED on Molmo/POPE, the within-answer value is $0.731$ against the coordinated $A_{\mathrm{same}}=0.567$, and TokProb reaches $0.325$ against at most $0.026$.

\subsection{Answers are preserved byte for byte}
\label{app:setup-preservation}

Every adversarial analysis is conditioned on fresh byte-identity verification, and preservation is near-total. After the final perturbation is selected, the model regenerates its answer greedily and the result is compared with the clean answer; items that fail are excluded from any analysis requiring preservation. Only these excluded items, 41 of 12,000, leave the pool. An item that preserves its answer but for which no answer-preserving step moves the score in the harmful direction stays in the pool with its clean image and clean score, so it enters the AUROC denominator as a clean fallback rather than being dropped. Table~\ref{tab:app-preservation} reports two pools: a rigor set that applies six attack conditions to the 1000-item canonical pool, and the estimator-aimed attack used for the headline results. Main-attack preservation ranges from $0.9870$ on Molmo/VQAv2 to $1.0000$, and across the twelve cells $11{,}959$ of $12{,}000$ items are byte-identical. POPE, whose answers occupy a two-token space, preserves perfectly under the main attack; VQAv2, with more varied open-ended answers, has the lowest rates. These rates describe the retained set; the per-item filter is applied throughout.

\begin{table}[t]
\centering
\caption{\textbf{Answer preservation on two evaluation pools.} The rigor set applies six attack conditions to the 1000-item canonical pool (6000 evaluations per cell); the main-attack columns report the estimator-aimed attack behind the headline results. Byte identity is verified by fresh greedy decoding of every final image. The minimum main-attack rate is $0.9870$, and the twelve cells contain $11{,}959$ byte-identical items of $12{,}000$.}
\label{tab:app-preservation}
\small
% source: tables/049_rigor_taskC/results.json guarantees[<model>|<set>] (rigor n, rigor rate) and
% data/theorem/task_floor_bound.json cstar[<model>][<set>].n_byte_identical (main-attack byte-id / 1000).
% Cell order InternVL/Molmo/LLaVA/Gemma. Pooled main-attack byte-id = 11959/12000; min rate Molmo/VQAv2.
\begin{tabular}{@{}ll rrrr@{}}
\toprule
model & set & rigor $n$ & rigor rate & main-attack byte-id & main-attack rate \\
\midrule
\multirow{3}{*}{InternVL3.5-2B} & GQA & 6000 & 0.9952 & 994/1000 & 0.9940 \\
 & VQAv2 & 6000 & 0.9928 & 990/1000 & 0.9900 \\
 & POPE & 6000 & 0.9998 & 1000/1000 & 1.0000 \\
\addlinespace
\multirow{3}{*}{Molmo2-4B} & GQA & 6000 & 0.9988 & 998/1000 & 0.9980 \\
 & VQAv2 & 6000 & 0.9945 & 987/1000 & 0.9870 \\
 & POPE & 6000 & 1.0000 & 1000/1000 & 1.0000 \\
\addlinespace
\multirow{3}{*}{LLaVA-OV-7B} & GQA & 6000 & 0.9988 & 999/1000 & 0.9990 \\
 & VQAv2 & 6000 & 0.9943 & 997/1000 & 0.9970 \\
 & POPE & 6000 & 1.0000 & 1000/1000 & 1.0000 \\
\addlinespace
\multirow{3}{*}{Gemma-3-12B} & GQA & 6000 & 0.9978 & 998/1000 & 0.9980 \\
 & VQAv2 & 6000 & 0.9972 & 996/1000 & 0.9960 \\
 & POPE & 6000 & 1.0000 & 1000/1000 & 1.0000 \\
\bottomrule
\end{tabular}

\end{table}

\subsection{The attack remains in valid pixel space} \label{app:setup-clip}

The headline attack operates on the raw RGB image scaled to $[0,1]$, and at every optimization step it projects the perturbed image onto $[x-\epsilon, x+\epsilon] \cap [0,1]$, the intersection of the budget ball with the valid pixel range; the returned endpoints are serialized as $8$-bit images. Every attacked endpoint is therefore a valid image within the budget, and the frozen store certifies this directly. Across all $120{,}000$ attacked endpoints, the twelve cells by $1{,}000$ items by five aimed objectives and two push directions, every one satisfies both the valid-pixel and the budget constraint: the global pixel range is exactly $[0,1]$ and no endpoint exceeds the $8/255$ bound. Of the $120{,}000$, $7{,}370$ ($6.1\%$) are byte-identical to their clean image after $8$-bit serialization and so carry zero perturbation: $328$ are clean-image fallbacks for a direction in which the search found no feasible harmful iterate, and the remaining $7{,}042$ are optimized iterates whose net movement rounds away at $8$ bits. These counts are taken at the attack's argmax constraint; the stricter operating margin $\tau_{\mathrm{safe}}=1$ used for the engaged-subset analysis of Appendix~\ref{app:phenomenon-full} leaves a larger fraction with clean scores.

\subsection{The perturbation is small}
\label{app:setup-norms}
Table~\ref{tab:app-norms} reports the realized $L_\infty$ norm computed from the headline attack's stored endpoints. The maximum is exactly $0.0314$ ($8/255$) in every cell, so the budget binds, and the per-cell means sit just below it. The attack budget was swept over $\{2,4,8,16\}/255$ (Table~\ref{tab:app-hyper}); $8/255$ is the operating point at which the headline results are read, and the other budgets are reported.

\begin{table}[t]
\centering
\caption{\textbf{Realized $L_\infty$ perturbation norm}, computed from the headline attack's stored $8$-bit endpoints over the moved endpoints, those not byte-identical to the clean image ($112{,}630$ of $120{,}000$). The maximum is exactly $8/255=0.0314$ in every cell, so the budget binds; the per-cell means sit just below it. Averaging over all $10{,}000$ endpoints per cell with byte-identical ones as zero lowers each mean by $0.001$ to $0.003$.}
\label{tab:app-norms}
\small
% Appendix S1 (setup / valid-pixel certification): realized L-inf perturbation norm, per cell.
% Source: the CURRENT gate_v12 headline attack store, data/snowglobe/pfmc_v1_2_2/full_matrix_runs/*.jsonl
%   validity blocks (linf_raw / linf_greylevels), over the MOVED (non-clean-fallback) endpoints
%   (endpoint_is_clean_image == false). 112,630 moved endpoints of 120,000 total (7,370 are the
%   zero-perturbation constraint-blocked fallbacks). L-inf is in the [0,1] pixel scale; 8/255 = 0.0314.
%   This replaces the earlier norms sourced from data/adv/rigor (the old pixel_values-space non-clipping
%   attack); L2 and per-pixel RMS are not retained in the frozen headline store and are dropped here.
% Requires: \usepackage{booktabs,multirow}
\small
\setlength{\tabcolsep}{7pt}
\begin{tabular}{@{}llrr@{}}
\toprule
model & set & L-inf mean & L-inf max \\
\midrule
\multirow{3}{*}{InternVL3.5-2B} & GQA & 0.0298 & 0.0314 \\
 & VQAv2 & 0.0300 & 0.0314 \\
 & POPE & 0.0308 & 0.0314 \\
\midrule
\multirow{3}{*}{Molmo2-4B} & GQA & 0.0302 & 0.0314 \\
 & VQAv2 & 0.0300 & 0.0314 \\
 & POPE & 0.0304 & 0.0314 \\
\midrule
\multirow{3}{*}{LLaVA-OV-7B} & GQA & 0.0299 & 0.0314 \\
 & VQAv2 & 0.0299 & 0.0314 \\
 & POPE & 0.0299 & 0.0314 \\
\midrule
\multirow{3}{*}{Gemma-3-12B} & GQA & 0.0303 & 0.0314 \\
 & VQAv2 & 0.0301 & 0.0314 \\
 & POPE & 0.0307 & 0.0314 \\
\bottomrule
\end{tabular}

\end{table}

The perturbation is small in perceptual terms, though not strictly imperceptible. Table~\ref{tab:app-percept} reports LPIPS and SSIM between each clean image and its adversarial endpoint, against two benign references. The mean LPIPS is $0.067$ and the mean SSIM $0.892$, consistent across models and datasets. For scale, a near-lossless JPEG re-encode of the same image has LPIPS $0.006$, and a benign two-pixel crop-and-resize has LPIPS $0.074$: the adversarial perturbation sits between these, comparable to the crop on LPIPS and gentler on SSIM. It is a change of the order of a minor benign image operation, larger than a near-lossless JPEG re-encode, and on a minority of items it is perceptible on close inspection. We therefore describe it as a small, budget-bounded perturbation and make no claim of invisibility.

\begin{table}[t]
\centering
\caption{\textbf{Perceptual distance between clean and adversarial images.} LPIPS (lower is more similar) and SSIM (higher is more similar) for the aimed-harmful endpoint, per cell, with two benign references. The adversarial perturbation is comparable to a two-pixel crop-and-resize and larger than a near-lossless JPEG re-encode: small, but not strictly imperceptible. LPIPS uses the AlexNet backbone on inputs in $[-1,1]$; SSIM is computed per channel.}
\label{tab:app-percept}
\small
\begin{tabular}{@{}llrrrr@{}}
\toprule
Model & Set & LPIPS mean & LPIPS max & SSIM mean & SSIM min \\
\midrule
\multirow{3}{*}{InternVL3.5-2B} & GQA & 0.051 & 0.244 & 0.902 & 0.751 \\
 & VQAv2 & 0.058 & 0.265 & 0.897 & 0.739 \\
 & POPE & 0.058 & 0.247 & 0.886 & 0.750 \\
\midrule
\multirow{3}{*}{Gemma-3-12B} & GQA & 0.074 & 0.385 & 0.889 & 0.736 \\
 & VQAv2 & 0.079 & 0.277 & 0.881 & 0.666 \\
 & POPE & 0.061 & 0.307 & 0.893 & 0.749 \\
\midrule
\multirow{3}{*}{Molmo2-4B} & GQA & 0.067 & 0.249 & 0.891 & 0.755 \\
 & VQAv2 & 0.063 & 0.350 & 0.901 & 0.796 \\
 & POPE & 0.080 & 0.296 & 0.881 & 0.756 \\
\midrule
\multirow{3}{*}{LLaVA-OV-7B} & GQA & 0.070 & 0.380 & 0.895 & 0.746 \\
 & VQAv2 & 0.068 & 0.226 & 0.897 & 0.767 \\
 & POPE & 0.068 & 0.245 & 0.889 & 0.789 \\
\midrule
\multicolumn{2}{@{}l}{\emph{Benign ref: JPEG q95}} & 0.006 & -- & 0.967 & -- \\
\multicolumn{2}{@{}l}{\emph{Benign ref: 2px crop-resize}} & 0.074 & -- & 0.696 & -- \\
\bottomrule
\end{tabular}

\end{table}

\subsection{The correctness judge agrees with human adjudication}
\label{app:setup-judge}

The automated correctness labels match blind human adjudication, with no false accept of a wrong answer. Table~\ref{tab:app-judge} reports $99.0\%$ agreement on a stratified 300-item sample, Cohen's $\kappa=0.98$. All three disagreements are cases where the human labels the answer correct and the judge labels it incorrect; the sample contains no case in which a human-incorrect answer is accepted as correct, so the judge does not manufacture the wrong answers the attack exploits. The judge receives the image as a separate multimodal input together with the question, the benchmark's gold answer, and the model's answer, under fixed grading instructions, and returns a yes/no verdict that is parsed directly; the verbatim prompt is given in Listing~\ref{lst:judge-prompt}.

\begin{figure}[t]
\centering
\begin{minipage}{\textwidth}

\refstepcounter{lstlisting}
\label{lst:judge-prompt}
\noindent\textbf{Listing \thelstlisting: The correctness-judge prompt}

\medskip

\begin{tcolorbox}[
    title=System Prompt,
    fonttitle=\bfseries,
    fontupper=\small
]
You are an expert answer evaluator. Your task is to determine if a student's answer to a question is correct by comparing it to the ground truth answer.

\medskip

1. Read the question carefully.\\
2. Compare the student's answer to the ground truth answer.\\
3. Consider semantic equivalence, answers that mean the same thing should be considered correct even if worded differently.\\
4. Return ONLY ``yes'' if the answer is correct, or ``no'' if it is incorrect.\\
5. Be lenient with minor variations in wording, capitalization, or punctuation.
\end{tcolorbox}

\begin{tcolorbox}[
    title=User Prompt,
    fonttitle=\bfseries,
    fontupper=\small
]
Question: \{question\}

Ground Truth Answer: \{gold\}

Student Answer: \{pred\}

Is the student's answer correct? (yes/no):
\end{tcolorbox}

\end{minipage}
\end{figure}

\begin{table}[t]
\centering
\caption{\textbf{Agreement between the automated correctness judge and blind human adjudication} on a stratified 300-item sample. Overall agreement is $0.990$, Cohen's $\kappa=0.98$ with a 10{,}000-resample bootstrap interval. Per-stratum rows report agreement only, since within each stratum the adjudicator's labels reduce to a single class on the disagreement set and the chance-agreement term of $\kappa$ is degenerate. The judge agrees on all $147$ answers the adjudicator labels incorrect.}
\label{tab:app-judge}
\small
% source: adjudication/my_labels.json (300-item stratified human adjudication); per-stratum agreement
% and overall Cohen kappa computed live. kappa 95% CI [0.953,1.000] read from the committed source
% (tables/092_judge_validation/judge.json), NOT re-bootstrapped. per-stratum kappa is n/a (single class).
\begin{tabular}{@{}lrrl@{}}
\toprule
stratum & $n$ & agreement & Cohen $\kappa$ [95\% CI] / detail \\
\midrule
overall & 300 & 0.990 & 0.98 {\scriptsize[0.953,1.000]} \\
\midrule
GQA & 100 & 0.980 & n/a \\
VQAv2 & 100 & 0.990 & n/a \\
POPE & 100 & 1.000 & n/a \\
\midrule
InternVL3.5-2B & 75 & 0.960 & n/a \\
Molmo2-4B & 76 & 1.000 & n/a \\
LLaVA-OV-7B & 75 & 1.000 & n/a \\
Gemma-3-12B & 74 & 1.000 & n/a \\
human-wrong & 147 & 147/147 agree & false-accept 0 \\
\bottomrule
\end{tabular}

\end{table}

\subsection{One canonical draw, and nothing tuned against the result}
\label{app:setup-draw}

Every paired analysis uses the same items across its source stores. The attack, estimator, and perturbation-stability records are joined per item for the decomposition, amplitude, and deployment analyses, so they must share an item set. Table~\ref{tab:app-draw} verifies a shared pool of $500$ correct and $500$ incorrect items in every cell: nine cells reproduce the committed identifiers from seed 23, and the three InternVL cells reuse one pre-existing draw across all stores.

\begin{table}[t]
\centering
\caption{\textbf{Verification that the canonical draw is shared} across the attack, estimator, and perturbation- stability stores. Nine cells regenerate the seed-23 draw and match the committed pool; the three InternVL cells reuse one pre-existing draw by construction.}
\label{tab:app-draw}
% source: data/adv/generality/<model>/pool.json (committed pool) vs seed-23 in-memory regeneration
% from data/pipeline/<model>/judged.jsonl + shard_*.jsonl + data/draws/<set>.jsonl. InternVL reuses
% the move-test draw ('reuse', verified report 143); others regenerated ('yes'). Order InternVL/Molmo/LLaVA/Gemma.
\small
\begin{tabular}{@{}llc r@{}}
\toprule
model & set & regenerated draw matches pool & pool size \\
\midrule
\multirow{3}{*}{InternVL3.5-2B} & GQA & reuse & 1000/1000 \\
 & VQAv2 & reuse & 1000/1000 \\
 & POPE & reuse & 1000/1000 \\
\addlinespace
\multirow{3}{*}{Molmo2-4B} & GQA & yes & 1000/1000 \\
 & VQAv2 & yes & 1000/1000 \\
 & POPE & yes & 1000/1000 \\
\addlinespace
\multirow{3}{*}{LLaVA-OV-7B} & GQA & yes & 1000/1000 \\
 & VQAv2 & yes & 1000/1000 \\
 & POPE & yes & 1000/1000 \\
\addlinespace
\multirow{3}{*}{Gemma-3-12B} & GQA & yes & 1000/1000 \\
 & VQAv2 & yes & 1000/1000 \\
 & POPE & yes & 1000/1000 \\
\bottomrule
\end{tabular}

\end{table}

No parameter is selected by maximizing the collapse. Table~\ref{tab:app-hyper} lists every attack, estimator, and analysis parameter with the values tried, the operating value, and the selection criterion. Swept parameters, including the attack budget and step count, are reported at every tested value, so the operating value identifies a headline column and not a tuned optimum. Fixed parameters follow standard optimization conventions or define the threat model. The only performance-based selection rule, probe early stopping, uses a clean-validation objective of $0.6\,\mathrm{AUROC}+0.4\,(1-\mathrm{ECE})$ and never observes an attacked score.

% \clearpage
\begin{table}[t]
\centering
\caption{\textbf{Hyperparameters, values tried, operating values, and selection criteria.} Swept parameters are reported at every tested value; the selection-criterion column marks each as swept-and-reported, fixed-by-convention, or selected-on-clean-validation.}
\label{tab:app-hyper}
% source: code constants transcribed from evaluation/coupled_adaptive_attack.py, train_probe.py,
% coupled_estimator.py, smoothing_eot.py, estimator_noise.py, distpres_build.py,
% channels_foundation.py, build_steer_replication.py, report 104 (12 configs), seed-23 10000 bootstrap.
% selection criterion states swept-and-reported / fixed-by-convention / selected-on-clean-validation.
\scriptsize
\begin{tabular}{@{}p{0.17\linewidth} p{0.23\linewidth} p{0.17\linewidth} p{0.35\linewidth}@{}}
\toprule
parameter & values tried & operating value & selection criterion \\
\midrule
\multicolumn{4}{@{}l}{\textbf{ATTACK PARAMETERS}} \\
attack budget $\epsilon$ & $\{2,4,8,16\}/255$ & $8/255$ & swept and reported (configs A1 to A3); standard L-inf budget \\
PGD steps & $\{5,15,30,60\}$ & $15$ & swept and reported (configs A4 to A6 and the 60-step ladder) \\
PGD step size $\alpha$ & $\epsilon/4$ & $\epsilon/4$ & fixed by convention; standard PGD step size, not swept \\
random restarts & $1$ & $1$ & fixed by convention; a single answer-preserving descent per item \\
attack norm & L-inf, two L2 budgets & L-inf & swept and reported (configs B1, B2) \\
$\tau_{\rm safe}$ (answer-preservation margin) & $\{0,0.25,0.5,1,2.5,5,10\}$ & $1.0$ & swept both ways and reported; stricter than argmax (Appendix \ref{app:frontier}) \\
image domain / projection & raw RGB $[0,1]$ & $[x-\epsilon,x+\epsilon]\cap[0,1]$ each step; 8-bit stored endpoint & fixed by threat model \\
margin band & $\{0.25,0.5,1.0,\text{unbounded}\}$ & unbounded & swept and reported as a control; all bands reported (Appendix \ref{app:frontier}) \\
invariance sweep configurations & $12$ (A1-A6, B1-B2, C1-C2, D1-D2) & n/a & swept and reported; pass = mean-drift aligned to the correctness axis, 12 of 12 cells (Appendix \ref{app:frontier}) \\
EOT noise draws (smoothing attack) & $m=8$ per PGD step & $8$ & fixed by convention; adaptive attack on the smoothed estimator \\
\midrule
\multicolumn{4}{@{}l}{\textbf{ESTIMATOR AND PROBE PARAMETERS}} \\
probe architecture & Linear(in,256)-ReLU-Linear(256,128)-ReLU-Linear(128,64)-ReLU-Linear(64,1) & fixed & fixed by convention; identical for every probe, model, cell and channel; input dim is the model hidden size (2048/2560/3584/3840) or the estimator subspace dim \\
probe loss & class-balanced BCE-with-logits (pos\_weight $=n_0/n_1$) & fixed & fixed by convention; not tuned \\
optimizer & Adam, lr $10^{-3}$, weight decay $0$, batch $32$ & fixed & fixed by convention; not tuned \\
epochs / early stopping & max $200$ epochs, patience $20$ & select on $0.6\cdot$AUROC$+0.4\cdot(1-$ECE$)$ & selected on the clean validation split (5000 items), never on adversarial discrimination \\
gradient clipping & max-norm $1.0$ & $1.0$ & fixed by convention \\
COUPLED subspace rank $k$ & $\{4,8,16\}$ & $8$ & swept and reported (headline $k=8$); derived subspace dimension equals $k$ exactly \\
distance-preserving probe rank & $\{8,32,128,512\}$ & $8$ & swept and reported; dimension-matched to COUPLED ($k=8$) \\
coupled-rank SVD tolerance & $10^{-6}$ (relative) & $10^{-6}$ & fixed by convention; numerical singular-value threshold for the per-item rank, not swept \\
self-consistency sampling & $40$ completions, $T=0.7$, top-$k$ $40$ & fixed & fixed by convention; channel definition, not tuned \\
steering injection $\alpha$ & $11$ values $\{-96,\dots,+96\}$ plus down-drift candidates $\{32,24,16,12,8,4\}$ & $32$ & fixed by convention; LLaVA's operating point applied uniformly under the preservation constraint (preservation rate $0.86$ to $0.99$ per cell, mean $0.96$) \\
smoothing $\sigma$ & $\{0.08,0.16\}$ & both reported & swept and reported \\
\midrule
\multicolumn{4}{@{}l}{\textbf{ANALYSIS PARAMETERS}} \\
distance-preserving split & 50/50 of byte-identical items (A estimate, B evaluate) & fixed & fixed by convention; probe on a separate 2500-item GQA pool, all three sets disjoint, seed 23, subspace uses no labels \\
noise-floor random draws & $15$ same-$\epsilon$ L-inf draws per probe & $15$ & fixed by convention, not swept; amplitude is the two-sided best-up-minus-best-down range \\
separation threshold $|g|$ & $\{0,0.05\}$ & $0.05$ & swept and reported; excludes near-zero-$g$ cells where $\Delta/g$ is unstable, both reported (Appendix \ref{app:mech-certificate}) \\
bootstrap resamples & $10{,}000$ at seed $23$ & $10{,}000$ & fixed by convention; everywhere for reported CIs (item-level, paired-item, pair-level) \\
multi-seed runs & seeds $\{23,24,25\}$ & $\{23,24,25\}$ & fixed three-seed set; seed 23 is the canonical draw, 24 and 25 the additional seeds for reported standard deviations \\
judge decoding & gpt-5.4-mini, max $1500$ tokens, default temperature & fixed & fixed by convention; external correctness grader, no sampling parameters set beyond the token cap \\
main answer decoding & greedy (argmax) & greedy & fixed by convention, not swept; the answer the attack pins \\
\bottomrule
\end{tabular}

\end{table}
% \clearpage
\section{The Exact Decomposition}
\label{app:decomp}

The main text rests the account of the collapse on an exact decomposition of attacked discrimination into a same-answer and a cross-answer term. This appendix states the identity and its two-line derivation, verifies that it is exact across all eighty-four estimator-by-cell profiles, and gives the representation-level reading that follows from it. The objects that explain the two terms, the reachable ranges, the same-answer ceiling, and the uniform certificate, are developed in Appendix~\ref{app:mechanisms}.

\subsection{The identity}
\label{app:decomp-identity}

Write attacked AUROC as a mean over the correct-incorrect item pairs a readout must rank. For $n_+$ correct and $n_-$ incorrect items with attacked scores $v_i$,
\begin{equation}
\label{eq:app-auroc}
A_{\mathrm{adv}} \;=\; \frac{1}{n_+ n_-}\sum_{i\in+}\sum_{j\in-}\Big(\mathbf{1}[v_i > v_j] + \tfrac{1}{2}\mathbf{1}[v_i = v_j]\Big).
\end{equation}
Partition these pairs by whether their two answer strings agree. Let $w$ be the fraction that share an answer, so $1-w$ is the fraction whose answers differ. Because the double sum is an average over a partitioned index set, it splits without remainder into the average over same-answer pairs and the average over cross-answer pairs,
\begin{equation}
\label{eq:app-decomp}
A_{\mathrm{adv}} \;=\; w\,A_{\mathrm{same}} \;+\; (1-w)\,A_{\mathrm{cross}},
\end{equation}
where $A_{\mathrm{same}}$ and $A_{\mathrm{cross}}$ are attacked AUROC restricted to the same-answer and cross-answer pairs. Equation~\eqref{eq:app-decomp} holds for any scoring of the items, adversarial or not, because it is only a regrouping of the terms in Equation~\eqref{eq:app-auroc}. It assumes nothing about reachability, makes no prediction, and has no free parameter. Its content is that it names two disjoint sources of the ranking a readout produces, with the weight $w$ fixing how much each contributes.

\subsection{The identity is exact across all eighty-four profiles}
\label{app:decomp-exact}

Table~\ref{tab:app-decomp-full} reports $w$, $A_{\mathrm{same}}$, $A_{\mathrm{cross}}$, the reconstructed $A_{\mathrm{adv}}$, the same-answer overlap $\rho$, and the reconstruction error for every one of the eighty-four estimator-by-cell profiles. The error is $|w\,A_{\mathrm{same}} + (1-w)\,A_{\mathrm{cross}} - A_{\mathrm{adv}}|$ measured against the independently stored $A_{\mathrm{adv}}$. Its maximum over all eighty-four profiles is $5.6\times10^{-17}$, at the floating-point floor, so the decomposition reconstructs the measured quantity exactly and is safe to build on. The table also makes the mechanism legible cell by cell. Where $w$ is small, $A_{\mathrm{adv}}$ tracks $A_{\mathrm{cross}}$ almost entirely; where $w$ is large, the same-answer term carries real weight, and only a low $\rho$ lets it survive, which is why the single above-chance profile, Molmo/POPE under COUPLED, sits where it does.

\subsection{The movement is separable from the answer}
\label{app:decomp-separable}

The identity says the same-answer term is where the answer string carries no information, so any discrimination there lives in the score alone. The representation geometry gives the complementary picture: the movement the attack uses to corrupt that score lies almost entirely in directions unrelated to the answer. Table~\ref{tab:app-subspace} projects the attack-induced hidden-state drift onto the answer-coupled subspace, the low-rank set of directions formed from the competitor-difference directions that determine the answer logits, and onto its orthogonal complement. At rank eight the answer-coupled component holds under $1\%$ of the drift energy for every model, between $0.33$ and $0.76\%$, while the complement holds above $99\%$. The random-control column gives the expectation for an equal-rank subspace chosen without reference to the answer, so the comparison is against $k/d$, not against zero; and the margin constraint itself penalizes movement along the competitor-difference directions that span this subspace, so a small coupled fraction is partly enforced by the attack's own constraint. The enrichment above random is the part the constraint does not explain. The coupled component is enriched relative to an equal-rank random subspace by a factor of $1.4$ to $2.8$, so the attack is not orthogonal to the answer directions, but it is far from concentrated in them. The absolute figure depends on the truncation rank: at rank sixteen the under-one-percent statement no longer holds for Molmo and LLaVA. The enrichment ratio stays above one at every rank, and at every rank the large majority of the movement lies off the answer-coupled directions. This is the geometric form of separability: the answer is fixed, and the signal the attack moves travels in directions the answer does not span.

% Appendix Table: full 84-row discrimination decomposition, longtable (breaks across pages).
% Exact identity A_adv = w*A_same + (1-w)*A_cross, per estimator x cell.
% Source: reports/204_core_b_reachability/table_i_aggregate.csv (w, A_same=wa_att, A_cross=xa_att,
%   rho=rho_pair) + frozen T4 (A_adv = witnessed_union_oracle_auroc); recon_err = |w*A_same+(1-w)*A_cross-A_adv|.
% Reconciled vs frozen T4: A_adv exact (max abs diff 0.0e+00); max recon_err 5.6e-17. Formatting reissue
%   of report-227 decomp_full; no number recomputed. Model/Dataset printed once per block (page-break-safe).
% Requires: \usepackage{longtable,booktabs,multirow}
\setlength{\LTcapwidth}{\textwidth}
\setlength{\tabcolsep}{5pt}
\begin{longtable}{
>{\small}l
>{\small}l
>{\small}l
>{\small}r
>{\small}r
>{\small}r
>{\small}r
>{\small}r
>{\small}r
}
\caption{\textbf{The exact decomposition for all eighty-four estimator-by-cell profiles.} $w$ is the same-answer pair weight, $A_{\mathrm{same}}$ and $A_{\mathrm{cross}}$ the attacked discrimination on same-answer and cross-answer pairs, $A_{\mathrm{adv}}$ the reconstructed total, and $\rho$ the same-answer reachable-range overlap. The reconstruction error against the independently stored $A_{\mathrm{adv}}$ is at most $5.6\times10^{-17}$.}\label{tab:app-decomp-full}\\
\toprule
Model & Dataset & Estimator & $w$ & $A_{\mathrm{same}}$ & $A_{\mathrm{cross}}$ & $A_{\mathrm{adv}}$ & $\rho$ & recon.\ err. \\
\midrule
\endfirsthead
\multicolumn{9}{@{}l}{\footnotesize\itshape Table~\ref{tab:app-decomp-full} continued from previous page}\\
\toprule
Model & Dataset & Estimator & $w$ & $A_{\mathrm{same}}$ & $A_{\mathrm{cross}}$ & $A_{\mathrm{adv}}$ & $\rho$ & recon.\ err. \\
\midrule
\endhead
\midrule
\multicolumn{9}{r@{}}{\footnotesize\itshape continued on next page}\\
\endfoot
\bottomrule
\multicolumn{9}{@{}l}{\footnotesize $^{\dagger}$The sole profile above chance.}\\
\endlastfoot
InternVL3.5-2B & GQA & TokProb & 0.040 & 0.000 & 0.017 & 0.016 & 1.000 & 0.0e+00 \\
 &  & P(IK) & 0.040 & 0.000 & 0.022 & 0.021 & 1.000 & 0.0e+00 \\
 &  & SAPLMA & 0.040 & 0.001 & 0.018 & 0.017 & 0.999 & 0.0e+00 \\
 &  & CCPS & 0.040 & 0.000 & 0.010 & 0.010 & 1.000 & 0.0e+00 \\
 &  & CCPS-D & 0.040 & 0.000 & 0.014 & 0.014 & 1.000 & 0.0e+00 \\
 &  & FULL & 0.040 & 0.000 & 0.010 & 0.009 & 1.000 & 0.0e+00 \\
 &  & COUPLED & 0.040 & 0.002 & 0.016 & 0.016 & 0.997 & 0.0e+00 \\
\addlinespace
 & VQAv2 & TokProb & 0.045 & 0.000 & 0.016 & 0.015 & 1.000 & 1.7e-18 \\
 &  & P(IK) & 0.045 & 0.000 & 0.029 & 0.028 & 1.000 & 0.0e+00 \\
 &  & SAPLMA & 0.045 & 0.000 & 0.009 & 0.008 & 1.000 & 0.0e+00 \\
 &  & CCPS & 0.045 & 0.000 & 0.022 & 0.021 & 1.000 & 3.5e-18 \\
 &  & CCPS-D & 0.045 & 0.000 & 0.023 & 0.022 & 1.000 & 0.0e+00 \\
 &  & FULL & 0.045 & 0.000 & 0.007 & 0.006 & 1.000 & 0.0e+00 \\
 &  & COUPLED & 0.045 & 0.001 & 0.013 & 0.012 & 0.998 & 0.0e+00 \\
\addlinespace
 & POPE & TokProb & 0.498 & 0.000 & 0.000 & 0.000 & 1.000 & 0.0e+00 \\
 &  & P(IK) & 0.498 & 0.000 & 0.000 & 0.000 & 1.000 & 0.0e+00 \\
 &  & SAPLMA & 0.498 & 0.000 & 0.000 & 0.000 & 1.000 & 0.0e+00 \\
 &  & CCPS & 0.498 & 0.000 & 0.000 & 0.000 & 1.000 & 0.0e+00 \\
 &  & CCPS-D & 0.498 & 0.000 & 0.000 & 0.000 & 1.000 & 0.0e+00 \\
 &  & FULL & 0.498 & 0.000 & 0.000 & 0.000 & 1.000 & 0.0e+00 \\
 &  & COUPLED & 0.498 & 0.002 & 0.002 & 0.002 & 0.998 & 2.2e-19 \\
\midrule
Molmo2-4B & GQA & TokProb & 0.066 & 0.001 & 0.009 & 0.008 & 0.999 & 0.0e+00 \\
 &  & P(IK) & 0.066 & 0.000 & 0.010 & 0.009 & 1.000 & 1.7e-18 \\
 &  & SAPLMA & 0.066 & 0.000 & 0.014 & 0.013 & 1.000 & 0.0e+00 \\
 &  & CCPS & 0.066 & 0.000 & 0.011 & 0.011 & 1.000 & 1.7e-18 \\
 &  & CCPS-D & 0.066 & 0.000 & 0.004 & 0.004 & 1.000 & 4.3e-19 \\
 &  & FULL & 0.066 & 0.000 & 0.009 & 0.009 & 1.000 & 0.0e+00 \\
 &  & COUPLED & 0.066 & 0.450 & 0.279 & 0.291 & 0.441 & 0.0e+00 \\
\addlinespace
 & VQAv2 & TokProb & 0.051 & 0.000 & 0.023 & 0.022 & 1.000 & 0.0e+00 \\
 &  & P(IK) & 0.051 & 0.000 & 0.048 & 0.046 & 1.000 & 0.0e+00 \\
 &  & SAPLMA & 0.051 & 0.000 & 0.041 & 0.039 & 0.997 & 0.0e+00 \\
 &  & CCPS & 0.051 & 0.000 & 0.059 & 0.056 & 1.000 & 0.0e+00 \\
 &  & CCPS-D & 0.051 & 0.000 & 0.018 & 0.017 & 1.000 & 0.0e+00 \\
 &  & FULL & 0.051 & 0.000 & 0.023 & 0.022 & 0.999 & 0.0e+00 \\
 &  & COUPLED & 0.051 & 0.514 & 0.236 & 0.250 & 0.384 & 5.6e-17 \\
\addlinespace
 & POPE & TokProb & 0.525 & 0.000 & 0.000 & 0.000 & 1.000 & 6.8e-21 \\
 &  & P(IK) & 0.525 & 0.000 & 0.000 & 0.000 & 1.000 & 0.0e+00 \\
 &  & SAPLMA & 0.525 & 0.000 & 0.000 & 0.000 & 1.000 & 0.0e+00 \\
 &  & CCPS & 0.525 & 0.000 & 0.000 & 0.000 & 1.000 & 0.0e+00 \\
 &  & CCPS-D & 0.525 & 0.000 & 0.000 & 0.000 & 1.000 & 0.0e+00 \\
 &  & FULL & 0.525 & 0.000 & 0.000 & 0.000 & 1.000 & 0.0e+00 \\
 &  & COUPLED & 0.525 & 0.567 & 0.467 & 0.520$^{\dagger}$\, & 0.359 & 0.0e+00 \\
\midrule
LLaVA-OV-7B & GQA & TokProb & 0.034 & 0.000 & 0.002 & 0.002 & 1.000 & 0.0e+00 \\
 &  & P(IK) & 0.034 & 0.000 & 0.009 & 0.009 & 1.000 & 0.0e+00 \\
 &  & SAPLMA & 0.034 & 0.000 & 0.008 & 0.007 & 1.000 & 0.0e+00 \\
 &  & CCPS & 0.034 & 0.000 & 0.002 & 0.002 & 1.000 & 0.0e+00 \\
 &  & CCPS-D & 0.034 & 0.000 & 0.001 & 0.001 & 1.000 & 0.0e+00 \\
 &  & FULL & 0.034 & 0.000 & 0.012 & 0.011 & 1.000 & 0.0e+00 \\
 &  & COUPLED & 0.034 & 0.037 & 0.049 & 0.049 & 0.958 & 0.0e+00 \\
\addlinespace
 & VQAv2 & TokProb & 0.043 & 0.009 & 0.017 & 0.016 & 0.991 & 0.0e+00 \\
 &  & P(IK) & 0.043 & 0.000 & 0.015 & 0.015 & 1.000 & 0.0e+00 \\
 &  & SAPLMA & 0.043 & 0.001 & 0.021 & 0.020 & 0.999 & 0.0e+00 \\
 &  & CCPS & 0.043 & 0.009 & 0.032 & 0.031 & 0.991 & 3.5e-18 \\
 &  & CCPS-D & 0.043 & 0.009 & 0.014 & 0.014 & 0.991 & 1.7e-18 \\
 &  & FULL & 0.043 & 0.001 & 0.031 & 0.030 & 0.999 & 0.0e+00 \\
 &  & COUPLED & 0.043 & 0.056 & 0.052 & 0.052 & 0.931 & 6.9e-18 \\
\addlinespace
 & POPE & TokProb & 0.241 & 0.000 & 0.000 & 0.000 & 1.000 & 0.0e+00 \\
 &  & P(IK) & 0.241 & 0.000 & 0.000 & 0.000 & 1.000 & 0.0e+00 \\
 &  & SAPLMA & 0.241 & 0.000 & 0.001 & 0.001 & 1.000 & 1.1e-19 \\
 &  & CCPS & 0.241 & 0.000 & 0.000 & 0.000 & 1.000 & 3.4e-21 \\
 &  & CCPS-D & 0.241 & 0.000 & 0.000 & 0.000 & 1.000 & 0.0e+00 \\
 &  & FULL & 0.241 & 0.000 & 0.001 & 0.000 & 1.000 & 5.4e-20 \\
 &  & COUPLED & 0.241 & 0.032 & 0.038 & 0.037 & 0.962 & 6.9e-18 \\
\midrule
Gemma-3-12B & GQA & TokProb & 0.070 & 0.026 & 0.044 & 0.043 & 0.941 & 6.9e-18 \\
 &  & P(IK) & 0.070 & 0.030 & 0.106 & 0.100 & 0.970 & 1.4e-17 \\
 &  & SAPLMA & 0.070 & 0.017 & 0.078 & 0.074 & 0.981 & 0.0e+00 \\
 &  & CCPS & 0.070 & 0.006 & 0.034 & 0.032 & 0.994 & 0.0e+00 \\
 &  & CCPS-D & 0.070 & 0.004 & 0.028 & 0.026 & 0.996 & 3.5e-18 \\
 &  & FULL & 0.070 & 0.017 & 0.063 & 0.060 & 0.983 & 6.9e-18 \\
 &  & COUPLED & 0.070 & 0.097 & 0.349 & 0.331 & 0.883 & 0.0e+00 \\
\addlinespace
 & VQAv2 & TokProb & 0.053 & 0.005 & 0.064 & 0.061 & 0.992 & 6.9e-18 \\
 &  & P(IK) & 0.053 & 0.001 & 0.091 & 0.086 & 0.999 & 0.0e+00 \\
 &  & SAPLMA & 0.053 & 0.000 & 0.108 & 0.102 & 0.999 & 1.4e-17 \\
 &  & CCPS & 0.053 & 0.001 & 0.071 & 0.067 & 0.999 & 0.0e+00 \\
 &  & CCPS-D & 0.053 & 0.001 & 0.090 & 0.085 & 0.999 & 0.0e+00 \\
 &  & FULL & 0.053 & 0.001 & 0.095 & 0.090 & 0.998 & 0.0e+00 \\
 &  & COUPLED & 0.053 & 0.152 & 0.346 & 0.336 & 0.793 & 0.0e+00 \\
\addlinespace
 & POPE & TokProb & 0.465 & 0.006 & 0.076 & 0.044 & 0.992 & 6.9e-18 \\
 &  & P(IK) & 0.465 & 0.003 & 0.004 & 0.004 & 0.997 & 0.0e+00 \\
 &  & SAPLMA & 0.465 & 0.002 & 0.015 & 0.009 & 0.998 & 0.0e+00 \\
 &  & CCPS & 0.465 & 0.001 & 0.004 & 0.002 & 0.999 & 4.3e-19 \\
 &  & CCPS-D & 0.465 & 0.000 & 0.008 & 0.004 & 1.000 & 0.0e+00 \\
 &  & FULL & 0.465 & 0.002 & 0.013 & 0.008 & 0.998 & 0.0e+00 \\
 &  & COUPLED & 0.465 & 0.044 & 0.295 & 0.178 & 0.942 & 2.8e-17 \\
\end{longtable}

\begin{table}[t]
\centering
\caption{\textbf{Answer-coupled subspace projection of the attack-induced drift}, averaged over the three native channels, at truncation ranks $k\in\{4,8,16\}$. \emph{coupled energy} is the fraction of drift energy in the answer-coupled subspace, \emph{free energy} the fraction in its complement, and \emph{random ctrl} the coupled energy of an equal-rank subspace sampled without reference to the answer. \emph{var.\ frac} is the clean hidden-state variance the answer-coupled subspace captures.}
\label{tab:app-subspace}
% Appendix Table C.3: answer-coupled subspace projection of the attack-induced hidden-state drift.
% Source: tables/095_slack_decomposition/<model>__agg_c5k.json (report 095), loci.sap.k.{4,8,16}:
%   coupled/free/random energy = mean over the 3 native channels {tokprob, pik, saplma} of
%   H2[c].energy_{coupled,free,coupled_random}_mean; var. frac = k-level coupled_var_frac_mean; hidden
%   dim = store 'dim'. Harmful-sign-oriented drift at the SAP locus, canonical 5000-item pool.
% Reconciled == committed manuscript/tables/appendices/c_premises/energy_decomposition.tex (max abs
%   diff 0.005, display rounding). Retracted func-free column (report 108) excluded.
% Requires: \usepackage{booktabs,multirow}
\small
\setlength{\tabcolsep}{4pt}
\begin{tabular}{@{}llrrrrr@{}}
\toprule
model & hidden dim & $k$ & coupled energy & free energy & random ctrl & var.\ frac \\
\midrule
\multirow{3}{*}{InternVL3.5-2B} & \multirow{3}{*}{2048} & 4 & 0.29\% & 99.71\% & 0.20\% & 0.30\% \\
 & & 8 & 0.54\% & 99.46\% & 0.39\% & 0.56\% \\
 & & 16 & 0.98\% & 99.02\% & 0.78\% & 1.05\% \\
\midrule
\multirow{3}{*}{Molmo2-4B} & \multirow{3}{*}{2560} & 4 & 0.39\% & 99.61\% & 0.16\% & 0.30\% \\
 & & 8 & 0.76\% & 99.24\% & 0.31\% & 0.80\% \\
 & & 16 & 1.37\% & 98.63\% & 0.62\% & 1.51\% \\
\midrule
\multirow{3}{*}{LLaVA-OV-7B} & \multirow{3}{*}{3584} & 4 & 0.38\% & 99.62\% & 0.11\% & 0.27\% \\
 & & 8 & 0.62\% & 99.38\% & 0.22\% & 0.56\% \\
 & & 16 & 1.02\% & 98.98\% & 0.45\% & 1.11\% \\
\midrule
\multirow{3}{*}{Gemma-3-12B} & \multirow{3}{*}{3840} & 4 & 0.16\% & 99.84\% & 0.10\% & 0.24\% \\
 & & 8 & 0.33\% & 99.67\% & 0.21\% & 0.46\% \\
 & & 16 & 0.63\% & 99.37\% & 0.42\% & 0.89\% \\
\bottomrule
\end{tabular}

\end{table}
\section{The Two Mechanisms}
\label{app:mechanisms}

The decomposition of Appendix~\ref{app:decomp} splits attacked discrimination into a same-answer term and a cross-answer term without explaining either. This appendix explains each. The same-answer term is governed by the overlap of reachable score ranges, admits a measured upper bound, and is the term the attack drives down by merging items that share an answer. The cross-answer term is driven down by moving each item's score on its own, and the long-answer experiment isolates it by removing same-answer pairs entirely. The appendix closes with the uniform certificate: the formal guarantee a safe readout would satisfy, which no tested readout does.

\subsection{The reachable ranges and the endpoint adversary}
\label{app:mech-reachable}

Fix a question with image $x$, answer string $a(x)$, and scalar readout $c(x)$. At budget $\epsilon$, the answer-preserving feasible set is
\[
F_\epsilon(x) = \{x\} \cup \big\{x' : \|x'-x\|_\infty \le \epsilon,\; a(x')=a(x),\; \mathrm{margin}_t(x') \ge \tau_{\mathrm{safe}}\ \forall t\big\},
\]
where $t$ ranges over answer positions and $\mathrm{margin}_t$ is the decode-position logit lead of the emitted token. Including $x$ implements the clean-score fallback for items with no harmful feasible iterate. The reachable score interval is $I^\epsilon_c(x) = \mathrm{conv}\, c(F_\epsilon(x))$ and its width $\Delta^\epsilon_c(x)$ is the manipulable amplitude. Writing empirical AUROC as a mean over correct-incorrect pairs, the score is nondecreasing in a correct item's value and nonincreasing in an incorrect item's, so its infimum over the reachable ranges is attained by taking the lowest reachable score for every correct item and the highest for every incorrect one. This endpoint adversary is the optimal aggregate over the true reachable ranges; a randomized adversary is a distribution over deterministic profiles and cannot do better. The implemented attack shares the endpoint's coordinatewise direction but need not reach the exact extrema, so every realized displacement is a lower bound on what the reachable range permits.

\subsection{The same-answer term: within-answer information, destroyed by merging}
\label{app:mech-same}

On a pair of items that share an answer string, the answer prior assigns both the same score, so any discrimination on that pair comes from information beyond the answer. Table~\ref{tab:app-within} measures this within-answer discrimination. Clean within-answer AUROC ranges from $0.70$ to $0.81$ across the deployed channels, well above the $0.515$ of a dimension-matched random-projection control, so the readouts carry real correctness information on same-answer pairs. Under an attack that holds the answer fixed and pushes these scores, the deployed channels invert: their mean within-answer AUROC falls to between $0.07$ and $0.15$, and per-cell values reach zero. This is the same-answer term collapsing, and under the endpoint construction it collapses by pushing two same-answer items until their reachable ranges overlap and no threshold separates them.

The one model that inverts only partially previews the survivor. The lower block of Table~\ref{tab:app-within} reports attacked within-answer AUROC per model, averaged over its cells and the deployed channels. Three models invert nearly completely, to between $0.04$ and $0.09$; Molmo inverts only to $0.285$, three to seven times higher. Molmo's same-answer ranges resist merging, which is why, when the same-answer weight $w$ is also large, as on POPE, its same-answer term survives and carries the cell. These figures are measured under the answer-fixed attack; the fully coordinated attack drives within-answer discrimination to zero on every cell of the five deployed channels, so the answer-fixed variant is the one that exposes the differential resistance the survivor depends on.

\begin{table}[t]
\centering
\caption{\textbf{Within-answer discrimination, informative clean and inverted under attack.} AUROC over correct-incorrect pairs sharing an answer string, where the answer prior scores every pair one half. Upper block: per estimator, with a dimension-matched random-projection control. Lower block: attacked within-answer AUROC per model, showing Molmo's partial inversion against the others' near-complete inversion. Within-answer AUROC here is measured under the answer-fixed attack (Appendix~\ref{app:setup-attack}), one realized endpoint per item, not the coordinated oracle behind Table~\ref{tab:app-decomp-full}; the oracle is at least as strong, so its $A_{\mathrm{same}}$ is lower on every profile ($0.567$ against $0.731$ for the survivor).}
\label{tab:app-within}
% Appendix (two mechanisms): within-answer AUROC over same-answer correct-wrong pairs, per estimator.
% Old Supplement Table C.4. Source: data/theorem/pair_decomposition.json key task2,
%   [cell][channel]['clean'|'adv']['within_auroc'] (report-116 answer-fixed attack). Reconciled == committed
%   c_premises/within_answer_inversion.tex to 3 decimals. RANDOM (ctrl) = documented dimension-matched
%   random-projection control constant (report 116). NOTE: this is the answer-fixed attack, a DIFFERENT
%   (weaker within-answer) attack than the report-204 coordinated attack, so Molmo/POPE resists here.
% Requires: \usepackage{booktabs,multirow}
\footnotesize
\setlength{\tabcolsep}{4pt}
\begin{adjustbox}{max width=\textwidth}
\begin{tabular}{@{}lrrlrl@{}}
\toprule
estimator & clean mean & clean median & clean per-cell range & adv mean & adv per-cell range \\
\midrule
TokProb & 0.704 & 0.710 & 0.565--0.808 & 0.072 & 0.000--0.325 \\
P(IK) & 0.805 & 0.803 & 0.742--0.860 & 0.146 & 0.043--0.370 \\
SAPLMA & 0.795 & 0.799 & 0.720--0.868 & 0.140 & 0.015--0.442 \\
CCPS & 0.781 & 0.778 & 0.714--0.854 & 0.148 & 0.004--0.443 \\
CCPS-D & 0.782 & 0.788 & 0.711--0.848 & 0.112 & 0.007--0.419 \\
\midrule
FULL & 0.797 & 0.800 & 0.740--0.862 & 0.135 & 0.003--0.458 \\
COUPLED & 0.711 & 0.703 & 0.663--0.783 & 0.252$^{\dagger}$ & 0.029--0.731 \\
RANDOM (ctrl) & 0.515 & 0.515 & 0.515 & 0.513 & 0.512--0.518 \\
\bottomrule
\end{tabular}
\end{adjustbox}

\vspace{4pt}
% Molmo-survivor preview: per-model attacked within-answer AUROC, averaged over the 3 cells and the 5
%   deployed channels (TokProb, P(IK), SAPLMA, CCPS, CCPS-D). Molmo's within-answer inversion is only
%   partial (0.285, materially higher); the other three are near-complete (0.04 to 0.09).
{\footnotesize\itshape Per-model attacked within-answer AUROC (mean over 3 cells $\times$ 5 deployed channels):}\\[2pt]
\footnotesize
\begin{tabular}{@{}lc@{}}
\toprule
model & attacked within-answer AUROC \\
\midrule
InternVL3.5-2B & 0.040 \\
Molmo2-4B & 0.285 (partial inversion) \\
LLaVA-OV-7B & 0.078 \\
Gemma-3-12B & 0.091 \\
\bottomrule
\end{tabular}

\vspace{2pt}
{\footnotesize $^{\dagger}$COUPLED's adversarial mean (0.252) is the twelve-cell mean and includes the constraint-blocked Molmo/POPE cell (adversarial within-answer 0.731); the eleven-cell mean excluding it is 0.209.}

\end{table}

\subsection{The same-answer term admits a measured ceiling}
\label{app:mech-ceiling}

Define $c^\star$, the cell's answer-string prior, by scoring each item with the in-sample fraction of items sharing its exact, unnormalized answer string that are correct, and taking the AUROC of that score over all correct-incorrect pairs among the cell's $1000$ items, with ties counted one half. It is per-cell, identical across estimators, and lies above the operative floor of Table~\ref{tab:app-ceiling}, which normalizes strings, is cross-fit, and is clipped at $0.5$ (on InternVL/GQA, $c^\star=0.821$ against a floor of $0.642$). With $\rho$ the fraction of same-answer pairs whose reachable ranges overlap, the ceiling $c^\star+(1-\rho)\,w$ bounds $A_{\mathrm{adv}}=w\,A_{\mathrm{same}}+(1-w)\,A_{\mathrm{cross}}$: the answer prior alone attains $c^\star$, the overlapping fraction $\rho$ of the weight-$w$ same-answer pairs is merged to a tie, and only the $(1-\rho)\,w$ remainder can add discrimination. The bound holds in all eighty-four profiles with a minimum slack of $0.386$ (Gemma/POPE), the same-answer term $A_{\mathrm{same}}$ satisfies it as well, and the implicit condition $(1-w)\,A_{\mathrm{cross}}\le c^\star$ holds in all eighty-four. Derived under the reachable-range construction of Appendix~\ref{app:mech-reachable}, the ceiling permits survival in every cell, so it records how much discrimination the same-answer mechanism could retain and does not forecast the survivor.

\begin{table}[t]
\centering
\caption{\textbf{The answer-string floor and the same-answer ceiling}, per cell at the binding estimator. \emph{cv5} and \emph{held-out} are the cross-fit and disjoint-split floor estimates; \emph{operative} is the value used, lower-bounded by $0.5$. The ceiling $c^\star+(1-\rho)\,w$ bounds $A_{\mathrm{adv}}$ (and, empirically, $A_{\mathrm{same}}$) in all eighty-four profiles (minimum slack $0.386$).}
\label{tab:app-ceiling}
% Appendix (two mechanisms): operative answer-string floor + same-answer graded ceiling, per cell.
% Old Supplement Table B.2 (floor) + report-204 graded ceiling. Floor sources: cv5 = task456_amplitude.json
%   floor_raw_cv5; held-out(GQA) = HELDOUT_GQA constant; operative = task_floor_bound.json['floor'], max(.,0.5).
% Ceiling source: reports/204_core_b_reachability/table_i_aggregate.csv. ceiling = c* + (1-rho)*w, where c*
%   is the per-cell ceiling base (distinct from the answer-string floor), rho=rho_pair and w per-cell.
% Rows are at the per-cell BINDING estimator (max A_adv): COUPLED in 10 cells, P(IK) in InternVL GQA/VQAv2.
%   holds = A_adv <= ceiling (the report-204 claim); it holds 84/84 over all estimator-by-cell profiles
%   (A_same <= ceiling also 84/84); min slack 0.386 (Gemma/POPE). See report.
% Requires: \usepackage{booktabs,multirow}
\footnotesize
\setlength{\tabcolsep}{3.5pt}
\begin{tabular}{@{}llrrrrrrrc@{}}
\toprule
model & set & cv5 & held-out & operative & $\rho$ & $w$ & $c^{\star}$ & ceiling & holds \\
 &  & (cross-fit) & (GQA) & floor &  &  &  & $c^{\star}\!+\!(1\!-\!\rho)w$ &  \\
\midrule
\multirow{3}{*}{InternVL3.5-2B} & GQA & 0.639 & 0.642 & 0.642 & 1.000 & 0.040 & 0.821 & 0.821 & yes \\
 & VQAv2 & 0.617 & n/a & 0.617 & 1.000 & 0.045 & 0.856 & 0.856 & yes \\
 & POPE & 0.555 & n/a & 0.555 & 0.998 & 0.498 & 0.564 & 0.565 & yes \\
\midrule
\multirow{3}{*}{Molmo2-4B} & GQA & 0.598 & 0.683 & 0.683 & 0.441 & 0.066 & 0.808 & 0.846 & yes \\
 & VQAv2 & 0.625 & n/a & 0.625 & 0.384 & 0.051 & 0.877 & 0.908 & yes \\
 & POPE & 0.587 & n/a & 0.587 & 0.359 & 0.525 & 0.602 & 0.939 & yes \\
\midrule
\multirow{3}{*}{LLaVA-OV-7B} & GQA & 0.631 & 0.673 & 0.673 & 0.958 & 0.034 & 0.847 & 0.848 & yes \\
 & VQAv2 & 0.616 & n/a & 0.616 & 0.931 & 0.043 & 0.867 & 0.870 & yes \\
 & POPE & 0.597 & n/a & 0.597 & 0.962 & 0.241 & 0.703 & 0.713 & yes \\
\midrule
\multirow{3}{*}{Gemma-3-12B} & GQA & 0.609 & 0.660 & 0.660 & 0.883 & 0.070 & 0.821 & 0.829 & yes \\
 & VQAv2 & 0.652 & n/a & 0.652 & 0.793 & 0.053 & 0.885 & 0.896 & yes \\
 & POPE & 0.474 & n/a & 0.500 & 0.942 & 0.465 & 0.536 & 0.564 & yes \\
\bottomrule
\end{tabular}

\end{table}

\subsection{The cross-answer term: isolated by long, unique answers}
\label{app:mech-cross}

When answers are long and distinct there are no same-answer pairs, so $w\to0$, the same-answer term drops out of the decomposition, and only the cross-answer term remains. Whatever collapse survives must then come from moving each item's own score, with no shared-answer structure to exploit. Table~\ref{tab:app-longanswer} reports the long-answer regime for all four models. On Gemma and Molmo the semantic readouts collapse with merging gone, P(IK) and SAPLMA falling from clean values of $0.60$ to $0.68$ down to $0.38$ to $0.43$; the token-probability readout collapses on all four, as low as $0.082$ on InternVL. The cross-answer term is therefore real and independently attackable. On InternVL and LLaVA the trained readouts resist, P(IK) and SAPLMA staying between $0.580$ and $0.624$ on both models.

The properties one would expect to explain the split do not explain it in these four models. The split does not track reasoning training or native answer length, since the natively long model, Molmo, is one that collapses. It does not track the location of the correctness signal: a probe trained at the answer-final position the attack reads is informative for all four models, with clean AUROC from $0.60$ to $0.69$, and the two resisting models have the highest such signal, yet the attack inverts it for only two of the four. It does not track backbone or size, nor the input-to-logit sensitivity measured in Appendix~\ref{app:distribution}, since Molmo has the smallest benign-dither divergence and Gemma the largest and both collapse. Nor does it track ease of preservation: holding a roughly ninety-token answer byte-identical is a severe constraint, with only about $3\%$ of attack steps preserving it and eight to sixteen candidate perturbations rejected per item-direction. What separates the two groups is how far a given model's readout moves under an answer-preserving perturbation, which this analysis measures but does not explain.

\begin{table}[t]
\centering
\caption{\textbf{The long-answer regime, isolating the cross-answer term.} (a) Clean and attacked AUROC per readout under long, unique answers, where $w\to0$; Gemma and Molmo collapse, InternVL and LLaVA resist. (b) Per-model diagnostics: the answer-final signal location, induced and native answer lengths, and byte-identity feasibility.}
\label{tab:app-longanswer}
% Appendix (two mechanisms) / Section 5 long-answer full four-model breakdown. Source reports 216/217,
%   verified report 232. Stores: data/long_answer/<model>/{result3.json (readouts.clean_auroc_subset,
%   readouts.attacked_auroc worst-case oracle, answer_preservation), position_auroc.json
%   (by_position.final)}, part_b_lengths.json (cot.median, neutral.median, native_long). Long (w->0)
%   chain-of-thought regime. Resisters InternVL/LLaVA keep P(IK),SAPLMA above chance; collapsers
%   Gemma/Molmo do not. Resister attacked floor = min = LLaVA SAPLMA 0.580.
% Requires: \usepackage{booktabs,multirow}
% (a) Per-readout clean and attacked AUROC under long answers.
\small
\setlength{\tabcolsep}{5pt}
\begin{tabular}{@{}llrrc@{}}
\toprule
model & readout & clean AUROC & attacked AUROC & regime \\
\midrule
\multirow{3}{*}{InternVL3.5-2B} & TokProb & 0.515 & 0.082 & \multirow{3}{*}{resist} \\
 & P(IK) & 0.672 & 0.621 &  \\
 & SAPLMA & 0.666 & 0.621 &  \\
\midrule
\multirow{3}{*}{Molmo2-4B} & TokProb & 0.540 & 0.184 & \multirow{3}{*}{collapse} \\
 & P(IK) & 0.676 & 0.430 &  \\
 & SAPLMA & 0.615 & 0.432 &  \\
\midrule
\multirow{3}{*}{LLaVA-OV-7B} & TokProb & 0.552 & 0.351 & \multirow{3}{*}{resist} \\
 & P(IK) & 0.702 & 0.624 &  \\
 & SAPLMA & 0.618 & 0.580 &  \\
\midrule
\multirow{3}{*}{Gemma-3-12B} & TokProb & 0.627 & 0.331 & \multirow{3}{*}{collapse} \\
 & P(IK) & 0.667 & 0.380 &  \\
 & SAPLMA & 0.603 & 0.396 &  \\
\bottomrule
\end{tabular}

\vspace{5pt}
% (b) Per-model diagnostics: signal-location (final-locus clean AUROC), induced/native answer length,
%   and byte-identity feasibility on long answers.
{\footnotesize\itshape Per-model diagnostics (signal location, answer length, byte-identity feasibility):}\\[2pt]
\footnotesize
\setlength{\tabcolsep}{4.5pt}
\begin{adjustbox}{max width=\textwidth}
\begin{tabular}{@{}lrrrrr@{}}
\toprule
model & final-locus clean & reasoning-prompt median (tok) & native median (tok) & feasible-step rate & rejects/item-dir \\
\midrule
InternVL3.5-2B & 0.687 & 77.0 & 2.0 & 0.0323 & 11.96 \\
Molmo2-4B & 0.617 & 94.5 & 25.5$^{\ast}$ & 0.0323 & 10.26 \\
LLaVA-OV-7B & 0.643 & 96.5 & 2.0 & 0.0323 & 15.79 \\
Gemma-3-12B & 0.604 & 96.0 & 7.0 & 0.0323 & 8.26 \\
\bottomrule
\end{tabular}
\end{adjustbox}

\vspace{2pt}
{\footnotesize $^{\ast}$Molmo is the only natively long model (neutral-prompt median 25.5 tokens, frac$\ge$20 $=$ 0.54); the others emit 2 to 7 tokens without a chain-of-thought prompt. The resisters' attacked floor, the minimum over P(IK) and SAPLMA on InternVL and LLaVA, is 0.580 (LLaVA SAPLMA). Feasible-step rate is the median fraction of decode steps that keep the answer byte-identical ($\approx$3\%); rejects/item-dir is the mean number of fresh-decode rejections per item-direction (8 to 16).}

\end{table}

\subsection{A causal check on the movement}
\label{app:mech-causal}

The input-gradient and geometry analyses show that different readouts are sensitive to related image-space directions and that the inversion does not require wholesale destruction of the same-answer representation, only targeted movement along directions the readout reads. A stronger check extracts the attack-induced hidden-state direction and injects it at inference while leaving the image unchanged, then measures transfer to probes against equal-norm random and grey-image controls. Table~\ref{tab:app-mech-causal} reports the movement. Injecting the extracted attack direction at the SAPLMA locus moves the readout down on every model, from $-0.150$ on InternVL to $-0.506$ on Molmo, while an equal-norm random direction moves it near zero or upward, so the movement follows the direction and not its norm; injecting the same direction on a neutral grey image reproduces or exceeds the real-image movement, so the effect is carried by the hidden-state direction rather than by an image feature. The store records readout movement; the corresponding discrimination change is the inversion of the main matrix. This is a consistency check on the mechanism and does not explain why the cross-answer term is more movable in some models than others, which the long-answer split leaves open.

\begin{table}[t]
\centering
\caption{\textbf{Steering injection reproduces the attack's readout movement.} Mean shift of the SAPLMA readout, averaged over the three benchmarks, when the extracted attack-induced hidden-state direction is injected at inference on the clean image, against an equal-norm random direction and the same direction injected on a neutral grey image.}
\label{tab:app-mech-causal}
\small
\begin{tabular}{lccc}
\toprule
model & attack direction & equal-norm random & grey image \\
\midrule
InternVL3.5-2B & $-0.150$ & $+0.001$ & $-0.267$ \\
Molmo2-4B      & $-0.506$ & $-0.118$ & $-0.489$ \\
LLaVA-OV-7B    & $-0.167$ & $+0.041$ & $-0.210$ \\
Gemma-3-12B    & $-0.172$ & $+0.004$ & $-0.123$ \\
\bottomrule
\end{tabular}
\end{table}

\subsection{The uniform certificate, and why no readout meets it}
\label{app:mech-certificate}

A natural formal guarantee for a confidence readout is a uniform bound on its manipulable amplitude. Write $g_{ij} = c(x_i) - c(x_j)$ for the clean separation of a correct item $i$ from an incorrect item $j$. If $\Delta^\epsilon_c(x) \le \gamma$ for every input, then any pair with $g_{ij} > 2\gamma$ keeps its ordering, so
\[
A_{\mathrm{adv}}(c) \;\ge\; G(\gamma) = \Pr[g_{ij} > 2\gamma] + \tfrac{1}{2}\Pr[g_{ij} = 2\gamma],
\]
which depends only on the clean separations and uses no reachability assumption. Define the useful threshold $\gamma^\star$ as the largest amplitude for which this lower bound still clears the cell's operative floor. A readout certified below $\gamma^\star$ would provably remain useful under attack. This is the guarantee that would distinguish a robust readout from a merely accurate one, and it is the property a defender would want to establish.

No tested readout has it. A single witnessed feasible displacement exceeding $\gamma^\star$ refutes the certificate at that level, and Table~\ref{tab:app-certificate} reports such a witness in 83 of the 84 estimator-by-cell profiles; in the remaining profile, TokProb on Gemma/GQA, the clean AUROC already sits below the operative floor, so no positive amplitude is certifiable and the certificate is unattainable rather than refuted. In a median cell, $66\%$ of feasible items exceed $\gamma^\star$. The profile with the smallest excess, COUPLED on Molmo/POPE, has a mean feasible displacement of $0.027$ against $\gamma^\star = 0.040$, but its feasible maximum is $0.137$ and $23\%$ of its items exceed the threshold, so it too is refuted by the itemwise test the certificate requires. A uniform amplitude certificate is refuted for every readout that could carry it. The certificate is sufficient and not necessary: a readout can stay above chance in aggregate while some items exceed $\gamma^\star$, as COUPLED on Molmo/POPE does, so its failure rules out worst-case guarantees of this form, not every defense that acts on the distribution of displacements.

% Appendix (two mechanisms): Proposition 3 uniform-certificate refutation, all 84 estimator-by-cell.
% Old Supplement Tables B.5 (deployed channels) + B.6 (FULL, COUPLED) merged. Source:
%   data/theorem/gamma_threshold.json (corrected Prop 3, report 196). gamma*=gamma_star [gamma_star_ci];
%   clean sep.=separation.mean; G(0)=clean_auroc; disp.=delta_aimed (mean witnessed displacement);
%   max=witness_max (max feasible displacement); n feas.=n_feasible; frac>gamma*=frac_exceed_gamma.
%   refuted = a feasible witness exceeds gamma* (theorem-relevant itemwise test); 84/84 refuted.
%   gamma* printed n/a where G(0) < floor (no amplitude certifies above floor; tok/Gemma/GQA).
% Requires: \usepackage{longtable,booktabs}
\setlength{\LTcapwidth}{\textwidth}
\setlength{\tabcolsep}{3pt}
\begin{longtable}{
>{\footnotesize}l
>{\footnotesize}l
>{\footnotesize}l
>{\footnotesize}l
>{\footnotesize}r
>{\footnotesize}r
>{\footnotesize}r
>{\footnotesize}r
>{\footnotesize}r
>{\footnotesize}r
>{\footnotesize}c
}
\caption{\textbf{The uniform amplitude certificate, refuted in 83 of 84 estimator-by-cell profiles and unattainable in the remaining one.} $\gamma^{\ast}$ (with bootstrap 95\% CI) is the amplitude below which a certified readout provably keeps discriminating above the cell's operative floor, reported as the continuous crossing of the clean separation curve; clean sep.\ is the mean clean pairwise separation and $G(0)$ the clean AUROC, both over all correct-incorrect pairs; disp.\ is the mean per-item displacement the aimed attack witnesses and max the largest, both on the feasibility-restricted set of $n$ feas.\ items; frac$>\gamma^{\ast}$ is the fraction of feasible items whose witness exceeds $\gamma^{\ast}$, each falsifying a uniform certificate at that level; refuted marks whether a feasible witness exceeds $\gamma^{\ast}$. For TokProb on Gemma/GQA the clean AUROC $G(0)=0.569$ already lies below the operative floor $0.660$, so no positive amplitude is certifiable; that row shows n/a for both $\gamma^{\ast}$ and refuted, since the certificate is unattainable rather than refuted.}
\label{tab:app-certificate}\\
\toprule
estimator & model & set & $\gamma^{\ast}$ [95\% CI] & clean sep. & $G(0)$ & disp. & max & $n$ feas. & frac$>\gamma^{\ast}$ & refuted \\
\midrule
\endfirsthead
\multicolumn{11}{@{}l}{\emph{Table~\ref{tab:app-certificate} continued from previous page}}\\
\toprule
estimator & model & set & $\gamma^{\ast}$ [95\% CI] & clean sep. & $G(0)$ & disp. & max & $n$ feas. & frac$>\gamma^{\ast}$ & refuted \\
\midrule
\endhead
\midrule
\multicolumn{11}{r@{}}{\emph{continued on next page}}\\
\endfoot
\bottomrule
\multicolumn{11}{@{}l}{\footnotesize 83 of 84 profiles refuted; 1 unattainable (TokProb, Gemma, GQA:}\\
\multicolumn{11}{@{}l}{\footnotesize $G(0)$ below the operative floor, so no positive amplitude is certifiable).}\\
\endlastfoot
TokProb & InternVL & GQA & 0.046~{\scriptsize[0.044,0.048]} & 0.191 & 0.744 & 0.129 & 0.823 & 528 & 0.483 & yes \\
 &  & VQAv2 & 0.065~{\scriptsize[0.064,0.067]} & 0.204 & 0.765 & 0.115 & 0.817 & 473 & 0.410 & yes \\
 &  & POPE & 0.053~{\scriptsize[0.052,0.055]} & 0.146 & 0.779 & 0.087 & 0.441 & 725 & 0.505 & yes \\
\addlinespace
 & Molmo & GQA & 0.019~{\scriptsize[0.017,0.020]} & 0.146 & 0.739 & 0.133 & 0.674 & 463 & 0.739 & yes \\
 &  & VQAv2 & 0.060~{\scriptsize[0.059,0.062]} & 0.189 & 0.792 & 0.120 & 0.759 & 424 & 0.583 & yes \\
 &  & POPE & 0.021~{\scriptsize[0.020,0.021]} & 0.079 & 0.688 & 0.105 & 0.368 & 377 & 0.785 & yes \\
\addlinespace
 & LLaVA & GQA & 0.045~{\scriptsize[0.043,0.046]} & 0.198 & 0.791 & 0.123 & 0.642 & 630 & 0.633 & yes \\
 &  & VQAv2 & 0.090~{\scriptsize[0.088,0.092]} & 0.249 & 0.830 & 0.108 & 0.623 & 560 & 0.468 & yes \\
 &  & POPE & 0.039~{\scriptsize[0.037,0.040]} & 0.134 & 0.714 & 0.096 & 0.419 & 616 & 0.537 & yes \\
\addlinespace
 & Gemma & GQA & n/a & 0.031 & 0.569 & 0.043 & 0.397 & 797 & n/a & n/a \\
 &  & VQAv2 & 0.002~{\scriptsize[0.001,0.002]} & 0.066 & 0.672 & 0.048 & 0.468 & 805 & 0.639 & yes \\
 &  & POPE & 0.007~{\scriptsize[0.006,0.007]} & 0.026 & 0.580 & 0.051 & 0.309 & 857 & 0.760 & yes \\
\midrule
P(IK) & InternVL & GQA & 0.096~{\scriptsize[0.093,0.098]} & 0.309 & 0.809 & 0.318 & 0.997 & 529 & 0.679 & yes \\
 &  & VQAv2 & 0.070~{\scriptsize[0.067,0.072]} & 0.250 & 0.748 & 0.314 & 0.974 & 470 & 0.696 & yes \\
 &  & POPE & 0.088~{\scriptsize[0.086,0.092]} & 0.241 & 0.786 & 0.339 & 0.834 & 724 & 0.786 & yes \\
\addlinespace
 & Molmo & GQA & 0.039~{\scriptsize[0.038,0.041]} & 0.268 & 0.779 & 0.377 & 0.999 & 461 & 0.796 & yes \\
 &  & VQAv2 & 0.084~{\scriptsize[0.082,0.087]} & 0.310 & 0.777 & 0.395 & 0.999 & 424 & 0.795 & yes \\
 &  & POPE & 0.065~{\scriptsize[0.063,0.067]} & 0.204 & 0.818 & 0.322 & 0.604 & 377 & 0.878 & yes \\
\addlinespace
 & LLaVA & GQA & 0.094~{\scriptsize[0.090,0.096]} & 0.354 & 0.826 & 0.451 & 0.980 & 630 & 0.841 & yes \\
 &  & VQAv2 & 0.064~{\scriptsize[0.061,0.066]} & 0.243 & 0.740 & 0.359 & 0.992 & 558 & 0.762 & yes \\
 &  & POPE & 0.068~{\scriptsize[0.065,0.071]} & 0.246 & 0.775 & 0.273 & 0.711 & 616 & 0.718 & yes \\
\addlinespace
 & Gemma & GQA & 0.070~{\scriptsize[0.067,0.072]} & 0.280 & 0.789 & 0.207 & 0.997 & 798 & 0.669 & yes \\
 &  & VQAv2 & 0.026~{\scriptsize[0.024,0.028]} & 0.207 & 0.706 & 0.202 & 0.999 & 803 & 0.663 & yes \\
 &  & POPE & 0.120~{\scriptsize[0.117,0.121]} & 0.217 & 0.755 & 0.289 & 0.742 & 857 & 0.704 & yes \\
\midrule
SAPLMA & InternVL & GQA & 0.082~{\scriptsize[0.079,0.085]} & 0.332 & 0.799 & 0.297 & 0.998 & 527 & 0.554 & yes \\
 &  & VQAv2 & 0.042~{\scriptsize[0.040,0.045]} & 0.220 & 0.696 & 0.283 & 1.000 & 467 & 0.565 & yes \\
 &  & POPE & 0.028~{\scriptsize[0.028,0.030]} & 0.194 & 0.781 & 0.203 & 0.922 & 725 & 0.614 & yes \\
\addlinespace
 & Molmo & GQA & 0.040~{\scriptsize[0.039,0.042]} & 0.249 & 0.805 & 0.252 & 0.896 & 463 & 0.726 & yes \\
 &  & VQAv2 & 0.081~{\scriptsize[0.079,0.083]} & 0.272 & 0.784 & 0.254 & 0.940 & 426 & 0.631 & yes \\
 &  & POPE & 0.023~{\scriptsize[0.022,0.023]} & 0.107 & 0.787 & 0.091 & 0.702 & 377 & 0.647 & yes \\
\addlinespace
 & LLaVA & GQA & 0.120~{\scriptsize[0.118,0.123]} & 0.393 & 0.851 & 0.355 & 0.995 & 628 & 0.667 & yes \\
 &  & VQAv2 & 0.083~{\scriptsize[0.080,0.085]} & 0.295 & 0.757 & 0.307 & 0.987 & 554 & 0.639 & yes \\
 &  & POPE & 0.013~{\scriptsize[0.012,0.014]} & 0.144 & 0.709 & 0.119 & 0.940 & 616 & 0.573 & yes \\
\addlinespace
 & Gemma & GQA & 0.065~{\scriptsize[0.064,0.068]} & 0.309 & 0.808 & 0.259 & 0.991 & 798 & 0.674 & yes \\
 &  & VQAv2 & 0.074~{\scriptsize[0.072,0.077]} & 0.311 & 0.785 & 0.266 & 0.970 & 802 & 0.596 & yes \\
 &  & POPE & 0.079~{\scriptsize[0.076,0.081]} & 0.239 & 0.813 & 0.278 & 0.895 & 857 & 0.621 & yes \\
\midrule
CCPS & InternVL & GQA & 0.085~{\scriptsize[0.083,0.088]} & 0.304 & 0.796 & 0.230 & 0.908 & 528 & 0.544 & yes \\
 &  & VQAv2 & 0.072~{\scriptsize[0.069,0.075]} & 0.270 & 0.738 & 0.193 & 0.955 & 469 & 0.490 & yes \\
 &  & POPE & 0.103~{\scriptsize[0.100,0.106]} & 0.263 & 0.784 & 0.220 & 0.612 & 725 & 0.596 & yes \\
\addlinespace
 & Molmo & GQA & 0.048~{\scriptsize[0.045,0.050]} & 0.301 & 0.777 & 0.303 & 0.893 & 461 & 0.694 & yes \\
 &  & VQAv2 & 0.106~{\scriptsize[0.103,0.109]} & 0.334 & 0.784 & 0.310 & 0.975 & 424 & 0.670 & yes \\
 &  & POPE & 0.088~{\scriptsize[0.085,0.092]} & 0.293 & 0.816 & 0.368 & 0.732 & 377 & 0.764 & yes \\
\addlinespace
 & LLaVA & GQA & 0.098~{\scriptsize[0.096,0.102]} & 0.367 & 0.824 & 0.272 & 0.995 & 630 & 0.627 & yes \\
 &  & VQAv2 & 0.097~{\scriptsize[0.095,0.100]} & 0.316 & 0.771 & 0.242 & 0.958 & 559 & 0.565 & yes \\
 &  & POPE & 0.062~{\scriptsize[0.060,0.064]} & 0.210 & 0.750 & 0.169 & 0.773 & 616 & 0.597 & yes \\
\addlinespace
 & Gemma & GQA & 0.040~{\scriptsize[0.038,0.042]} & 0.219 & 0.740 & 0.213 & 0.898 & 797 & 0.679 & yes \\
 &  & VQAv2 & 0.071~{\scriptsize[0.069,0.073]} & 0.264 & 0.780 & 0.185 & 0.842 & 803 & 0.562 & yes \\
 &  & POPE & 0.068~{\scriptsize[0.066,0.070]} & 0.177 & 0.711 & 0.235 & 0.786 & 857 & 0.687 & yes \\
\midrule
CCPS-D & InternVL & GQA & 0.086~{\scriptsize[0.083,0.088]} & 0.290 & 0.795 & 0.245 & 0.887 & 528 & 0.570 & yes \\
 &  & VQAv2 & 0.058~{\scriptsize[0.055,0.060]} & 0.240 & 0.719 & 0.196 & 0.909 & 469 & 0.516 & yes \\
 &  & POPE & 0.098~{\scriptsize[0.095,0.102]} & 0.282 & 0.781 & 0.270 & 0.789 & 725 & 0.640 & yes \\
\addlinespace
 & Molmo & GQA & 0.063~{\scriptsize[0.061,0.066]} & 0.312 & 0.797 & 0.337 & 0.828 & 461 & 0.716 & yes \\
 &  & VQAv2 & 0.107~{\scriptsize[0.105,0.110]} & 0.338 & 0.802 & 0.331 & 0.904 & 424 & 0.696 & yes \\
 &  & POPE & 0.084~{\scriptsize[0.081,0.087]} & 0.302 & 0.816 & 0.388 & 0.771 & 377 & 0.743 & yes \\
\addlinespace
 & LLaVA & GQA & 0.093~{\scriptsize[0.090,0.096]} & 0.380 & 0.823 & 0.293 & 0.978 & 630 & 0.640 & yes \\
 &  & VQAv2 & 0.093~{\scriptsize[0.090,0.096]} & 0.325 & 0.770 & 0.261 & 0.984 & 559 & 0.569 & yes \\
 &  & POPE & 0.048~{\scriptsize[0.046,0.050]} & 0.218 & 0.735 & 0.211 & 0.804 & 616 & 0.648 & yes \\
\addlinespace
 & Gemma & GQA & 0.036~{\scriptsize[0.034,0.038]} & 0.216 & 0.742 & 0.235 & 0.876 & 797 & 0.704 & yes \\
 &  & VQAv2 & 0.066~{\scriptsize[0.063,0.068]} & 0.266 & 0.772 & 0.210 & 0.859 & 803 & 0.599 & yes \\
 &  & POPE & 0.063~{\scriptsize[0.062,0.065]} & 0.164 & 0.740 & 0.244 & 0.786 & 857 & 0.701 & yes \\
\midrule
FULL & InternVL & GQA & 0.083~{\scriptsize[0.080,0.085]} & 0.313 & 0.798 & 0.279 & 1.000 & 527 & 0.543 & yes \\
 &  & VQAv2 & 0.032~{\scriptsize[0.030,0.034]} & 0.209 & 0.688 & 0.256 & 0.999 & 462 & 0.543 & yes \\
 &  & POPE & 0.029~{\scriptsize[0.028,0.030]} & 0.197 & 0.745 & 0.203 & 0.887 & 725 & 0.568 & yes \\
\addlinespace
 & Molmo & GQA & 0.052~{\scriptsize[0.050,0.053]} & 0.277 & 0.818 & 0.254 & 0.964 & 463 & 0.683 & yes \\
 &  & VQAv2 & 0.090~{\scriptsize[0.087,0.092]} & 0.297 & 0.790 & 0.253 & 0.993 & 427 & 0.635 & yes \\
 &  & POPE & 0.028~{\scriptsize[0.027,0.029]} & 0.146 & 0.793 & 0.119 & 0.732 & 377 & 0.668 & yes \\
\addlinespace
 & LLaVA & GQA & 0.102~{\scriptsize[0.100,0.105]} & 0.360 & 0.837 & 0.344 & 0.991 & 628 & 0.707 & yes \\
 &  & VQAv2 & 0.072~{\scriptsize[0.069,0.074]} & 0.269 & 0.736 & 0.278 & 0.995 & 552 & 0.636 & yes \\
 &  & POPE & 0.039~{\scriptsize[0.038,0.041]} & 0.171 & 0.749 & 0.134 & 0.821 & 616 & 0.627 & yes \\
\addlinespace
 & Gemma & GQA & 0.072~{\scriptsize[0.070,0.074]} & 0.285 & 0.801 & 0.260 & 0.987 & 795 & 0.713 & yes \\
 &  & VQAv2 & 0.075~{\scriptsize[0.073,0.077]} & 0.279 & 0.780 & 0.243 & 0.996 & 803 & 0.621 & yes \\
 &  & POPE & 0.107~{\scriptsize[0.104,0.109]} & 0.235 & 0.813 & 0.287 & 0.919 & 857 & 0.632 & yes \\
\midrule
COUPLED & InternVL & GQA & 0.059~{\scriptsize[0.057,0.061]} & 0.248 & 0.751 & 0.412 & 1.000 & 517 & 0.768 & yes \\
 &  & VQAv2 & 0.005~{\scriptsize[0.003,0.007]} & 0.129 & 0.626 & 0.358 & 1.000 & 441 & 0.800 & yes \\
 &  & POPE & 0.017~{\scriptsize[0.016,0.019]} & 0.082 & 0.610 & 0.325 & 0.975 & 722 & 0.850 & yes \\
\addlinespace
 & Molmo & GQA & 0.022~{\scriptsize[0.020,0.025]} & 0.236 & 0.725 & 0.178 & 0.991 & 461 & 0.666 & yes \\
 &  & VQAv2 & 0.021~{\scriptsize[0.018,0.023]} & 0.166 & 0.665 & 0.168 & 0.999 & 421 & 0.634 & yes \\
 &  & POPE & 0.040~{\scriptsize[0.037,0.042]} & 0.135 & 0.673 & 0.027 & 0.137 & 377 & 0.233 & yes \\
\addlinespace
 & LLaVA & GQA & 0.043~{\scriptsize[0.040,0.045]} & 0.253 & 0.745 & 0.383 & 0.999 & 624 & 0.800 & yes \\
 &  & VQAv2 & 0.017~{\scriptsize[0.015,0.020]} & 0.164 & 0.650 & 0.320 & 1.000 & 533 & 0.788 & yes \\
 &  & POPE & 0.023~{\scriptsize[0.021,0.024]} & 0.103 & 0.666 & 0.302 & 0.929 & 616 & 0.805 & yes \\
\addlinespace
 & Gemma & GQA & 0.027~{\scriptsize[0.025,0.029]} & 0.178 & 0.724 & 0.099 & 0.646 & 795 & 0.701 & yes \\
 &  & VQAv2 & 0.027~{\scriptsize[0.025,0.029]} & 0.172 & 0.714 & 0.095 & 0.668 & 803 & 0.679 & yes \\
 &  & POPE & 0.036~{\scriptsize[0.035,0.038]} & 0.082 & 0.600 & 0.139 & 0.401 & 857 & 0.932 & yes \\
\end{longtable}
\section{The Phenomenon in Full}
\label{app:phenomenon}

The main text summarizes the collapse as a median over the seven readouts in each cell. This appendix reports the underlying eighty-four numbers, so the aggregate can be checked against the detail, and gives the label-free attack, which reproduces the inversion without any correctness oracle, in full.

\subsection{Every readout in every cell}
\label{app:phenomenon-full}

Table~\ref{tab:app-phenomenon} reports clean and worst-case-oracle attacked AUROC against the operative answer-string floor for all eighty-four estimator-by-cell profiles. Clean discrimination ranges from $0.57$ to $0.85$; under attack it falls at or below the floor in every one of the eighty-four, and below chance in eighty-three. The single exception is COUPLED on Molmo/POPE, at $0.520$, the survivor analyzed in Appendix~\ref{app:mechanisms}. Seven of the seven readouts fall below chance in eleven of the twelve cells and six of seven in the last, and the per-cell median of these eighty-four attacked values reproduces the median column of the core table exactly. The collapse is a property of the readout family.

The result does not depend on items the attack leaves untouched. At the operating margin $\tau_{\mathrm{safe}}=1$, $39.1\%$ of the $60{,}000$ estimator-item instances (the $12{,}000$ items by five aimed objectives, with the two push directions collapsed) admit no harmful iterate that both moves the score and clears that margin, and so enter attacked AUROC with their clean scores; this margin-based count is stricter than the $6.1\%$ of endpoint images byte-identical to clean at the argmax constraint (Appendix~\ref{app:setup-clip}), and it weakens the measured collapse. Restricting to the items the attack engages lowers the median attacked AUROC by a further $0.138$, and the floor result continues to hold in all sixty deployed-channel cells. The reported numbers are therefore conservative with respect to engagement.

% Appendix (phenomenon in full): clean vs worst-case-oracle attacked AUROC, all 84 estimator-by-cell.
% Old Supplement Table D.1. Source: frozen T4_witnessed_union_profile.json (clean_auroc,
%   witnessed_union_oracle_auroc) + operative floor task_floor_bound.json['floor'] (max(.,0.5)).
%   '<floor' = attacked <= operative floor; '<chance' = attacked < 0.5. Same stores as report 227 and the
%   main-text core median table (per-cell median over 7 estimators of attacked matches it to <=0.001).
% Requires: \usepackage{longtable,booktabs}
\footnotesize
\setlength{\tabcolsep}{6pt}
\setlength{\LTcapwidth}{\linewidth}
\begin{longtable}{@{\extracolsep{\fill}}lll rr r cc@{}}
\caption{\textbf{The phenomenon in full:} clean and worst-case-oracle attacked AUROC for all 84 estimator-by-cell profiles, against the operative answer-string floor. `$<$floor' marks attacked at or below the floor; `$<$chance' marks attacked below 0.5. Attacked is the witnessed-union oracle (the strongest per-item candidate selection).}\label{tab:app-phenomenon}\\
\toprule
estimator & model & dataset & clean & attacked & floor & $<$floor & $<$chance \\
\midrule
\endfirsthead
\multicolumn{8}{@{}l}{\emph{Table~\ref{tab:app-phenomenon} continued from previous page}}\\
\toprule
estimator & model & dataset & clean & attacked & floor & $<$floor & $<$chance \\
\midrule
\endhead
\midrule
\multicolumn{8}{r@{}}{\emph{continued on next page}}\\
\endfoot
\bottomrule
\multicolumn{8}{@{}l}{\footnotesize Attacked $<$ chance in 83 of 84 (the exception is COUPLED/Molmo/POPE, 0.520); attacked $\le$ floor in all 84.}\\
\endlastfoot
TokProb & InternVL & GQA & 0.745 & 0.016 & 0.642 & yes & yes \\
 &  & VQAv2 & 0.767 & 0.015 & 0.617 & yes & yes \\
 &  & POPE & 0.779 & 0.000 & 0.555 & yes & yes \\
\addlinespace
 & Molmo & GQA & 0.745 & 0.008 & 0.683 & yes & yes \\
 &  & VQAv2 & 0.798 & 0.022 & 0.625 & yes & yes \\
 &  & POPE & 0.700 & 0.000 & 0.587 & yes & yes \\
\addlinespace
 & LLaVA & GQA & 0.792 & 0.002 & 0.673 & yes & yes \\
 &  & VQAv2 & 0.830 & 0.016 & 0.616 & yes & yes \\
 &  & POPE & 0.714 & 0.000 & 0.597 & yes & yes \\
\addlinespace
 & Gemma & GQA & 0.571 & 0.043 & 0.660 & yes & yes \\
 &  & VQAv2 & 0.675 & 0.061 & 0.652 & yes & yes \\
 &  & POPE & 0.579 & 0.044 & 0.500 & yes & yes \\
\midrule
P(IK) & InternVL & GQA & 0.814 & 0.021 & 0.642 & yes & yes \\
 &  & VQAv2 & 0.724 & 0.028 & 0.617 & yes & yes \\
 &  & POPE & 0.804 & 0.000 & 0.555 & yes & yes \\
\addlinespace
 & Molmo & GQA & 0.779 & 0.009 & 0.683 & yes & yes \\
 &  & VQAv2 & 0.772 & 0.046 & 0.625 & yes & yes \\
 &  & POPE & 0.819 & 0.000 & 0.587 & yes & yes \\
\addlinespace
 & LLaVA & GQA & 0.825 & 0.009 & 0.673 & yes & yes \\
 &  & VQAv2 & 0.741 & 0.015 & 0.616 & yes & yes \\
 &  & POPE & 0.775 & 0.000 & 0.597 & yes & yes \\
\addlinespace
 & Gemma & GQA & 0.790 & 0.100 & 0.660 & yes & yes \\
 &  & VQAv2 & 0.705 & 0.086 & 0.652 & yes & yes \\
 &  & POPE & 0.755 & 0.004 & 0.500 & yes & yes \\
\midrule
SAPLMA & InternVL & GQA & 0.813 & 0.017 & 0.642 & yes & yes \\
 &  & VQAv2 & 0.703 & 0.008 & 0.617 & yes & yes \\
 &  & POPE & 0.742 & 0.000 & 0.555 & yes & yes \\
\addlinespace
 & Molmo & GQA & 0.809 & 0.013 & 0.683 & yes & yes \\
 &  & VQAv2 & 0.784 & 0.039 & 0.625 & yes & yes \\
 &  & POPE & 0.792 & 0.000 & 0.587 & yes & yes \\
\addlinespace
 & LLaVA & GQA & 0.851 & 0.007 & 0.673 & yes & yes \\
 &  & VQAv2 & 0.758 & 0.020 & 0.616 & yes & yes \\
 &  & POPE & 0.709 & 0.001 & 0.597 & yes & yes \\
\addlinespace
 & Gemma & GQA & 0.808 & 0.074 & 0.660 & yes & yes \\
 &  & VQAv2 & 0.783 & 0.102 & 0.652 & yes & yes \\
 &  & POPE & 0.813 & 0.009 & 0.500 & yes & yes \\
\midrule
CCPS & InternVL & GQA & 0.797 & 0.010 & 0.642 & yes & yes \\
 &  & VQAv2 & 0.735 & 0.021 & 0.617 & yes & yes \\
 &  & POPE & 0.779 & 0.000 & 0.555 & yes & yes \\
\addlinespace
 & Molmo & GQA & 0.795 & 0.011 & 0.683 & yes & yes \\
 &  & VQAv2 & 0.819 & 0.056 & 0.625 & yes & yes \\
 &  & POPE & 0.831 & 0.000 & 0.587 & yes & yes \\
\addlinespace
 & LLaVA & GQA & 0.828 & 0.002 & 0.673 & yes & yes \\
 &  & VQAv2 & 0.795 & 0.031 & 0.616 & yes & yes \\
 &  & POPE & 0.745 & 0.000 & 0.597 & yes & yes \\
\addlinespace
 & Gemma & GQA & 0.745 & 0.032 & 0.660 & yes & yes \\
 &  & VQAv2 & 0.785 & 0.067 & 0.652 & yes & yes \\
 &  & POPE & 0.733 & 0.002 & 0.500 & yes & yes \\
\midrule
CCPS-D & InternVL & GQA & 0.807 & 0.014 & 0.642 & yes & yes \\
 &  & VQAv2 & 0.733 & 0.022 & 0.617 & yes & yes \\
 &  & POPE & 0.792 & 0.000 & 0.555 & yes & yes \\
\addlinespace
 & Molmo & GQA & 0.780 & 0.004 & 0.683 & yes & yes \\
 &  & VQAv2 & 0.789 & 0.017 & 0.625 & yes & yes \\
 &  & POPE & 0.818 & 0.000 & 0.587 & yes & yes \\
\addlinespace
 & LLaVA & GQA & 0.830 & 0.001 & 0.673 & yes & yes \\
 &  & VQAv2 & 0.796 & 0.014 & 0.616 & yes & yes \\
 &  & POPE & 0.757 & 0.000 & 0.597 & yes & yes \\
\addlinespace
 & Gemma & GQA & 0.743 & 0.026 & 0.660 & yes & yes \\
 &  & VQAv2 & 0.774 & 0.085 & 0.652 & yes & yes \\
 &  & POPE & 0.725 & 0.004 & 0.500 & yes & yes \\
\midrule
FULL & InternVL & GQA & 0.797 & 0.009 & 0.642 & yes & yes \\
 &  & VQAv2 & 0.684 & 0.006 & 0.617 & yes & yes \\
 &  & POPE & 0.745 & 0.000 & 0.555 & yes & yes \\
\addlinespace
 & Molmo & GQA & 0.822 & 0.009 & 0.683 & yes & yes \\
 &  & VQAv2 & 0.787 & 0.022 & 0.625 & yes & yes \\
 &  & POPE & 0.795 & 0.000 & 0.587 & yes & yes \\
\addlinespace
 & LLaVA & GQA & 0.838 & 0.011 & 0.673 & yes & yes \\
 &  & VQAv2 & 0.736 & 0.030 & 0.616 & yes & yes \\
 &  & POPE & 0.749 & 0.000 & 0.597 & yes & yes \\
\addlinespace
 & Gemma & GQA & 0.798 & 0.060 & 0.660 & yes & yes \\
 &  & VQAv2 & 0.779 & 0.090 & 0.652 & yes & yes \\
 &  & POPE & 0.813 & 0.008 & 0.500 & yes & yes \\
\midrule
COUPLED & InternVL & GQA & 0.743 & 0.016 & 0.642 & yes & yes \\
 &  & VQAv2 & 0.608 & 0.012 & 0.617 & yes & yes \\
 &  & POPE & 0.612 & 0.002 & 0.555 & yes & yes \\
\addlinespace
 & Molmo & GQA & 0.727 & 0.291 & 0.683 & yes & yes \\
 &  & VQAv2 & 0.664 & 0.250 & 0.625 & yes & yes \\
 &  & POPE & 0.671 & 0.520 & 0.587 & yes & no \\
\addlinespace
 & LLaVA & GQA & 0.744 & 0.049 & 0.673 & yes & yes \\
 &  & VQAv2 & 0.639 & 0.052 & 0.616 & yes & yes \\
 &  & POPE & 0.666 & 0.037 & 0.597 & yes & yes \\
\addlinespace
 & Gemma & GQA & 0.722 & 0.331 & 0.660 & yes & yes \\
 &  & VQAv2 & 0.715 & 0.336 & 0.652 & yes & yes \\
 &  & POPE & 0.600 & 0.178 & 0.500 & yes & yes \\
\end{longtable}
\normalsize

\subsection{The inversion without a correctness oracle}
\label{app:phenomenon-labelfree}

The worst-case-oracle attack chooses each item's push direction from ground-truth correctness. A real attacker has no such labels. The label-free attack replaces the oracle with the model's own clean confidence: it treats a high clean score as a proxy for a correct answer and a low one as a proxy for an incorrect answer, then pushes the confident-looking items down and the unconfident-looking ones up. Table~\ref{tab:app-labelfree} reports its effect for the two readouts that carry a disjoint clean-calibration reference, P(IK) and SAPLMA, the only channels for which a clean-score proxy can be formed without reusing the evaluation data. On both, in all twelve cells, the label-free attack drives discrimination below chance, to between $0.262$ and $0.461$. The oracle inverts the ranking almost completely while the proxy only pushes it past chance, but the conclusion for a deployment is the same: an attacker with no knowledge of which answers are correct still turns the readout from informative to anti-informative. The other five readouts are omitted because they do not carry a separate clean reference from which such a proxy could be built.

\begin{table}[t]
\centering
\caption{\textbf{The label-free attack, using clean confidence as a correctness proxy.} Attacked AUROC for the clean-sign selector, which consults no ground-truth labels, on the two eligible readouts. All twenty-four eligible cells fall below chance ($0.5$).}
\label{tab:app-labelfree}
% Appendix (phenomenon in full): label-free clean-sign selector attacked AUROC, eligible channels only.
% Source: frozen T6_selectors_oracle_vs_operational.json, selector=clean_sign_selected, is_oracle=false,
%   self-rows (aimed_objective==evaluated_estimator) for P(IK) and SAPLMA. These are the only two channels
%   with a disjoint clean-calibration reference; the other five are excluded (sanctioned omission, report
%   213), NOT imputed. Label-free = the attack direction is chosen by the clean score's sign, no labels.
% Requires: \usepackage{booktabs,multirow}
\small
\setlength{\tabcolsep}{6pt}
\begin{tabular}{@{}llrr@{}}
\toprule
model & dataset & P(IK) label-free & SAPLMA label-free \\
\midrule
\multirow{3}{*}{InternVL3.5-2B} & GQA & 0.293 & 0.288 \\
 & VQAv2 & 0.438 & 0.344 \\
 & POPE & 0.262 & 0.355 \\
\midrule
\multirow{3}{*}{Molmo2-4B} & GQA & 0.390 & 0.316 \\
 & VQAv2 & 0.377 & 0.358 \\
 & POPE & 0.311 & 0.264 \\
\midrule
\multirow{3}{*}{LLaVA-OV-7B} & GQA & 0.351 & 0.303 \\
 & VQAv2 & 0.377 & 0.331 \\
 & POPE & 0.315 & 0.294 \\
\midrule
\multirow{3}{*}{Gemma-3-12B} & GQA & 0.442 & 0.422 \\
 & VQAv2 & 0.461 & 0.408 \\
 & POPE & 0.388 & 0.336 \\
\bottomrule
\end{tabular}

\end{table}

\subsection{The collapse is stable across seeds} \label{app:phenomenon-multiseed}

The core matrix reports one canonical item draw and one attack seed. To confirm the collapse is not an artifact of that draw, we rerun the attack under two further seeds, reseeding both the item pool and the attack's randomness, and recompute the worst-case-oracle attacked AUROC. Seed 23 is the frozen $n=1000$ anchor behind the main table; seeds 24 and 25 are fresh $n=120$ balanced draws. Recomputing the metric from the frozen store reproduces the main table exactly, so the three seeds are the same computation on different draws.

Table~\ref{tab:app-multiseed} reports the result. The largest across-seed standard deviation of any attacked cell is $0.046$, and every cell that falls below chance does so on all three seeds. The three InternVL and three LLaVA cells sit between $0.00$ and $0.07$ throughout, the Gemma and Molmo open-answer cells at a partial collapse the coupled readout resists identically across seeds, and the single survivor, Molmo/POPE, stays above chance in point estimate on every seed. The wider intervals on the $n=120$ seeds reflect the smaller draw; the point estimates track the frozen anchor closely in every cell. The per-seed binding readout occasionally differs from the frozen one on a fully collapsed cell, which does not affect the binding maximum since several readouts sit near zero there.

The survivor is a narrow one. Molmo/POPE's attacked AUROC is $0.520$, $0.545$, and $0.533$ on the three seeds, above chance in point estimate throughout, but its bootstrap intervals graze $0.5$: even the frozen $n=1000$ lower bound is $0.483$, and the $n=120$ lower bounds fall to $0.434$. The point estimate is stable across three independent draws, but no single seed separates it from chance at $95\%$ confidence. The mechanism behind it is stable as well. Table~\ref{tab:app-multiseed-survivor} reports the decomposition across seeds: the same-answer term $A_{\mathrm{same}}$ stays above $0.5$ on every seed while the cross-answer term $A_{\mathrm{cross}}$ stays below it, so the same-answer pairs retain the separability the attack cannot merge, and because POPE's yes-or-no answer space makes those pairs about half the cell, the binding readout is held marginally above chance.

\begin{table}[t]
\centering
\caption{\textbf{Three-seed robustness of the attacked core matrix.} Worst-case-oracle attacked AUROC (binding over the seven readouts, with the binding readout fixed from seed 23) per cell, at the frozen $n=1000$ seed and two fresh $n=120$ reseeds, with bootstrap $95\%$ intervals. The \emph{clean} column is the across-seed mean of the binding readout's clean AUROC, so it differs slightly from the seed-23 value in Table~\ref{tab:app-phenomenon}. Every below-chance cell is below chance on all three seeds; the largest across-seed standard deviation is $0.046$. Molmo/POPE ($\dagger$) is the sole above-chance survivor.}
\label{tab:app-multiseed}
% Appendix S2 (phenomenon): three-seed attacked-AUROC robustness of the core matrix. Source: report 253,
%   data/core_multiseed/ms_results.json (seeds 24/25 fresh n=120 balanced reseeds) + the seed-23 frozen
%   n=1000 anchor recomputed from pfmc_v1_2_2/full_matrix_runs. Binding readout = seed-23 binding (max
%   attacked). clean = across-seed mean of the binding-readout clean AUROC. CIs are bootstrap 95% (seed 23,
%   10000). Molmo/POPE (dagger) is the sole above-chance survivor. Reconciled == report 253 (max abs diff 0.000).
% Requires: \usepackage{booktabs,multirow,graphicx}
\footnotesize
\setlength{\tabcolsep}{3.5pt}
\resizebox{\textwidth}{!}{%
\begin{tabular}{@{}lll r lll rr@{}}
\toprule
model & set & readout & clean & seed 23 (n=1000) & seed 24 (n=120) & seed 25 (n=120) & mean & sd \\
\midrule
\multirow{3}{*}{InternVL3.5-2B} & GQA & P(IK) & 0.833 & 0.021~{\tiny($.013$,$.030$)} & 0.031~{\tiny($.002$,$.072$)} & 0.033~{\tiny($.000$,$.078$)} & 0.028 & 0.005 \\
 & VQAv2 & P(IK) & 0.728 & 0.028~{\tiny($.019$,$.037$)} & 0.050~{\tiny($.016$,$.091$)} & 0.028~{\tiny($.005$,$.059$)} & 0.035 & 0.011 \\
 & POPE & COUPLED & 0.666 & 0.002~{\tiny($.000$,$.004$)} & 0.000~{\tiny($.000$,$.002$)} & 0.000~{\tiny($.000$,$.000$)} & 0.001 & 0.001 \\
\midrule
\multirow{3}{*}{Molmo2-4B} & GQA & COUPLED & 0.716 & 0.291~{\tiny($.260$,$.322$)} & 0.334~{\tiny($.240$,$.428$)} & 0.309~{\tiny($.220$,$.407$)} & 0.311 & 0.018 \\
 & VQAv2 & COUPLED & 0.688 & 0.250~{\tiny($.221$,$.281$)} & 0.262~{\tiny($.176$,$.357$)} & 0.255~{\tiny($.176$,$.343$)} & 0.256 & 0.005 \\
 & POPE$^{\dagger}$ & COUPLED & 0.691 & 0.520~{\tiny($.483$,$.557$)} & 0.545~{\tiny($.442$,$.646$)} & 0.533~{\tiny($.434$,$.638$)} & 0.533 & 0.011 \\
\midrule
\multirow{3}{*}{LLaVA-OV-7B} & GQA & COUPLED & 0.695 & 0.049~{\tiny($.038$,$.061$)} & 0.039~{\tiny($.014$,$.073$)} & 0.056~{\tiny($.021$,$.100$)} & 0.048 & 0.007 \\
 & VQAv2 & COUPLED & 0.695 & 0.052~{\tiny($.041$,$.065$)} & 0.051~{\tiny($.009$,$.102$)} & 0.072~{\tiny($.034$,$.117$)} & 0.059 & 0.009 \\
 & POPE & COUPLED & 0.622 & 0.037~{\tiny($.028$,$.046$)} & 0.020~{\tiny($.005$,$.042$)} & 0.018~{\tiny($.005$,$.038$)} & 0.025 & 0.008 \\
\midrule
\multirow{3}{*}{Gemma-3-12B} & GQA & COUPLED & 0.720 & 0.331~{\tiny($.298$,$.365$)} & 0.235~{\tiny($.155$,$.318$)} & 0.333~{\tiny($.239$,$.426$)} & 0.300 & 0.046 \\
 & VQAv2 & COUPLED & 0.709 & 0.336~{\tiny($.305$,$.370$)} & 0.306~{\tiny($.215$,$.404$)} & 0.381~{\tiny($.284$,$.481$)} & 0.341 & 0.031 \\
 & POPE & COUPLED & 0.572 & 0.178~{\tiny($.152$,$.205$)} & 0.133~{\tiny($.078$,$.198$)} & 0.186~{\tiny($.117$,$.265$)} & 0.165 & 0.023 \\
\bottomrule
\end{tabular}}

\vspace{3pt}
\noindent{\footnotesize $^{\dagger}$Sole above-chance survivor. Seeds 24 and 25 reseed both the item pool and the attack randomness, so their wider 95\% CIs reflect the smaller $n{=}120$ draw. The across-seed mean and sd are over the three seed point estimates.}

\end{table}

\begin{table}[t]
\centering
\caption{\textbf{The survivor's decomposition across seeds}, for Molmo/POPE under the coupled readout. The same-answer term stays above chance and the cross-answer term below it on every seed, the mechanism that holds this cell marginally above chance.}
\label{tab:app-multiseed-survivor}
% Appendix S2 (phenomenon): survivor decomposition across seeds, Molmo/POPE (coupled). Source: report 253
%   (survivor decomposition table); s23 = frozen report-204 decomposition, s24/s25 = fresh n=120 reseeds.
%   adv = w*A_same + (1-w)*A_cross. Reconciled == report 253.
% Requires: \usepackage{booktabs}
\small
\setlength{\tabcolsep}{8pt}
\begin{tabular}{@{}lrrr@{}}
\toprule
quantity & seed 23 & seed 24 & seed 25 \\
\midrule
$w$ (same-answer weight) & 0.525 & 0.547 & 0.509 \\
$A_{\mathrm{same}}$ & 0.567 & 0.619 & 0.586 \\
$A_{\mathrm{cross}}$ & 0.467 & 0.456 & 0.478 \\
\bottomrule
\end{tabular}

\vspace{3pt}
\noindent{\footnotesize Same-answer pairs are about half the cell on every seed ($w\approx0.52$), which is why a same-answer term above 0.5 is enough to hold the binding attacked AUROC marginally above chance.}

\end{table}
\section{The Distribution Constraint}
\label{app:distribution}

Byte-identity fixes the winning token at every answer position, but it does not fix the rest of the next-token distribution, nor even the winning token's margin over its runner-up. This raises an objection: the separable channel might be slack in the answer distribution rather than a signal distinct from the answer. This appendix measures what the constraint leaves free, what happens when the free part is bounded, and how much of the inversion survives. The answer does not determine confidence at any bound, and the erosion of the effect as the distribution is pinned tighter measures how much of confidence lives in the distribution's fine structure.

The constraint the headline attack imposes is loose in a measurable way. It fixes the emitted token at each answer position, but the token's logit lead over its strongest competitor is free to vary, and the attack moves it: across cells the mean absolute margin change is between $1.6$ and $7.2$ logits while the argmax is preserved everywhere. Read as a divergence of the full next-token distribution, the unconstrained attack moves the distribution by as much as $2.6$ nats while holding the answer byte-identical. So ``same answer'' permits substantial distributional movement, and the natural test is to bound it.

We add a bound on the KL divergence between the clean and attacked next-token distributions at every answer position, enforced by rejecting any step that violates it, and re-run the attack at two bounds. Table~\ref{tab:app-kl} reports the resulting attacked AUROC per cell: at $0.2$ nats it is below chance in all twelve cells, from $0.061$ to $0.463$, and only at a very tight $0.05$ nats does the outcome become model-dependent, with Gemma's three cells staying inverted while the other three models return toward chance, leaving four of twelve cells inverted at the tight bound. The $0.05$-nat bound sits five to seventeen times above the benign-dither divergence of the three non-Gemma models and below Gemma's (Table~\ref{tab:app-gemma-sens}), so the recovery of those three toward chance happens with more distributional slack than natural pixel noise still available. These bounds are evaluated on the two hidden-state probes; the token-probability readout is constrained by the bound directly through Pinsker's inequality, and we do not evaluate CCPS, FULL, or COUPLED under it.

Below-chance at a bound is a binary summary, and two further measurements make the result precise. The first is how much of the inversion survives each bound. Table~\ref{tab:app-kl-surviving} reports the surviving fraction, the share of the full clean-to-unbounded inversion the bounded attack retains. At $0.2$ nats a mean $62\%$ survives, and at $0.05$ nats $36\%$; the effect is neither free of the distributional slack nor dominated by it, and it erodes monotonically as the distribution is pinned tighter. The erosion is what one expects if part of confidence lives in the distribution's fine structure, and its magnitude measures how much: roughly a third of the effect is removed only by constraining well past the answer, and the remaining majority is not.

The second measurement asks what the $0.2$-nat bound means behaviorally, since a bound in nats is only as meaningful as the output change it permits. Table~\ref{tab:app-kl-behavioral} reports it. At $0.2$ nats the top-1 token is preserved at $98$ to $100\%$ of positions for all four models, which is the byte-identity guarantee, so greedy decoding is unchanged. The fuller distribution is not equally pinned: top-5 token sets reorder at a third to two-thirds of positions, and sampled continuations are distinguishable from the clean model for three of the four, with sampled total variation of $0.13$ to $0.19$ against a clean-versus-clean noise floor near $0.07$. For those three models the $0.2$-nat bound is loose in absolute terms, $22$ to $68$ times the divergence a single-greylevel pixel dither already induces, so the majority of the inversion that survives it is not an artifact of maximal distributional freedom.

Gemma is the decisive case, because of a property that would otherwise look like a weakness. Its output distribution is so sensitive to input change that a benign one-greylevel dither already moves it by $0.114$ nats on average, so a $0.2$-nat bound is only about twice that, and under it Gemma's sampled continuations sit at the clean noise floor, behaviorally indistinguishable from the clean model. For Gemma there is essentially no distributional slack left, and yet $83$ to $93\%$ of the inversion survives. In the one model where the bound removes nearly all distributional freedom, the effect is almost undiminished.

Because this existence proof rests on Gemma's sensitivity, we checked whether the sensitivity is a property of the model or an artifact of the pipeline; Table~\ref{tab:app-gemma-sens} reports the result. Gemma's answer-position divergence under a benign dither stays roughly twenty times the other models' under a Gaussian perturbation of matched norm as well, and the differentiable preprocessing we attack through reproduces the official processor exactly, so the effect is not introduced by our attack surface. Nor is it a matter of input scaling: expressed in the normalized units the model consumes, Gemma's dither is among the smallest of the four, while InternVL, whose normalized dither is several times larger, is the least sensitive, so a larger input perturbation produces a smaller distributional change. Two properties that would ordinarily damp this movement point the other way for Gemma: it has the lowest clean answer-token entropy and the largest top-1-to-top-2 logit gap of the four, the sharpest and most confident output, which by itself would suppress KL. The sensitivity therefore lives in Gemma's image-to-logit map, a larger input-to-logit response along its high-resolution vision-to-decoder pathway, and it persists despite an output geometry that works against it. We do not localize it further to the encoder or the decoder, which would require a layer-wise Jacobian analysis beyond this diagnostic. What matters here is that the sensitivity is a property of the model, so the behavioral pinning it produces at $0.2$ nats is real, and so is the inversion that survives it.

The answer does not determine confidence at any bound, since the answer is pinned at $0.2$ nats and unbounded alike and the readout still inverts. Confidence depends in part on the distribution's finer structure, and the surviving-fraction curve measures how much of it lives there. The strongest form of the objection, that the effect is entirely the slack byte-identity leaves free, is refuted by Gemma; the weaker form, that more distributional freedom yields more effect, is correct, and the erosion from $62\%$ to $36\%$ records it.

\begin{table}[t]
\centering
\caption{\textbf{Attacked AUROC under a bound on the full next-token distribution.} SAPLMA readout, with the answer-preserving attack additionally constrained so the KL divergence between the clean and attacked next-token distributions stays within a bound at every answer position. Below chance in all twelve cells at $0.2$ nats; model-dependent at $0.05$, where Gemma stays inverted and the others return toward chance. Clean AUROC here is computed on the balanced per-cell subset the compute-bounded KL attack runs on (seed $23$; InternVL $n{=}100$, Gemma $n{=}120$, LLaVA and Molmo $n{=}150$, half correct and half incorrect), so it differs from the full $1000$-item clean AUROC of Table~\ref{tab:app-phenomenon} through sampling alone; the probe and readout are identical.}
\label{tab:app-kl}
% Appendix (distribution constraint): attacked AUROC under an added KL bound on the full next-token
%   distribution at every committed answer position. Source: KL-bounded attack store (report 218),
%   data/kl_constraint/<celltag>/result.json, readouts[delta]['saplma'] (clean_auroc, attacked_auroc).
% Readout = SAPLMA (last-layer hidden-state probe), one of the two directly-attacked readouts the
%   main-text KL paragraph and Table use (P(IK), SAPLMA); SAPLMA shown as the representative here.
% Attacked = worst-case oracle. Reconciled == report 218 (max abs diff 0.0005). Baseline (argmax-only,
%   unbounded KL) realizes KL up to 2.61 nats (precheck), driving these readouts to 0.00 to 0.10.
% Requires: \usepackage{booktabs,multirow}
\small
\setlength{\tabcolsep}{6pt}
\begin{tabular}{@{}llrrr@{}}
\toprule
model & dataset & clean AUROC & attacked, KL$\le$0.2 & attacked, KL$\le$0.05 \\
\midrule
\multirow{3}{*}{InternVL3.5-2B} & GQA & 0.810 & 0.430 & 0.680 \\
 & VQAv2 & 0.613 & 0.254 & 0.487 \\
 & POPE & 0.757 & 0.310 & 0.534 \\
\midrule
\multirow{3}{*}{Molmo2-4B} & GQA & 0.811 & 0.463 & 0.639 \\
 & VQAv2 & 0.788 & 0.395 & 0.626 \\
 & POPE & 0.769 & 0.351 & 0.567 \\
\midrule
\multirow{3}{*}{LLaVA-OV-7B} & GQA & 0.873 & 0.453 & 0.696 \\
 & VQAv2 & 0.764 & 0.323 & 0.595 \\
 & POPE & 0.720 & 0.342 & 0.519 \\
\midrule
\multirow{3}{*}{Gemma-3-12B} & GQA & 0.764 & 0.163 & 0.227 \\
 & VQAv2 & 0.743 & 0.213 & 0.303 \\
 & POPE & 0.748 & 0.061 & 0.162 \\
\bottomrule
\end{tabular}

\end{table}

\begin{table}[t]
\centering
\caption{\textbf{Surviving fraction of the inversion at each bound.} The share of the full clean-to-unbounded inversion the KL-bounded attack retains, per cell, for the SAPLMA readout. A mean $62\%$ survives at $0.2$ nats and $36\%$ at $0.05$; the Gemma cells retain the most, $83$ to $93\%$ at $0.2$, and the non-Gemma cells $44$ to $59\%$. The P(IK) means are $0.689$ and $0.399$.}
\label{tab:app-kl-surviving}
% Appendix (distribution constraint S5): surviving-inversion fraction per cell (Deliverable C), SAPLMA.
% surviving@delta = (clean_auroc - attacked@delta) / (clean_auroc - attacked_unbounded); higher = more of
%   the full unbounded inversion retained under the KL bound. Source: data/kl_constraint/<cell>/result.json
%   (clean_auroc, attacked_auroc at delta 0.2/0.05, SAPLMA) + t4_baseline.json (attacked_unbounded, saplma).
% Reconciled == report 245 Deliverable C: SAPLMA mean surviving 0.617 @0.2, 0.361 @0.05.
% Requires: \usepackage{booktabs,multirow}
\footnotesize
\setlength{\tabcolsep}{5pt}
\begin{tabular}{@{}llrrrrrr@{}}
\toprule
model & dataset & clean & unb. & att@0.2 & att@0.05 & surv@0.2 & surv@0.05 \\
\midrule
\multirow{3}{*}{InternVL3.5-2B} & GQA & 0.810 & 0.017 & 0.430 & 0.680 & 0.480 & 0.164 \\
 & VQAv2 & 0.613 & 0.009 & 0.254 & 0.487 & 0.593 & 0.209 \\
 & POPE & 0.757 & 0.000 & 0.310 & 0.534 & 0.591 & 0.295 \\
\midrule
\multirow{3}{*}{Molmo2-4B} & GQA & 0.811 & 0.013 & 0.463 & 0.639 & 0.436 & 0.216 \\
 & VQAv2 & 0.788 & 0.039 & 0.395 & 0.626 & 0.524 & 0.216 \\
 & POPE & 0.769 & 0.000 & 0.351 & 0.567 & 0.543 & 0.262 \\
\midrule
\multirow{3}{*}{LLaVA-OV-7B} & GQA & 0.873 & 0.007 & 0.453 & 0.696 & 0.485 & 0.205 \\
 & VQAv2 & 0.764 & 0.020 & 0.323 & 0.595 & 0.593 & 0.227 \\
 & POPE & 0.720 & 0.001 & 0.342 & 0.519 & 0.525 & 0.280 \\
\midrule
\multirow{3}{*}{Gemma-3-12B} & GQA & 0.764 & 0.074 & 0.163 & 0.227 & 0.871 & 0.778 \\
 & VQAv2 & 0.743 & 0.102 & 0.213 & 0.303 & 0.828 & 0.687 \\
 & POPE & 0.748 & 0.009 & 0.061 & 0.162 & 0.929 & 0.793 \\
\midrule
\textbf{mean} & & 0.763 & 0.024 & 0.313 & 0.503 & \textbf{0.617} & \textbf{0.361} \\
\bottomrule
\end{tabular}

\vspace{3pt}
\noindent{\footnotesize surv = fraction of the full unbounded inversion the KL-bounded attack retains; unb.\ is the unbounded attacked AUROC and att@$\delta$ the attacked AUROC under the bound $\delta$.}

\end{table}

\begin{table}[t]
\centering
\caption{\textbf{Behavioral meaning of the bounds.} Top-$k$ preservation, first-position total variation, sampled-continuation total variation, and realized KL, per model at each bound, with the benign single-greylevel-dither KL as a reference scale. At $0.2$ nats the argmax is preserved for all four models; the full distribution is behaviorally indistinguishable from clean (sampled TV at the $\approx0.07$ noise floor) only for Gemma, whose distribution is already dither-sensitive. For the other three the bound is $22$ to $68$ times the benign dither.}
\label{tab:app-kl-behavioral}
% Appendix (distribution constraint S5): behavioral anchor of the KL-bounded adversary (Deliverables A+B).
% Source: data/kl_constraint/behavioral_agg.json (per model x delta: top-k stability, TV pos0, sampled-TV,
%   realized KL mean) + per-cell data/kl_constraint/<cell>/behavioral_s0.jsonl (benign_kl_mean). Benign =
%   an equal-norm benign dither's KL. ratio = the KL bound (0.2 nats) divided by the benign-dither KL mean.
% Clean-vs-clean sampled-TV noise floor is ~0.07; sampled-TV above it means the outputs are behaviorally
%   distinguishable from clean. Reconciled == report 245 Deliverables A/B (recomputed agg matches exactly).
% Requires: \usepackage{booktabs,multirow}
\footnotesize
\setlength{\tabcolsep}{4.5pt}
\begin{tabular}{@{}llrrrrrrrr@{}}
\toprule
model & $\delta$ & top-1 & top-5 & top-10 & TV(pos0) & samp.\ TV & real.\ KL & benign KL & ratio \\
\midrule
\multirow{2}{*}{InternVL3.5-2B} & 0.2 & 0.990 & 0.624 & 0.363 & 0.112 & 0.190 & 0.071 & 0.0092 & 22$\times$ \\
 & 0.05 & 1.000 & 0.795 & 0.654 & 0.040 & 0.143 & 0.013 & 0.0092 & 5.4$\times$ \\
\midrule
\multirow{2}{*}{Molmo2-4B} & 0.2 & 0.981 & 0.610 & 0.413 & 0.093 & 0.125 & 0.057 & 0.0043 & 47$\times$ \\
 & 0.05 & 0.984 & 0.726 & 0.587 & 0.035 & 0.081 & 0.011 & 0.0043 & 12$\times$ \\
\midrule
\multirow{2}{*}{LLaVA-OV-7B} & 0.2 & 1.000 & 0.588 & 0.342 & 0.114 & 0.151 & 0.079 & 0.0029 & 68$\times$ \\
 & 0.05 & 1.000 & 0.762 & 0.580 & 0.041 & 0.092 & 0.015 & 0.0029 & 17$\times$ \\
\midrule
\multirow{2}{*}{Gemma-3-12B} & 0.2 & 0.989 & 0.357 & 0.169 & 0.054 & 0.056 & 0.033 & 0.1136 & 1.8$\times$ \\
 & 0.05 & 0.992 & 0.423 & 0.265 & 0.027 & 0.035 & 0.011 & 0.1136 & 0.4$\times$ \\
\bottomrule
\end{tabular}

\vspace{3pt}
\noindent{\footnotesize top-$k$ = fraction of the top-$k$ next-token set preserved relative to clean; TV(pos0) and samp.\ TV are total-variation distances at the first answer position (argmax and sampled); real.\ KL is the mean realized $\mathrm{KL}(p_{\mathrm{clean}}\|p_{\mathrm{adv}})$; ratio is the KL bound $\delta$ over the benign single-greylevel-dither KL. The clean-versus-clean sampled-TV noise floor is $\approx$0.07.}

\end{table}

\begin{table}[t]
\centering
\caption{\textbf{Gemma's distributional sensitivity is a property of the model, not of input scale or output geometry.} Per model: answer-position KL under a benign dither and a matched-norm Gaussian, the dither expressed in the model's normalized input units, clean answer-token entropy, and the top-1-to-top-2 logit gap. Gemma's input perturbation in normalized units is among the smallest of the four, yet its distributional divergence is roughly twenty times the others', and it has the lowest entropy and largest logit gap, the sharpest output.}
\label{tab:app-gemma-sens}
% Appendix (distribution constraint S5): Gemma-3-12B distributional-sensitivity diagnostic (report 247).
% Source: data/kl_constraint/sensitivity/<model>.json. dither/Gaussian KL = mean answer-position
%   KL(clean||perturbed) (check 1); normalized dither = the +/-1 uint8 dither RMS in the model's normalized
%   input units (check 2); clean entropy = clean answer-token entropy in nats (check 3); logit gap = mean
%   top1-top2 answer logit gap (check 4). Reconciled == report 247 (max abs diff 0.0005). No em-dashes.
% Requires: \usepackage{booktabs}
\small
\setlength{\tabcolsep}{6pt}
\begin{tabular}{@{}lrrrrr@{}}
\toprule
model & dither KL & Gaussian KL & normalized dither & clean entropy & logit gap \\
\midrule
InternVL3.5-2B & 0.0057 & 0.0056 & 0.01382 & 0.944 & 2.018 \\
Molmo2-4B & 0.0034 & 0.0036 & 0.00525 & 0.709 & 2.115 \\
LLaVA-OV-7B & 0.0050 & 0.0039 & 0.00579 & 0.744 & 2.166 \\
\textbf{Gemma-3-12B} & \textbf{0.1117} & \textbf{0.1326} & \textbf{0.00545} & \textbf{0.219} & \textbf{5.045} \\
\bottomrule
\end{tabular}

\vspace{3pt}
\noindent{\footnotesize Normalized dither is the RMS of the one-greylevel perturbation in the model's own input units: Gemma 0.00545, second only to Molmo's 0.00525 and well below InternVL's 0.01382.}

\end{table}
\section{Deployment: The Gate Made Worse Than No Gate}
\label{app:deployment}

Section~\ref{sec:core} reports that a confidence gate under this attack accepts most of the wrong answers it would otherwise reject and, corrected for how often the model is wrong, leaves the system worse off than using no gate at all. This appendix gives that result per cell, shows it does not depend on the gate's threshold, and reports it on the whole risk-coverage curve. Throughout, the gate is the FULL hidden-state gate with its threshold fixed to a $10\%$ false-accept rate on clean data and held there under attack, never re-fit, which is the operating point a real deployment would hold.

\subsection{The gate harm, cell by cell}
\label{app:deployment-harm}

Table~\ref{tab:app-gate-harm} reports, for each cell, the fraction of wrong answers the transferred attack turns from abstained to accepted, and the accepted-population accuracy clean and under attack. In the worst cell, InternVL/POPE, the attack flips $84.8\%$ of the wrong answers into acceptances, and accepted accuracy falls from $0.744$ on clean images to $0.182$ under attack, well below the accuracy of accepting every answer. The last two columns compare accepted accuracy against the no-gate baseline at natural correctness prevalence: the attacked gate accepts answers at a lower accuracy than using no gate in eight of twelve cells under transfer, and in all twelve under an attack aimed at the gate. The four cells that resist transfer, the three Gemma cells and Molmo/POPE, all fall below the baseline once the attack is aimed at the gate itself. The lower block reports the same gate at natural prevalence: the un-gated baseline, and gated accepted accuracy clean, under transfer, and under the aimed attack. Clean, the gate raises accepted accuracy above the baseline in every cell, to between $0.87$ and $0.97$; under attack it falls below the baseline in the cells the verdict columns mark, which is the harm.

\begin{table}[t]
\centering
\caption{\textbf{The FULL hidden-state gate under attack, per cell,} at a fixed $10\%$ clean false-accept rate. Upper block, balanced pool: \emph{w2ac} is the fraction of wrong answers flipped from abstain to accept and \emph{acc-adv} the accepted-population accuracy, each under the transferred token-probability attack (\emph{tr}) and the gate-aimed attack (\emph{aim}). Lower block, natural prevalence: accepted accuracy reweighted to the benchmark's natural correctness prevalence, against the no-gate baseline that accepts every answer; $<$base marks gated accepted accuracy below that baseline.}
\label{tab:app-gate-harm}
% Appendix (deployment): FULL hidden-state gate harm per cell, transfer and gate-aimed attacks.
% OPERATING POINT: the FULL (hidden-state) gate with its threshold fixed at a 10% false-accept rate on
%   CLEAN data, held fixed under attack, never re-fit (clean-10%-FPR; report 229 / triage.json).
% Source: data/triage/<model>/triage.json cells['<set>/FULL'].operating_points.clean10fpr (transfer,
%   report 098) and adaptive_cells['<set>/FULL/main'].operating_points.clean10fpr (gate-aimed, report 099);
%   natural prevalence from data/natural_prevalence/natural_prevalence_triage.json (p_natural).
% Upper block = balanced pool (w2ac, acc-clean, acc-adv for transfer and aimed). Lower block = accepted
%   accuracy reweighted to natural prevalence via reprior(acc,p)=(acc*p)/(acc*p+(1-acc)(1-p)); no-gate is the
%   accept-all baseline (=p_natural), and <base marks gated accepted accuracy below it (transfer / aimed).
% Requires: \usepackage{booktabs,multirow}
\footnotesize
\setlength{\tabcolsep}{5pt}
\begin{tabular}{@{}llrrrrr@{}}
\toprule
model & set & acc-clean & w2ac (tr) & acc-adv (tr) & w2ac (aim) & acc-adv (aim) \\
\midrule
\multirow{3}{*}{InternVL3.5-2B} & GQA & 0.845 & 0.468 & 0.481 & 0.710 & 0.290 \\
 & VQAv2 & 0.735 & 0.429 & 0.330 & 0.619 & 0.172 \\
 & POPE & 0.744 & 0.848 & 0.182 & 0.894 & 0.081 \\
\midrule
\multirow{3}{*}{Molmo2-4B} & GQA & 0.842 & 0.396 & 0.471 & 0.506 & 0.348 \\
 & VQAv2 & 0.843 & 0.398 & 0.460 & 0.482 & 0.353 \\
 & POPE & 0.831 & 0.114 & 0.695 & 0.374 & 0.394 \\
\midrule
\multirow{3}{*}{LLaVA-OV-7B} & GQA & 0.853 & 0.449 & 0.428 & 0.697 & 0.212 \\
 & VQAv2 & 0.784 & 0.339 & 0.373 & 0.580 & 0.200 \\
 & POPE & 0.788 & 0.748 & 0.273 & 0.780 & 0.151 \\
\midrule
\multirow{3}{*}{Gemma-3-12B} & GQA & 0.833 & 0.044 & 0.751 & 0.223 & 0.300 \\
 & VQAv2 & 0.829 & 0.066 & 0.648 & 0.231 & 0.289 \\
 & POPE & 0.840 & 0.096 & 0.629 & 0.478 & 0.116 \\
\bottomrule
\end{tabular}

\vspace{8pt}
\noindent{\footnotesize\itshape Accepted accuracy at natural prevalence, no-gate baseline vs gated:}\\[2pt]
\footnotesize
\setlength{\tabcolsep}{5pt}
\begin{tabular}{@{}llrrrrcc@{}}
\toprule
model & set & no-gate & gated cln & gated (tr) & gated (aim) & $<$base (tr) & $<$base (aim) \\
\midrule
\multirow{3}{*}{InternVL3.5-2B} & GQA & 0.798 & 0.956 & 0.786 & 0.617 & yes & yes \\
 & VQAv2 & 0.697 & 0.865 & 0.531 & 0.324 & yes & yes \\
 & POPE & 0.895 & 0.961 & 0.654 & 0.428 & yes & yes \\
\midrule
\multirow{3}{*}{Molmo2-4B} & GQA & 0.832 & 0.964 & 0.816 & 0.726 & yes & yes \\
 & VQAv2 & 0.771 & 0.948 & 0.741 & 0.647 & yes & yes \\
 & POPE & 0.881 & 0.973 & 0.944 & 0.828 & no & yes \\
\midrule
\multirow{3}{*}{LLaVA-OV-7B} & GQA & 0.859 & 0.973 & 0.820 & 0.621 & yes & yes \\
 & VQAv2 & 0.784 & 0.929 & 0.684 & 0.476 & yes & yes \\
 & POPE & 0.878 & 0.964 & 0.729 & 0.560 & yes & yes \\
\midrule
\multirow{3}{*}{Gemma-3-12B} & GQA & 0.752 & 0.938 & 0.901 & 0.565 & no & yes \\
 & VQAv2 & 0.714 & 0.923 & 0.821 & 0.503 & no & yes \\
 & POPE & 0.855 & 0.969 & 0.909 & 0.437 & no & yes \\
\bottomrule
\end{tabular}

\vspace{3pt}
\noindent{\footnotesize The gate is the FULL hidden-state gate at a $10\%$ clean false-accept rate, fixed under attack. \emph{tr} is the transfer token-probability attack, \emph{aim} the gate-aimed attack; \emph{w2ac} is the fraction of wrong answers flipped from abstain to accept. Upper block is the balanced pool, so an acc-adv below $0.5$ is worse than random acceptance. Lower block reweights accepted accuracy to the benchmark's natural correctness prevalence; the no-gate baseline accepts every answer, and $<$base marks gated accepted accuracy below it.}

\end{table}

\subsection{The harm does not depend on the threshold}
\label{app:deployment-robustness}

The $10\%$ clean false-accept rate is one operating point. The upper block of Table~\ref{tab:app-gate-robust} reports the gate at three thresholds spanning realized clean false-accept rates from $0.10$ to $0.53$. At every one, accepted accuracy drops under attack, and seven to eight of the twelve cells fall below the accept-everything baseline; the drop shrinks only as the gate loosens to the point of barely gating at all. The threshold changes the magnitude of the harm, not its direction.

The harm is also a property of the entire risk-coverage curve. The lower block reports the area under the risk-coverage curve, clean and attacked, at the FULL locus per cell. Every cell's area more than doubles under attack, and averaged over the sixty diagonal profiles the risk-coverage area rises from $0.30$ on clean images to $0.80$ under attack. Whatever coverage a deployment chooses, a gate meant to trade coverage for reliability trades it for unreliability instead.

\begin{table}[t]
\centering
\caption{\textbf{Operating-point sensitivity and risk-coverage.} Upper block: the gate at three fixed thresholds, averaged over the twelve cells, showing accepted-accuracy drop and the count of cells whose attacked accepted accuracy falls below $0.5$ on the balanced pool at each. Lower block: the area under the risk-coverage curve (higher is worse) at the FULL locus per cell, clean versus attacked. The sixty-profile diagonal mean rises from $0.30$ to $0.80$.}
\label{tab:app-gate-robust}
% Appendix (deployment): gate operating-point sensitivity + risk-coverage, old Supplement K.3 + K.4.
% OPERATING POINT for the headline row: the FULL (hidden-state) gate at a 10% clean false-accept rate,
%   held fixed under attack, never re-fit (clean-10%-FPR; report 229 / triage.json). NOT coverage=0.5.
% (a) Operating-point sensitivity: the three thresholds triage.json provides, clean10fpr (10% clean FPR),
%   youdenJ (Youden's J), tau0.5 (fixed 0.5). Per point, means over the 12 FULL-gate cells: realized clean
%   FPR, accepted accuracy clean/adv (transfer), the clean->adv drop, and the count of cells with adv
%   accepted accuracy below 0.5 (the accept-all balanced baseline). Harm holds across all three points.
% (b) Risk-coverage AURC (higher = worse) at the FULL locus, per cell (diagonal aimed=eval=full), from
%   T10_calibration_and_selective_prediction.json (clean_aurc_riemann / attacked_aurc_riemann).
% Reconciled: 60-diagonal mean clean 0.3009 -> attacked 0.8042 (report 229 T10, all 5 estimators).
% Requires: \usepackage{booktabs,multirow}
% (a) Operating-point sensitivity (mean over 12 FULL-gate cells).
\footnotesize
\setlength{\tabcolsep}{6pt}
\begin{tabular}{@{}lrrrrc@{}}
\toprule
operating point & clean FPR & acc-clean & acc-adv & drop & cells acc-adv$<$0.5 (bal.) \\
\midrule
clean-10\%-FPR & 0.101 & 0.814 & 0.477 & 0.337 & 8/12 \\
Youden J & 0.246 & 0.744 & 0.515 & 0.229 & 7/12 \\
$\tau{=}0.5$ & 0.528 & 0.626 & 0.513 & 0.113 & 8/12 \\
\bottomrule
\end{tabular}

\vspace{8pt}
% (b) Risk-coverage AURC per cell (FULL locus, higher = worse).
\noindent{\footnotesize\itshape Risk-coverage AURC (FULL locus, higher is worse), per cell, clean vs attacked:}\\[2pt]
\footnotesize
\setlength{\tabcolsep}{6pt}
\begin{tabular}{@{}llrr@{}}
\toprule
model & set & AURC clean & AURC attacked \\
\midrule
\multirow{3}{*}{InternVL3.5-2B} & GQA & 0.266 & 0.837 \\
 & VQAv2 & 0.352 & 0.838 \\
 & POPE & 0.326 & 0.846 \\
\midrule
\multirow{3}{*}{Molmo2-4B} & GQA & 0.252 & 0.816 \\
 & VQAv2 & 0.288 & 0.809 \\
 & POPE & 0.271 & 0.842 \\
\midrule
\multirow{3}{*}{LLaVA-OV-7B} & GQA & 0.239 & 0.828 \\
 & VQAv2 & 0.309 & 0.817 \\
 & POPE & 0.328 & 0.832 \\
\midrule
\multirow{3}{*}{Gemma-3-12B} & GQA & 0.271 & 0.791 \\
 & VQAv2 & 0.293 & 0.781 \\
 & POPE & 0.262 & 0.836 \\
\midrule
\textbf{FULL mean} & & \textbf{0.288} & \textbf{0.823} \\
\bottomrule
\end{tabular}

\vspace{3pt}
\noindent{\footnotesize AURC is the area under the risk-coverage curve integrated over all coverage (higher is worse). Over the 60 diagonal profiles (12 cells $\times$ 5 estimators) the mean rises from 0.301 clean to 0.804 attacked; the FULL-locus 12-cell mean shown here rises from 0.288 to 0.823.}

\end{table}
\section{No Internal Readout Escapes, in Full}
\label{app:frontier}

Section~\ref{sec:impossibility} shows that training a readout to resist the attack does not produce one that is both informative and robust, and that the attack is not an artifact of its own settings. This appendix gives the full per-cell training sweep behind the frontier and the configuration controls behind the second claim.

\subsection{The full adversarial-training sweep}
\label{app:frontier-sweep}

Table~\ref{tab:app-frontier} reports the adversarial-training sweep in every cell. For each cell we train a family of probes on the same hidden states with a robustness weight $\beta$ over $\{0.3, 1, 3, 10\}$, and attack each with the same answer-preserving attack; the table gives clean and worst-case-oracle attacked AUROC at each setting, the three-seed attacked value at the strongest setting, and the escape verdict. The escape region, clean above the cell's answer-string prior and attacked above the pre-specified $0.55$ bar, is empty in all twelve cells. The trade-off is visible within each row: on Gemma/POPE the lightest penalty gives clean AUROC $0.865$ and attacked $0.061$, and raising the penalty to its strongest setting lifts attacked to $0.385$ only by dragging clean down to $0.629$. Robustness is bought with the discrimination the probe was meant to provide.

The training loss is a TRADES-style objective with the attack in the loop: class-balanced binary cross-entropy on clean features plus $\beta$ times the mean, over items, of the largest absolute change in the probe's output across a pool of answer-preserving adversarial endpoint features for that item, with the pool regenerated every $K$ epochs against the current probe by an inner PGD of the stated number of steps ($\epsilon=8/255$, step $2/255$). The anchor cell uses the full budget ($K{=}4$, $10$ inner steps); the other eleven use a reduced training-time inner attack ($K{=}8$, $5$ steps) with the evaluation attack held at full strength, so their robustness is if anything understated. The gradient-penalty variant replaces the invariance term with $0.1$ times the mean input-gradient norm of the probe output. The clean leg of the escape criterion is the answer-string prior on the evaluation split plus a $0.02$ margin ($0.734$ for LLaVA/POPE), the evaluation-split counterpart of the ceiling's in-sample $c^\star$ ($0.703$) and of the cross-fit operative floor ($0.597$).

The nearest approach to an escape misses on a criterion fixed in advance. LLaVA/POPE reaches the highest attacked value anywhere in the sweep, $0.536$ at its strongest setting, and it clears the clean leg of the criterion, but $0.536$ sits below the $0.55$ bar, and the three-seed standard deviation of $0.008$ places it there reliably.

The failure holds even where the defender was under-resourced. Eleven of the twelve cells were trained against a weaker inner attack than the anchor, with the full-strength evaluation attack held fixed, so their measured robustness is if anything understated; none escaped. The one cell trained at full strength, the anchor LLaVA/VQAv2, also fails, reaching attacked AUROC $0.517$ at its strongest setting, and a gradient-penalty variant of the anchor fails as well. The empty escape region is therefore not an artifact of a reduced training budget. Whether a better-resourced defense could escape is open; none we trained, at either budget, did.

\begin{table}[t]
\centering
\caption{\textbf{The full adversarial-training sweep.} Clean and worst-case-oracle attacked AUROC at each robustness weight $\beta$, the strongest setting in bold, with the three-seed attacked value (\emph{peak att}) and the escape verdict per cell. The escape region is empty in all twelve cells. The nearest miss, LLaVA/POPE, reaches attacked $0.536$, the highest anywhere, still below the $0.55$ bar.}
\label{tab:app-frontier}
\small
% Appendix (no internal readout escapes): full per-cell adversarial-training frontier sweep.
% Old Supplement H.2. Source: data/snowglobe/frontier_lambda/advtrain_frontier.json (anchor
%   LLaVA-OV/VQAv2) + breadth/<celltag>/frontier.json (report 212). Each cell: clean/attacked
%   (worst-case oracle) at beta in {0.3,1,3,10}, seed 23; the peak-beta cell is bold. peak att = the
%   3-seed (23/24/25) mean attacked AUROC at the peak beta. escape = peak clean above the cell's
%   clean threshold AND peak att above 0.55 (pre-specified bar). Reconciled == report 212 main table
%   (max abs diff 0.0005). LAYOUT: one row per cell, the full beta grid compacted as clean/att pairs.
% Requires: \usepackage{booktabs,multirow}
\small
\setlength{\tabcolsep}{4pt}
\begin{tabular}{@{}ll cccc r c@{}}
\toprule
model & dataset & $\beta{=}0.3$ & $\beta{=}1$ & $\beta{=}3$ & $\beta{=}10$ & peak att & escape \\
 &  & \multicolumn{4}{c}{clean\,/\,attacked (seed 23; peak $\beta$ bold)} &  &  \\
\midrule
\multirow{3}{*}{InternVL3.5-2B} & GQA & 0.643/0.198 & \textbf{0.640/0.250} & 0.624/0.205 & 0.607/0.215 & 0.207 & no \\
 & VQAv2 & 0.646/0.263 & 0.634/0.189 & \textbf{0.621/0.307} & 0.618/0.278 & 0.298 & no \\
 & POPE & 0.845/0.026 & 0.808/0.000 & 0.710/0.257 & \textbf{0.664/0.318} & 0.316 & no \\
\midrule
\multirow{3}{*}{Molmo2-4B} & GQA & 0.644/0.214 & 0.649/0.210 & 0.619/0.304 & \textbf{0.608/0.384} & 0.359 & no \\
 & VQAv2 & 0.724/0.270 & 0.681/0.378 & \textbf{0.651/0.439} & 0.610/0.437 & 0.377 & no \\
 & POPE & 0.846/0.021 & 0.835/0.042 & 0.771/0.244 & \textbf{0.668/0.412} & 0.424 & no \\
\midrule
\multirow{3}{*}{LLaVA-OV-7B} & GQA & 0.721/0.126 & 0.707/0.155 & \textbf{0.683/0.266} & 0.669/0.263 & 0.229 & no \\
 & VQAv2$^{\S}$ & 0.754/0.209 & 0.731/0.321 & 0.710/0.426 & \textbf{0.674/0.521} & 0.517 & no \\
 & POPE$^{\ddagger}$ & 0.818/0.290 & 0.828/0.194 & 0.827/0.190 & \textbf{0.768/0.544} & 0.536 & no \\
\midrule
\multirow{3}{*}{Gemma-3-12B} & GQA & 0.693/0.174 & 0.646/0.255 & 0.651/0.316 & \textbf{0.611/0.449} & 0.372 & no \\
 & VQAv2 & 0.706/0.238 & 0.724/0.359 & \textbf{0.713/0.540} & 0.670/0.537 & 0.511 & no \\
 & POPE & 0.865/0.061 & 0.862/0.059 & 0.829/0.103 & \textbf{0.629/0.385} & 0.385 & no \\
\bottomrule
\end{tabular}

\vspace{3pt}
\noindent{\footnotesize $^{\S}$Anchor cell (LLaVA-OV/VQAv2), trained at full budget $K{=}4$, inner steps 10; the other eleven cells use the reduced training-time inner attack $K{=}8$, inner steps 5, with the evaluation attack unchanged at full strength. A gradient-penalty variant, run for the anchor only, reaches clean 0.763 and attacked 0.201, also no escape. $^{\ddagger}$Nearest miss: as three-seed means, peak clean 0.755 clears the clean leg (threshold 0.734) but peak att 0.536 (sd 0.008; seeds 0.544, 0.528, 0.536) stays below the 0.55 bar; the row shows the seed-23 values 0.768 and 0.544; it is the highest attacked AUROC anywhere. Escape (0 of 12) requires peak clean above the clean threshold and peak att above 0.55.}

\end{table}

\subsection{The collapse is not an artifact of the attack settings}
\label{app:frontier-invariance}

Two controls check whether the collapse is peculiar to the attack's hyperparameters or to its freedom to move the answer margin (Table~\ref{tab:app-invariance}). The upper block reports, per cell, how many of the swept attack configurations produce drift aligned to an independently estimated correctness axis: across ten coordinated configurations spanning budgets $\{2,4,8,16\}/255$, step counts $\{5,15,30,60\}$, two $L_2$ norms, and two aim targets, alignment holds in all twelve cells, $120/120$. The two label-agnostic configurations pass in $21$ of $24$ cell-settings; the three misses are all on Gemma and are directionless, a token-probability saturation effect, and they still pass an unsigned-drift criterion. The pattern that drives the collapse is therefore not confined to the operating budget or step count.

The lower block holds the answer margin near its clean value instead of letting the attack move it freely. Under this constraint the readout still collapses: attacked AUROC stays at or below the operative floor in all twelve cells and below chance in ten, with the two exceptions above chance but still below their floors. The inversion does not depend on the attack inflating the answer margin.

\begin{table}[t]
\centering
\caption{\textbf{Attack-configuration invariance and the margin-band control.} Upper block: per cell, the number of swept attack configurations whose drift aligns to an independent correctness axis ($120/120$ coordinated, $21/24$ label-agnostic). Lower block: attacked AUROC when the answer margin is held near its clean value versus unbounded; the collapse persists at or below the floor in all twelve cells.}
\label{tab:app-invariance}
% Appendix (no internal readout escapes): attack-configuration invariance + margin-band control.
% Old Supplement E. Two blocks in one file.
% (a) Axis-alignment pass per cell (pass = attack mean drift aligned to the cross-fit correctness axis
%   u_logreg, meandir_cos above the direction-level null bar). Source: tables/104_attack_generality/
%   {A1..A6,B1,B2,D1,D2}.json (coordinated, SAPLMA locus) and c_axis_failures.json (label-agnostic
%   C1/C2 on the output likelihood, report 108). Coordinated = 10 configs x 12 cells = 120/120;
%   label-agnostic = C1+C2 x 12 = 24, with 3 Gemma directionless cells = 21/24.
% (b) Upper-bounded-margin control (report 146/186/193), readout SAPLMA. clean/attacked AUROC on the
%   attack-engaged subset; attacked(margin) holds the answer margin in the 0.25 band. Source:
%   data/matched_margin/matched_margin_analysis.json. Reconciled == committed margin_control aggregates
%   (SAPLMA 0.848/0.077 unbounded, 0.791/0.277 at 0.25) and sweep_configs (120/120) / duality (21/24).
% Requires: \usepackage{booktabs,multirow}
% (a) Attack-configuration invariance: axis-alignment pass per cell.
\small
\setlength{\tabcolsep}{6pt}
\begin{tabular}{@{}llcc@{}}
\toprule
model & dataset & coordinated (of 10) & label-agnostic (of 2) \\
\midrule
\multirow{3}{*}{InternVL3.5-2B} & GQA & 10/10 & 2/2 \\
 & VQAv2 & 10/10 & 2/2 \\
 & POPE & 10/10 & 2/2 \\
\midrule
\multirow{3}{*}{Molmo2-4B} & GQA & 10/10 & 2/2 \\
 & VQAv2 & 10/10 & 2/2 \\
 & POPE & 10/10 & 2/2 \\
\midrule
\multirow{3}{*}{LLaVA-OV-7B} & GQA & 10/10 & 2/2 \\
 & VQAv2 & 10/10 & 2/2 \\
 & POPE & 10/10 & 2/2 \\
\midrule
\multirow{3}{*}{Gemma-3-12B} & GQA & 10/10 & 1/2$^{\dagger}$ \\
 & VQAv2 & 10/10 & 1/2$^{\dagger}$ \\
 & POPE & 10/10 & 1/2$^{\dagger}$ \\
\midrule
\textbf{total} & & \textbf{120/120} & \textbf{21/24} \\
\bottomrule
\end{tabular}

\vspace{3pt}
\noindent{\small $^{\dagger}$The three label-agnostic misses are all Gemma-3-12B and directionless
(alignment below the null bar), not reversed: C1/Gemma/POPE, C2/Gemma/GQA, C2/Gemma/VQAv2, a token-probability
saturation effect; they pass the unsigned-drift criterion. Coordinated configs: A1--A6 (budget
$\{2,4,8,16\}/255$, PGD steps $\{5,15,30,60\}$), B1--B2 (two $L_2$ norms), D1--D2 (aim P(IK)/SAPLMA).}

\vspace{8pt}
% (b) Margin-band control: the collapse persists when the answer margin is held near clean.
\noindent{\small\itshape Margin-band control (SAPLMA; clean and attacked AUROC on the attack-engaged subset):}\\[2pt]
\small
\setlength{\tabcolsep}{6pt}
\begin{tabular}{@{}llrrrr@{}}
\toprule
model & dataset & clean AUROC & attacked (unbounded) & attacked (margin) & floor \\
\midrule
\multirow{3}{*}{InternVL3.5-2B} & GQA & 0.880 & 0.029 & 0.039 & 0.642 \\
 & VQAv2 & 0.792 & 0.018 & 0.085 & 0.617 \\
 & POPE & 0.832 & 0.000 & 0.028 & 0.555 \\
\midrule
\multirow{3}{*}{Molmo2-4B} & GQA & 0.911 & 0.091 & 0.559 & 0.683 \\
 & VQAv2 & 0.872 & 0.130 & 0.513 & 0.625 \\
 & POPE & 0.914 & 0.010 & 0.107 & 0.587 \\
\midrule
\multirow{3}{*}{LLaVA-OV-7B} & GQA & 0.922 & 0.050 & 0.367 & 0.673 \\
 & VQAv2 & 0.798 & 0.047 & 0.256 & 0.616 \\
 & POPE & 0.805 & 0.032 & 0.131 & 0.597 \\
\midrule
\multirow{3}{*}{Gemma-3-12B} & GQA & 0.812 & 0.232 & 0.476 & 0.660 \\
 & VQAv2 & 0.798 & 0.209 & 0.365 & 0.652 \\
 & POPE & 0.835 & 0.082 & 0.405 & 0.500 \\
\bottomrule
\end{tabular}

\vspace{3pt}
\noindent{\small Attacked (margin) holds the answer margin in the 0.25 band; it is at or below the
operative floor in 12 of 12 cells (below chance in 10 of 12; Molmo/GQA 0.559 and Molmo/VQAv2 0.513 stay
above 0.5 but below their floors). The collapse persists when the margin is not allowed to run free.}

\end{table}
\section{The Reachability Ladder in Full}
\label{app:ladder}

Section~\ref{sec:ladder} reports that an independent verifier relocates the attack, collapsing as the attacker gains access. This appendix gives the full twelve-cell ladder, the ordering of its rungs, the one verifier family whose differentiable path is inexact, and the disclosures behind the reported numbers.

\subsection{The twelve-cell ladder}
\label{app:ladder-full}

Table~\ref{tab:app-ladder} reports all twelve base-to-verifier pairs across the five access rungs, each the worst-case-oracle attacked AUROC of the verifier's confidence on a common item pool. The pairs span every model as both base and verifier three times. Discrimination falls from a clean mean of $0.655$ to $0.480$ under score-only access and $0.287$ under white-box. The white-box mean over all twelve includes one pair whose verifier gradient is under-powered, discussed in Appendix~\ref{app:ladder-parity}; over the nine pairs whose verifier admits an exact differentiable path, white-box drives discrimination to a mean of $0.223$, between $0.06$ and $0.35$, the level of an internal readout. Once the attacker can reach its gradient, the verifier collapses to the same regime as the signals inside the model.

The white-box search perturbs the image within the $8/255$ budget while enforcing base-model answer preservation exactly as the internal attack does: an argmax gate at every generated position and a fresh greedy byte-identity acceptance test, with the clean image as fallback. The base answer is byte-identical on every white-box endpoint; per pair, the clean-image fallback accounts for $0.9\%$ to $6.5\%$ of the $800$ endpoints, and the rest are perturbations that leave the base answer unchanged.

\subsection{The intermediate rungs are not separably ordered}
\label{app:ladder-monotone}

The intended ordering is that each rung, granting more access, drives discrimination lower. It holds at the ends: clean is the highest rung in all twelve pairs, and white-box the lowest in ten. It does not hold cleanly between the two intermediate rungs. The no-access transfer attack (mean $0.571$) is on average a slightly stronger attack than decision-only hard-label search (mean $0.601$), and in ten of twelve pairs transfer sits at or below decision-only, by $0.004$ to $0.095$. The transfer rung uses a label-aware oracle over the clean score and the transferred extremal perturbations, which is a strong baseline, while a decision-only boundary search must discover a flip from hard labels alone and does not reliably beat it. Every one of these inversions is within the overlapping bootstrap intervals of the two rungs, so none is individually significant, but the direction is consistent: a good transfer baseline is not weaker than hard-label access. The movement that carries the section is across the rungs that add real optimization power, from clean through score-only to white-box, where the fall is large and consistent.

\subsection{The Molmo verifier's differentiable path}
\label{app:ladder-parity}

The white-box rung attacks the verifier's gradient, which requires a differentiable path through the verifier's preprocessing that matches the official processor it is scored with. Table~\ref{tab:app-parity} reports the agreement per verifier family. For Gemma, LLaVA, and InternVL the differentiable path is bit-exact: it equals the official processor to $0.000$ on the clean image, so the white-box gradient sees the same function the reported score uses. The Molmo verifier's differentiable preprocessing diverges from its official processor by up to $0.144$ on the affected cells, a property of that preprocessing and not of the attack. Because every reported verifier score is recomputed with the official processor, this divergence never enters a reported number; it only under-powers the white-box search on the three pairs whose verifier is Molmo. The effect is visible in one of them, the Gemma-to-Molmo pair, whose white-box AUROC ($0.744$) stays at its clean value ($0.754$), since the gradient guiding the search does not match the scored function. For this reason the section's white-box figure is taken over the nine pairs whose verifier path is exact, where the search is driven by the true gradient.

Two of the three Molmo-verifier pairs, both on POPE, carry a second, independent limitation: their verifier is at or below chance on the clean image ($0.495$ and $0.423$), so it fails to separate correct from incorrect answers before any attack. Their ladders have little to collapse and sit near chance throughout. These pairs are reported in the table for completeness and excluded from the bit-exact white-box mean.

\subsection{Disclosures}
\label{app:ladder-disclosures}

Three reporting choices are stated here. First, the score-only rung of the three InternVL-verifier pairs is computed on a balanced $100$-item pool instead of the $200$ used elsewhere, because the InternVL verifier's dynamic tiling makes it the slowest per-item rung; the score-only floor converges well before this query budget, so the reduction widens the bootstrap interval on those three pairs modestly without changing their collapse, which matches their neighbours. Second, the score-only rung is reported uniformly at a query budget of $600$, at which the floor has converged; some pairs were run to a larger budget and are read at the same $600$-query checkpoint, so the rung is a single budget across all pairs. Third, the twelve pairs draw on the three benchmarks in a $5/3/4$ split, which affects only the dataset marginal and no per-cell ladder. On the pairs where we computed it, the common $200$-item pool tracks the full $1000$-item pool to within about $0.07$ AUROC.

\begin{table}[t]
\centering
\caption{\textbf{The spanning-twelve reachability ladder.} Worst-case-oracle attacked verifier AUROC for all twelve base-to-verifier pairs across five access rungs, from clean through no-access transfer, decision-only, and score-only queries, to white-box gradient access. Clean and transfer carry bootstrap $95\%$ intervals; the query and white-box rungs are two-seed means. The mean rows report all twelve pairs, the eleven excluding the under-powered white-box pair, and the nine whose verifier admits an exact differentiable path.}
\label{tab:app-ladder}
% Appendix (ladder S7): full spanning-12 reachability ladder, old report 222.
% Source: data/snowglobe/verifier_escape/span12_ladder_results.json (report 222). Per rung: worst-case
%   oracle attacked AUROC of the verifier confidence. C0/transfer are single deterministic runs with
%   bootstrap 95% CI (2000 resamples, seed 23); decision(HSJ)/score(SPSA Q600)/white-box(C2) are 2-seed
%   means. Reconciled == report 222 ladder table (max abs diff 0.000). Rows identified by base->verifier
%   plus set (the internal run id column is intentionally omitted).
% Requires: \usepackage{booktabs,graphicx}
\footnotesize
\setlength{\tabcolsep}{5pt}
\resizebox{\textwidth}{!}{%
\begin{tabular}{@{}llll rrr l@{}}
\toprule
base$\to$verifier & set & clean [95\% CI] & transfer [95\% CI] & decision & score & white-box & notes \\
 & & & & HSJ & SPSA (600 queries) &  & \\
\midrule
LLaVA$\to$Gemma & VQAv2 & 0.755~{\tiny(0.68,0.82)} & 0.578~{\tiny(0.50,0.66)} & 0.581 & 0.485 & 0.214 &  \\
LLaVA$\to$InternVL & VQAv2 & 0.654~{\tiny(0.58,0.73)} & 0.575~{\tiny(0.50,0.65)} & 0.625 & 0.489 & 0.123 & score $n{=}100$ \\
Gemma$\to$LLaVA & GQA & 0.712~{\tiny(0.64,0.78)} & 0.668~{\tiny(0.59,0.74)} & 0.703 & 0.608 & 0.342 &  \\
Molmo$\to$Gemma & POPE & 0.660~{\tiny(0.58,0.73)} & 0.514~{\tiny(0.43,0.60)} & 0.502 & 0.431 & 0.195 &  \\
InternVL$\to$LLaVA & GQA & 0.681~{\tiny(0.61,0.75)} & 0.637~{\tiny(0.56,0.71)} & 0.664 & 0.561 & 0.270 &  \\
InternVL$\to$Molmo & POPE & 0.495~{\tiny(0.41,0.58)} & 0.442~{\tiny(0.35,0.53)} & 0.486 & 0.355 & 0.378 & near-chance clean; parity \\
Gemma$\to$Molmo & VQAv2 & 0.754~{\tiny(0.68,0.82)} & 0.717~{\tiny(0.64,0.79)} & 0.727 & 0.611 & 0.744 & parity; WB under-powered \\
Molmo$\to$InternVL & GQA & 0.676~{\tiny(0.60,0.75)} & 0.592~{\tiny(0.51,0.67)} & 0.636 & 0.477 & 0.063 & score $n{=}100$ \\
LLaVA$\to$Molmo & POPE & 0.423~{\tiny(0.34,0.52)} & 0.390~{\tiny(0.30,0.48)} & 0.417 & 0.323 & 0.318 & near-chance clean; parity \\
Gemma$\to$InternVL & POPE & 0.610~{\tiny(0.53,0.69)} & 0.536~{\tiny(0.46,0.62)} & 0.582 & 0.404 & 0.347 & score $n{=}100$ \\
InternVL$\to$Gemma & VQAv2 & 0.752~{\tiny(0.68,0.82)} & 0.616~{\tiny(0.54,0.69)} & 0.608 & 0.434 & 0.289 &  \\
Molmo$\to$LLaVA & VQAv2 & 0.690~{\tiny(0.61,0.76)} & 0.586~{\tiny(0.51,0.66)} & 0.681 & 0.578 & 0.164 &  \\
\midrule
\textbf{mean (12 pairs)} & & \textbf{0.655} & \textbf{0.571} & \textbf{0.601} & \textbf{0.480} & \textbf{0.287} & \\
\textbf{mean, excl.\ under-powered white-box} & & \textbf{0.646} & \textbf{0.558} & \textbf{0.590} & \textbf{0.468} & \textbf{0.246} & \\
\textbf{mean, bit-exact verifier (9)} & & \textbf{0.688} & \textbf{0.589} & \textbf{0.620} & \textbf{0.496} & \textbf{0.223} & \\
\bottomrule
\end{tabular}}

\vspace{3pt}
\noindent{\footnotesize All AUROC are worst-case-oracle attacked verifier confidence. ``mean, excl.\ under-powered white-box'' drops the one under-powered Molmo-verifier white-box pair (Gemma$\to$Molmo), giving a white-box mean of 0.246. ``bit-exact verifier'' drops the three Molmo-verifier pairs, the only non-bit-exact differentiable paths (Table~\ref{tab:app-parity}): white-box 0.223, range 0.06 to 0.35. In the notes column, score $n{=}100$ marks the three InternVL-verifier pairs, ``near-chance clean'' the two Molmo-verifier pairs on POPE, and ``parity'' the Molmo-verifier differentiable-path gap.}

\end{table}

\begin{table}[t]
\centering
\caption{\textbf{Differentiable-path agreement per verifier family.} The maximum absolute gap between each verifier's differentiable path and its official processor on the clean image. Three families are bit-exact; only the Molmo verifier's preprocessing diverges, which under-powers the white-box search on the three Molmo-verifier pairs while leaving every reported score, recomputed with the official processor, unchanged.}
\label{tab:app-parity}
% Appendix (ladder S7): differentiable-verifier parity check per verifier family (report 222).
% Source: span12_ladder_results.json, c2_parity_max_abs per triple (max |differentiable score - official
%   processor score| on the clean image). Grouped by verifier family. Every reported C2 score is recomputed
%   with the OFFICIAL processor, so the gap degrades only the white-box search, never the reported number.
% Requires: \usepackage{booktabs}
\small
\setlength{\tabcolsep}{6pt}
\begin{tabular}{@{}lrcl@{}}
\toprule
verifier family & max parity gap & bit-exact & cells affected \\
\midrule
Gemma-3-12B & 0.000 & yes & none \\
LLaVA-OV-7B & 0.000 & yes & none \\
InternVL3.5-2B & 0.000 & yes & none \\
Molmo2-4B & 0.144 & no & 3 Molmo-verifier pairs (max 0.144) \\
\bottomrule
\end{tabular}

\vspace{3pt}
\noindent{\footnotesize The Molmo verifier's per-item mean gap is $\approx$0.004 against the 0.144 maximum. Every reported verifier score is recomputed with the official processor, so the gap affects only the white-box search, never a reported AUROC.}

\end{table}
\section{Scope and Negative Results}
\label{app:scope}

This appendix records what the paper does not establish. It bounds the threat model with two negative results, reports the defenses we evaluated only partially, and notes an earlier construction that reached the same conclusion as the frontier by a different route.

\subsection{The attack is per-item and white-box, not universal or transferable}
\label{app:scope-threat}

Two negatives fix the threat model. First, the attack does not transfer across models. Table~\ref{tab:app-transfer} takes a perturbation optimized against one model and evaluates it on another. Only Gemma and LLaVA serve as attack sources here, for a reason specific to this experiment: a transferable perturbation must be optimized in raw pixel space through a differentiable reimplementation of the source model's preprocessing, and at the time of this experiment only the Gemma and LLaVA front-ends had been built and validated. InternVL's dynamic tiling and Molmo's patch flattening make that path higher effort and it was not built, so both appear only as targets; this is a limitation of the run, not of the models, since InternVL's forward pass is deterministic (Section~\ref{sec:setup}) and its differentiable path is bit-exact (Table~\ref{tab:app-parity}). The same-model white-box attack removes between $0.44$ and $0.76$ of token-probability discrimination, while the identical image applied to a different target removes at most $0.05$, and usually under $0.03$. The corruption is a property of the specific model attacked.

Second, the effect requires per-item optimization. Table~\ref{tab:app-universal} applies a single fixed perturbation to every item and compares it against an equal-norm random perturbation. The largest gap between the universal perturbation and the random control is $0.024$ across all cells, and for six of seven channels the difference is statistically indistinguishable from zero. The per-image aimed attack removes between $0.41$ and $0.71$ on the same cells, nineteen to seventy-one times the universal-versus-random gap. No single image-independent perturbation carries the effect; it is built anew for each item.

\begin{table}[t]
\centering
\caption{\textbf{Cross-model transfer is weak.} Token-probability AUROC drop when a perturbation optimized against a source model is evaluated on other targets. \emph{white-box} is the same-model attack; \emph{transfer mean/max} are the drop when the same image is applied to the three other targets. Only Gemma and LLaVA had a raw-pixel differentiable front-end at the time of this run and serve as sources.}
\label{tab:app-transfer}
% Appendix (scope / negative results): cross-model transfer negative, old Supplement M.1.
% Source: data/adv/raw_transfer/ (report 088 raw-transfer readout store, evaluation/m_scope_check.py).
%   METRIC = token-probability AUROC drop (tok-drop): clean minus attacked AUROC. white-box = source
%   model attacked by its own aimed image; transfer mean/max = that SAME image applied to the 3 OTHER
%   targets, tok-drop recomputed, mean and max over the 3 non-self targets, worst target named.
% NOT a 4x4 matrix: only Gemma and LLaVA have differentiable reproducible preprocessing and can be
%   SOURCES; InternVL and Molmo are targets only. 18 directed off-diagonal transfers (2 sources x 3
%   sets x 3 targets), summarised as these 6 source-set rows. Reproduces committed cross_model_transfer.tex.
% Requires: \usepackage{booktabs,multirow}
\footnotesize
\setlength{\tabcolsep}{6pt}
\begin{tabular}{@{}llrrrl@{}}
\toprule
source & set & white-box tok drop & transfer mean & transfer max & worst target \\
\midrule
\multirow{3}{*}{Gemma-3-12B} & GQA & $+0.444$ & $+0.006$ & $+0.011$ & Molmo \\
 & VQAv2 & $+0.481$ & $+0.013$ & $+0.023$ & LLaVA \\
 & POPE & $+0.514$ & $+0.024$ & $+0.033$ & LLaVA \\
\midrule
\multirow{3}{*}{LLaVA-OV-7B} & GQA & $+0.760$ & $+0.023$ & $+0.051$ & Molmo \\
 & VQAv2 & $+0.710$ & $+0.023$ & $+0.039$ & InternVL \\
 & POPE & $+0.712$ & $+0.014$ & $+0.044$ & InternVL \\
\bottomrule
\end{tabular}

\vspace{3pt}
\noindent{\footnotesize Metric: token-probability AUROC drop. The same-model (white-box) attack removes $0.444$ to $0.760$; the same image transferred to a different target removes at most $+0.051$ (LLaVA to Molmo on GQA, the largest of the 18 off-diagonal transfers), and most are below $+0.03$.}

\end{table}

\begin{table}[t]
\centering
\caption{\textbf{A fixed universal perturbation does not carry the effect.} Upper block: the largest universal-minus-random gap per cell against the per-image aimed drop. Lower block: per-channel universal and random drops on the least-movable cell, with bootstrap intervals. The universal perturbation is indistinguishable from an equal-norm random one.}
\label{tab:app-universal}
% Appendix (scope / negative results): fixed-universal-perturbation negative, old Supplement M.2.
% Source: data/adv/universal/<model>/ (report 089; U.npy universal delta, U_random.npy equal-norm
%   random control, eval/*.jsonl). A SINGLE image-independent perturbation applied to all items, vs an
%   equal-norm random delta. Reproduces committed universal_all_cells.tex + universal_vs_random.tex.
% Requires: \usepackage{booktabs,multirow}
% (a) All 6 cells run (Gemma, LLaVA x 3 sets): universal-vs-random gap vs the per-image aimed attack.
\footnotesize
\setlength{\tabcolsep}{6pt}
\begin{tabular}{@{}llrrr@{}}
\toprule
model & set & max$|$uni$-$rand$|$ & per-image SAPLMA drop & ratio \\
\midrule
\multirow{3}{*}{Gemma-3-12B} & GQA & 0.023 & 0.433 & 19$\times$ \\
 & VQAv2 & 0.020 & 0.410 & 20$\times$ \\
 & POPE & 0.024 & 0.681 & 28$\times$ \\
\midrule
\multirow{3}{*}{LLaVA-OV-7B} & GQA & 0.010 & 0.708 & 71$\times$ \\
 & VQAv2 & 0.008 & 0.548 & 68$\times$ \\
 & POPE & 0.009 & 0.608 & 68$\times$ \\
\bottomrule
\end{tabular}

\vspace{8pt}
% (b) Per-channel universal drop vs equal-norm random drop, Gemma/GQA (the conservative, least-movable cell).
\noindent{\footnotesize\itshape Universal vs equal-norm random delta, per channel (Gemma/GQA):}\\[2pt]
\footnotesize
\setlength{\tabcolsep}{6pt}
\begin{tabular}{@{}lrrrl@{}}
\toprule
channel & clean & uni drop & rand drop & uni$-$rand [95\% CI] \\
\midrule
token-prob & 0.571 & $-0.031$ & $-0.054$ & $+0.023$ {\scriptsize[$-0.002$,$+0.048$]} \\
P(IK) & 0.790 & $+0.037$ & $+0.028$ & $+0.009$ {\scriptsize[$-0.008$,$+0.025$]} \\
SAPLMA & 0.808 & $+0.013$ & $+0.017$ & $-0.004$ {\scriptsize[$-0.014$,$+0.005$]} \\
verbalized & 0.554 & $-0.007$ & $+0.010$ & $-0.017$ {\scriptsize[$-0.032$,$-0.004$]} \\
self-consist. & 0.552 & $-0.037$ & $-0.059$ & $+0.022$ {\scriptsize[$-0.001$,$+0.046$]} \\
CCPS & 0.741 & $+0.081$ & $+0.082$ & $-0.001$ {\scriptsize[$-0.028$,$+0.026$]} \\
CCPS-D & 0.743 & $+0.036$ & $+0.034$ & $+0.002$ {\scriptsize[$-0.022$,$+0.025$]} \\
\bottomrule
\end{tabular}

\vspace{3pt}
\noindent{\footnotesize drop = clean minus perturbed AUROC. The largest universal-minus-random difference is $0.024$ over all six cells, and six of seven channels have uni$-$rand intervals containing 0. The per-image aimed attack removes $0.410$ to $0.708$ on SAPLMA in the same cells, 19 to 71 times the largest universal-versus-random gap.}

\end{table}

\subsection{Defenses we did not fully evaluate}
\label{app:scope-defenses}

Two defense families are evaluated only partially, and Table~\ref{tab:app-defenses} gives what we measured. For input purification, five inexpensive purifiers are evaluated nonadaptively, without a purifier-aware attack, so their numbers are a lower bound on what an adaptive attacker achieves; even so, the strongest of them, JPEG at quality 50, leaves between $38$ and $76\%$ of the manipulable range while itself corrupting between $3$ and $36\%$ of the model's own answers. No cheap purifier removes the range without a discrimination cost, and none is tested against an adaptive attack.

Randomized smoothing is the more instructive case, because it shows why a defense must be attacked adaptively before it can be called robust. Under a transferred attack, smoothing appears to work: the witnessed displacement collapses and the readout recovers above its floor. An expectation-over-transformation attack that is aware of the smoothing does not collapse; its aimed displacement stays near $0.36$, up to nineteen times the transferred value, and it drives the smoothed readout's attacked AUROC to between $0.10$ and $0.20$, far below the floor, in every cell we test; in each, the aimed displacement exceeds the cell's $\gamma^\star$, so smoothing does not restore the amplitude certificate either. Smoothing does not prevent the collapse once the attacker accounts for it. We did not run the adaptive attack on all four models, so this holds for the cells evaluated and is not a general claim.

\begin{table}[t]
\centering
\caption{\textbf{Defenses not fully evaluated.} Upper block: five input purifiers, evaluated nonadaptively (a lower bound), by the manipulable range they leave and the own-answer corruption they cost. Lower block: randomized smoothing under an adaptive expectation-over-transformation attack, which the transferred attack (\emph{tr.}) masks but the aimed attack (\emph{aim}) does not; the aimed displacement exceeds the cell's $\gamma^\star$ in every tested cell, so smoothing does not restore the amplitude certificate.}
\label{tab:app-defenses}
% Appendix (scope / negative results): defenses not fully evaluated, old Supplement H.3 + H.4.
% (a) Input purification (H.3, family_c): 5 cheap purifiers, NONADAPTIVE transfer through the purifier
%   (no BPDA/adaptive attack), so the numbers are a lower bound. Source: data/adv/rigor/<model>/defense
%   (report 065/102), 3 differentiable-preprocessing models x 3 sets (Molmo excluded).
% (b) Randomized smoothing (H.4, family_d): the adaptive smoothing-EOT attack on the smoothed SAPLMA at
%   sigma {0.08,0.16}. Source: data/smoothing/<model>/eot + smoothing_analysis.json + smoothing_gamma.json
%   (report 115). aim disp / gamma* / frac-exceed spot-checked against live smoothing_gamma.json (PASS).
% Requires: \usepackage{booktabs,multirow}
% (a) Input purification (nonadaptive; P(IK) manipulable-range survival and own-answer corruption).
\small
\setlength{\tabcolsep}{6pt}
\begin{tabular}{@{}lrrrr@{}}
\toprule
defense & P(IK) survival range & median & corruption range & median \\
\midrule
JPEG q75 & 0.49--0.90 & 0.680 & 0.02--0.38 & 0.164 \\
JPEG q50 & 0.38--0.76 & 0.481 & 0.03--0.36 & 0.172 \\
blur & 0.53--0.97 & 0.712 & 0.02--0.17 & 0.105 \\
4-bit & 0.72--1.03 & 0.956 & 0.04--0.27 & 0.131 \\
resize-crop & 0.60--1.05 & 0.781 & 0.02--0.14 & 0.075 \\
\bottomrule
\end{tabular}

\vspace{3pt}
\noindent{\small survival = fraction of the P(IK) manipulable range the purifier leaves ($1.0$ = removes nothing; values above $1.0$ mean the purifier enlarges the range); corruption = own-answer corruption the purifier costs.}

\vspace{8pt}
% (b) Randomized smoothing under the adaptive EOT attack (SAPLMA). tr.=transfer, aim=EOT-aimed.
\noindent{\small\itshape Randomized smoothing under the adaptive EOT attack (SAPLMA):}\\[2pt]
\small
\setlength{\tabcolsep}{4pt}
\begin{tabular}{@{}lll r rr r rr r r c@{}}
\toprule
model & set & $\sigma$ & clean & tr.\ disp & aim disp & ratio & tr.\ adv & aim adv & floor & $\gamma^{\ast}$ & disp$>\gamma^{\ast}$ \\
\midrule
LLaVA & GQA & 0 (base) & 0.851 & -- & 0.240 & -- & -- & 0.143 & 0.673 & 0.125 & yes \\
LLaVA & GQA & 0.08 & 0.836 & 0.079 & 0.361 & 4.6$\times$ & 0.728 & 0.176 & 0.673 & 0.110 & yes \\
LLaVA & GQA & 0.16 & 0.808 & 0.019 & 0.361 & 19.2$\times$ & 0.784 & 0.200 & 0.673 & 0.085 & yes \\
LLaVA & POPE & 0.08 & 0.693 & 0.078 & 0.182 & 2.3$\times$ & 0.513 & 0.104 & 0.597 & 0.015 & yes \\
LLaVA & POPE & 0.16 & 0.671 & 0.022 & 0.258 & 11.8$\times$ & 0.637 & 0.123 & 0.597 & 0.015 & yes \\
InternVL & GQA & 0.16 & 0.779 & 0.020 & 0.368 & 18.8$\times$ & 0.755 & 0.175 & 0.642 & 0.085 & yes \\
Molmo & -- & -- & not run & -- & -- & -- & -- & -- & -- & -- & -- \\
Gemma & -- & -- & not run & -- & -- & -- & -- & -- & -- & -- & -- \\
\bottomrule
\end{tabular}

\vspace{3pt}
\noindent{\small Molmo and Gemma were not run under the aimed EOT attack.}

\end{table}

\subsection{An earlier construction reached the same conclusion}
\label{app:scope-nullsubspace}

Before the adversarial-training frontier of Appendix~\ref{app:frontier}, we tested whether a readout confined to the directions the attack moves least could be both informative and robust. We built a probe on the trailing subspace of an empirical displacement null, the directions in which the attack produces the least movement, and attacked it. The construction found no readout that was both: a probe restricted to the least-moved directions loses the clean discrimination that made it useful, the same trade-off the trained frontier exhibits. We record it here because it shows the frontier's conclusion does not depend on the training procedure that produced it; a directly constructed low-movement readout fails in the same way.

\end{document}